\documentclass[qqa4lart]{article}

\usepackage[body={5in,9in}]{geometry}

\usepackage{amsmath,amsfonts}

\def\stackunder#1#2{\mathrel{\mathop{#2}\limits_{#1}}}
\long\def\TeXButton#1#2{#2}%
\def\QTR#1#2{{\csname#1\endcsname #2}}
\def\proof{\paragraph{Proof. }}
\def\endproof{\mbox{\ $\Box$}}

\def\cents{\hbox{\rm\rlap/c}}%

\begin{document}

\author{Zhaohui Zhu$^1$\thanks{%
Corresponding author. E-mail address: Zhaohui@nuaa.edu.cn,
zhaohui.nuaa@gmail.com \quad (Zhaohui Zhu)}\quad Yan Zhang$^1$\quad Jinjin
Zhang$^2$ \\
1 College of Computer Science \\
Nanjing University of Aeronautics and Astronautics\\
Nanjing, P. R. China, 210016\\
2 School of Information Science\\
Nanjing Audit University\\
Nanjing, P. R. China, 211815}
\title{Merging Process Algebra and Action-based Computation Tree Logic\thanks{%
This work received financial support of the National Natural Science of
China(No. 60973045) and Fok Ying-Tung Education Foundation. }}
\maketitle

\begin{abstract}
Process algebra and temporal logic are two popular paradigms for the
specification, verification and systematic development of reactive and
concurrent systems. These two approaches take different standpoint for
looking at specifications and verifications, and offer complementary
advantages. In order to mix algebraic and logic styles of specification in a
uniform framework, the notion of a logic labelled transition system (LLTS)
has been presented and explored by L\"uttgen and Vogler. This paper intends
to propose a LLTS-oriented process calculus which, in addition to usual
process-algebraic operators, involves logic connectives ($conjunction$ and $%
disjunction$) and standard temporal operators ($always$ and $unless$). This
calculus preserves usual properties of these logic operators, allows one to
freely mix operational and logic operators, and supports compositional
reasoning. Moreover, the links between this calculus and Action-based
Computation Tree Logic (ACTL) including characteristic formulae of process
terms, characteristic processes of ACTL formulae and Galois connection are
explored.

{\bf Key Words} Process Calculus\quad Action-based Computation Tree
Logic\quad Ready Simulation\quad Logic Labelled Transition System\quad
Galois Connection
\end{abstract}

\section{Introduction}

\subsection{Two popular paradigms in formal method}

The dominant approaches for the specification, verification and systematic
development of reactive and concurrent systems are based on either states or
actions. For state-based approaches, an execution of a system is viewed as a
sequence of states, while another approach regards an execution as a
sequence of actions.

State-based approaches devote themselves to specifying and verifying
abstract properties of systems, which often involve formalisms in logic
style. Since the seminal work of Pnueli [53], logics have been adopted to
serve as useful tools for specifying and verifying of reactive and
concurrent systems. In such framework, a specification is expressed by a set
of formulae in some logic system and verification is a deductive or
model-checking activity.

Action-based approaches put attention to behavior of systems, which have
tended to use formalisms in algebraic style. These formalisms are referred
to as process algebra or process calculus [45, 36, 35, 10]. In such
paradigm, a specification and its implementation usually are formulated by
the same notations, which are terms (expressions) of a formal language built
from a number of operators, and the underlying semantics are often assigned
operationally. Intuitively, a specification describes the desired high-level
behavior, and an implementation provides lower-level details indicating how
this behavior is to be achieved. The verification amounts to compare terms,
which is often referred to as $implementation$ $verification$ or $%
equivalence $ $checking$ [2]. The comparison of a specification to an
implementation is based on behavioral relations. Such relations depend on
particular observation criterions, and are typically equivalences (or
preorders), which capture a notion of ``having the same observation''
(respectively, ``refinement''). At the present time, due to lack of
consensus on what constitutes an appropriate notion of observable behavior,
a variety of observation criterions and behavioral relations have been
proposed [25]. The correctness of an implementation may be verified in a
proof--theory oriented manner or in a semantics oriented manner. The former
is rooted in an axiomatization of the behavioral relation, while the later
appeals to coinduction technology which is considered as one of the most
important contributions of concurrency theory to computer science [55].

Since logic and algebraic frameworks take different standpoint for looking
at specifications and verifications, they offer complementary advantages:

On the logic side, there exist a number of logic systems, e.g., Linear
temporal logic [53], Computation tree logic [17], $\mu -$calculus [38] and
so on, in which the most common reasonable property of concurrent systems,
such as invariance (safety), liveness, etc., can be formulated without
referring operational details (see, e.g., [16, 57]). Moreover, one of
inherent advantage of logic approach is that it is ability to deal with
partial specifications: one can establish that a given system realizes a
particular property without involving its full specification. On the other
hand, the inclusion of classes of models is a natural refinement preorder on
logic specifications, hence refining a logic specification amounts to enrich
original one by adding new formulas consistently. However, logic approach
has been criticized for being global, non-modular and non-compositional. In
other words, we often are required to consider a given system as a whole
whenever formulating and verifying a logic property. For instance, it always
lacks a natural way to combine temporal properties, which are required
separately for subsystem P$_1$ and P$_2$, into a temporal specification for P%
$_1$ $\parallel $ P$_2$. Such deficiency has been indicated by Pnueli in
[53] where temporal logic is described as being endogenous, that is,
assuming the complete program as fixed context. Summarizing, a variety of
logics may serve as powerful tools for expressing and verifying a wide
spectrum of properties of concurrent systems, but, due to their global
perspective and abstract nature, it is difficult for them to describe the
link between the structure of implementation and that of specification, and
hence logic approaches often give little support for systematic development
of concurrent systems.

On the algebraic side, since systems are represented by terms in some
algebras, complex systems may be built up from existent systems using
algebraic operators. Moreover, the observable behavior of the complex system
does not change if an subsystem is replaced by one with the same behavior,
which is granted by the fact that behavioral relations considered in process
algebras are often required to be compatible with process operators, in
other words, these relations are (pre)congruence over terms. These features
cause the main advantage of algebraic paradigm, that is, it always supports
compositional constructing and reasoning. Such compositionality brings us
advantages in developing systems, such as, supporting modular design and
verification, avoiding verifying the whole system from scratch when its
parts are modified, allowing reusability of proofs and so on [5]. Thus
algebraic approaches offer significant support for rigorous systematic
development of reactive and concurrent systems. However, since algebraic
approaches specify a system by means of prescribing in detail how the system
should behave, it is often difficult for them to describe abstract
properties of systems, which is a major disadvantage of such approaches.

\subsection{Connections between process algebras and logics}

It is natural to wonder what the connection between the algebraic approach
and logic approach is. Based on structural operational semantics (SOS) in
Plotkin-style, terms in process algebras can be ``transformed'' into
labelled transition systems. The latter may be viewed as models (in the
model-theoretic sense) for suitable logic language. Hence this induces the
satisfiability relation $\models $ between process terms and formulas. Given
such satisfiability relation, three connections between a process algebra
and a logic deserve special mention, which are considered by Pnueli in [54]
and recalled in the following. Let ${\rm P}$ be a process algebra equipped
with a behavioral relation $\bowtie $, and ${\rm L}$ a logic language
associated with a satisfiability relation $\models $.

\begin{itemize}
\item  {\bf Adequacy of (}{\rm L}$,\models $){\rm \ }w.r.t ({\rm P}, $%
\bowtie $)
\end{itemize}

The logic {\rm L }is said to be adequate w.r.t ({\rm P}, $\bowtie $) if for
any process $p$ and $q$,

either

\begin{center}
$p\bowtie q$ iff $\forall \alpha \in ${\rm L}($p\models \alpha $ $%
\Leftrightarrow q\models \alpha $ ) (if $\bowtie $ is an equivalence)
\end{center}

or

\begin{center}
$p\bowtie q$ iff $\forall \alpha \in ${\rm L}($q\models \alpha $ $%
\Rightarrow p\models \alpha $ ) (if $\bowtie $ is a preorder)
\end{center}

This notion is considered by Hennesy and Milner in [34], where they prove
that Hennesy-Milner logic (HML) is adequate w.r.t bisimilarity for image
finite CCS terms. It is one of key evens that make Milner think that CCS is
definitely interesting enough\footnote{%
See: http://www.sussex.ac.uk/Users/mfb21/interviews/milner/}. Following
their work, the literature on concurrency theory offers a wealth of modal
characterizations for various behavioral relations. A good overview on this
subject may be found in [9]. In the realm of modal logic, more generalized
results concerning Hennesy-Milner property (class) have been established [8,
11, 27, 28, 37]. Recently, such issue is also considered in depth in the
framework of coalgebras\footnote{%
In this realm, a coalgebraic modal logic for $F-$coalgebras is said to be
adequate if behavioral equivalence implies logical equivalence, and it is
said to be expressive if the converse holds.} (see, e.g., [47, 51]).

As pointed out by Pnueli in [54], the requirement of adequacy is the weakest
one of compatibility between a process algebra and a logic. A symptom of its
weakness is that the same logic may be adequate for some process languages
with very different expressivity [54]. For instance, HML is adequate w.r.t
bisimilarity for both CCS and the fragment of CCS consisting of
recursion-free terms. Moreover, the Hennesy-Milner characterization is less
useful if one intend to check the equivalence of process terms using model
checking [2].

Stronger associations between processes algebras and logic systems involve
translating between them: characterizing a given process in terms of logic
formulae, and graphical representing a given logic formula by means of
process terms. Next we recall them in turns.

\begin{itemize}
\item  {\bf Expressivity of (}{\rm L}$,\models $){\rm \ }w.r.t ({\rm P}, $%
\bowtie $)
\end{itemize}

A stronger compatibility requirement involves expressivity. The logic {\rm L
}is said to be expressivity w.r.t ({\rm P}, $\bowtie $) if for any process $%
p $ in {\rm P}, there exists a formula $L(p)$ $\in {\rm L}$ such that

{\bf (E1)} $q\models L(p)$ iff $q\bowtie p$ for any process $q$ in {\rm P},
and

{\bf (E2)} $p\models \varphi $ iff $\models L(p)\rightarrow \varphi $ for
any formula $\varphi $ in ${\rm L}$.

Clearly, if such formula for a process can be algorithmically constructed,
implementation verification can be reduced to model checking according to
(E1), and the verification of an assertion $p\models \varphi $ can be
transformed into the validity problem within ${\rm L}$ by (E2). Graf and
Sifakis were probably the first to develop logics which are expressive for
process algebras. In [29], they present Synchronization Tree Logic (STL, for
short) for a process algebra with a congruence relation $\approx $. STL
contains process terms as formulae, and its semantic is defined so that both
(E1) and (E2) hold with the function $L=\lambda x.x$.

Given a process $p$, a formula $\phi _p$ is said to be a characteristic
formula of $p$ if it satisfies (E1). Such notion also provides a very
elegant link between process algebra and logic, and between implementation
verification and model checking [2]. Graf and Sifakis provide a method of
constructing characteristic formula modulo observational congruence for any
recursion-free CCS term [30]. Hitherto, over different structures, e.g.,
finite LTS, Kripke structures, time automata and so on, a number of examples
of characteristic-formula constructions for various behavioral relations
have been reported in the literature [4, 15, 20, 23, 39, 40, 46, 56, 59].
The underlying structures of these constructions are identical, that is,
characteristic formulae often are defined as fixed points of some functions.
Recently, ground on this phenomenon, L.Aceto et al. offer a general
framework for the constructions of characteristic formulae [2, 3].

\begin{itemize}
\item  {\bf Expressivity of }({\rm P}, $\bowtie $){\rm \ }w.r.t {\bf (}{\rm L%
}$,\models $)
\end{itemize}

Another stronger association between process algebras and logics involves an
inverse translation, which associates with each formula $\varphi \in {\rm L}$
a set $P(\varphi )$ that consists of all the processes satisfying $\varphi $%
. A process language is said to be expressive for ${\rm L}$ if such
translation is given in a syntactic manner. In order to obtain such
expressivity, additional operators that construct process sets are often
needed.

In a classic paper [14], Boudol and Larsen offer a process language $\Theta $
and a translation $\zeta (.)$ in a syntactic manner, and show that any HML
formula $\phi $ is representable by a finite set $\zeta (\phi )$ of terms in
$\Theta $. In particular, $\zeta (\phi )$ can be reduced to a singleton, say
$\{\phi ^{*}\}$, if and only if the given formula $\phi $ is consistent and
prime. Moreover, such term $\phi ^{*}$ satisfies the property below

\begin{center}
$t\models \phi \Leftrightarrow \phi ^{*}\sqsubseteq t$ for any term $t$ in $%
\Theta $.
\end{center}

Here $\sqsubseteq $ is a behavioral relation considered in [14]. In such
situation, the model checking problem can be reduced to implementation
verification. Clearly, $\phi ^{*}$ plays an analogous role of characteristic
formula in a contrary way. In fact, characteristic-formula construction and $%
\zeta (.)$ indeed induce a Galois connection between $(\Theta ,\sqsubseteq )$
and the set of consistent prime formulae augment with some preorder [14]. In
[1], L.Aceto et al. address the same issue, and show that, modulo the
covariant-contravariant simulation preorder, any consistent and prime
formula in the covariant-contravariant modal logic also admits a
representation by means of process terms.

\subsection{Background and motivation}

As mentioned above, logic approaches and algebraic approaches offer
complementary advantages when specifying systems. The former is good at
specifying abstract properties of systems, while the latter is applicable if
we intend to specify the system itself through describing its behavioral and
structural properties.

Impelled by taking advantage of both approaches when designing systems,
so-called heterogeneous specifications have been proposed, which uniformly
integrate these two specification styles. Among them, based on B\"uchi
automata and LTS augmented with a predicate, Cleaveland and L\"uttgen
provide a semantic framework for heterogenous system design [18, 19], where
must-testing preorder offered by Nicola and Hennessy [48] is adopted to
describe refinement relation. In addition to usual operational operators,
such framework also involves logic connectives. However, since must-testing
preorder is not a precongruence in such situation, this setting does not
support compositional reasoning. Moreover, the logic connective conjunction
in this framework lacks the desired property that $r$ is an implementation
of the specification $p\wedge q$ if and only if $r$ implements both $p$ and $%
q$.

Recently, L\"uttgen and Vogler introduce the notion of a Logic LTS (LLTS,
for short), which combines operational and logic styles of specification in
one unified framework [42, 43]. In order to handle logic conjunctions of
specifications, LLTS involves consideration of inconsistencies, which,
compared with usual LTS, is one distinguishing feature of it. Two kinds of
constructors over LLTSs are considered in [42, 43]: operational
constructors, e.g., CSP-style parallel composition, hiding and so on, and
logic connectives including conjunction and disjunction. Such framework
allows one to freely mix these two kinds of constructors, while most early
theories couple them loosely and do not allow for mixed specification.
Moreover, the drawbacks in [19, 18] mentioned above have been remedied by
adopting ready-tree semantics [42]. In order to support compositional
reasoning in the presence of the parallel constructor, a variant of the
usual notion of ready simulation is employed to characterize the refinement
relation [43]. Some standard modal operators in temporal logics, such as $%
always$ and $unless$, are also integrated into this framework [44].

Along the direction suggested by L\"uttgen and Vogler in [43], we propose a
process calculus called CLL in [60], which reconstructs their setting in
process algebraic style. In addition to prefix $\alpha .()$, external choice
$\Box $ and parallel operator $\parallel _A$, CLL contains logic operators $%
\wedge $ and $\vee $ over process terms, which correspond to the
constructors conjunction and disjunction over LLTSs respectively. The
language CLL is explored in detail from two different but equivalent angles.
Based on behavioral view, the notion of ready simulation is adopted to
formalize the refinement relation, and the behavioral theory is developed.
Based on proof-theoretic view, a sound and ground-complete axiomatic system
for CLL is provided. In effect, it gives an axiomatization of ready
simulation in the presence of logic operators.

However, due to lack of modal operators, CLL still does not afford
describing abstract properties of systems. This paper intends to enrich CLL
with temporal operators $always$ and $unless$ by two distinct approaches.
One approach is to introduce nonstandard process-algebraic operators $\sharp
$, $\varpi $, $\bigtriangleup $ and $\odot $ to capture L\"uttgen and
Vogler's constructions in [44] directly. The other is to provide graphical
representing of temporal operators $always$ and $unless$ in recursive
manner. The latter is independent of L\"uttgen and Vogler's constructions
but depends on the greatest fixed-point characterization obtained in this
paper. Moreover, the connections between the resulting calculus (that we
call CLLT) and ACTL [49] are explored from angles recalled in the preceding
subsection. These connections include characteristic formulae of process
terms, characteristic processes of formulae and Galois connection.

The remainder of this paper is organized as follows. The next section
presents some preliminaries. In Section~3, SOS rules of CLLT are introduced,
the existence and uniqueness of stable transition model for CLLT is
demonstrated, and a few of basic properties of the LTS associated with CLLT
are given. Section~4 and 5 are devoted to the study of temporal operators $%
always$ $\sharp $ and $unless$ $\varpi $ respectively. Section 6 establishes
a fixed-point characterization of the operator $\varpi $. Section 7 provides
a recursive approach to dealing with the temporal operator $\varpi $.
Section~8 explores the links between CLLT and ACTL. Finally, a brief
conclusion and discussion are given in Section~9.

\section{Preliminaries}

In this section, we shall set up notation and terminology and briefly sketch
the process calculus CLL.

\subsection{Logic LTS}

This subsection will introduce some useful notations and recall the notion
of a Logic LTS. Here we do not give examples motivating and illustrating the
use of such notion, which may be found in [43, 44].

Let $Act$ be a set of visible actions ranged over by letters $a$, $b$, etc.,
and let $Act_\tau $ denote $Act\cup \{\tau \}$ ranged over by $\alpha $ and $%
\beta $, where $\tau $ represents invisible actions. An LTS with a predicate
$F$ is a quadruple $(P,Act_\tau ,\rightarrow ,F)$, where $P$ is a set of
states, $\rightarrow \subseteq P\times Act_\tau \times P$ is the transition
relation and $F\subseteq P$. As usual, we write $p\stackrel{\alpha }{%
\rightarrow }q$ if $(p,\alpha ,q)\in \rightarrow $. A state $q$ is said to
be an $\alpha $-derivative of $p$ if $p\stackrel{\alpha }{\rightarrow }q$.
The assertion $p\stackrel{\alpha }{\rightarrow }$ holds if $p$ has a $\alpha
$-derivative, otherwise $p\stackrel{\alpha }{\not \rightarrow }$ holds.
Given a state $p$, the ready set of $p$, denoted by $I(p)$, is defined as $%
\{\alpha \in Act_\tau :p\stackrel{\alpha }{\rightarrow }\}$. A state $p$ is
said to be stable if it can not engage in any $\tau $-transition, i.e., $p%
\stackrel{\tau }{\not \rightarrow }$. Some useful decorated transition
relations are listed below.

$p\stackrel{\alpha }{\rightarrow }_Fq$ iff $p\stackrel{\alpha }{\rightarrow }%
q$ and $p,q\notin F$.

$p\stackrel{\varepsilon }{\Rightarrow }q$ iff $p(\stackrel{\tau }{%
\rightarrow })^{*}q$, where $(\stackrel{\tau }{\rightarrow })^{*}$ is the
transitive and reflexive closure of $\stackrel{\tau }{\rightarrow }$.

$p\stackrel{\alpha }{\Rightarrow }q$ iff $p\stackrel{\varepsilon }{%
\Rightarrow }r\stackrel{\alpha }{\rightarrow }s\stackrel{\varepsilon }{%
\Rightarrow }q$ for some $r,s\in P$.

$p\stackrel{\varepsilon }{\Rightarrow }|q$ (or, $p\stackrel{\alpha }{%
\Rightarrow }|q$) iff $p\stackrel{\varepsilon }{\Rightarrow }q\stackrel{\tau
}{\not \rightarrow }$ ($p\stackrel{\alpha }{\Rightarrow }q\stackrel{\tau }{%
\not \rightarrow }$, respectively).

$p\stackrel{\varepsilon }{\Rightarrow }_Fq$ iff there exists a sequence of $%
\tau -$labelled transitions from $p$ to $q$ such that all states along this
sequence, including $p$ and $q$, are not in $F$. The decorated transition $p%
\stackrel{\alpha }{\Rightarrow }_Fq$ may be defined similarly.

$p\stackrel{\varepsilon }{\Rightarrow }_F|q$ (or, $p\stackrel{\alpha }{%
\Rightarrow }_F|q$) iff $p\stackrel{\varepsilon }{\Rightarrow }_Fq\stackrel{%
\tau }{\not \rightarrow }$ ($p\stackrel{\alpha }{\Rightarrow }_Fq\stackrel{%
\tau }{\not \rightarrow }$, respectively).

\qquad

{\bf Remark 2.1} Notice that some notations above are slightly different
from ones adopted by L\"uttgen and Vogler. In [43, 44], the notation $p%
\stackrel{\varepsilon }{\Rightarrow }|q$ (or, $p\stackrel{\alpha }{%
\Rightarrow }|q$) has the same meaning as $p\stackrel{\varepsilon }{%
\Rightarrow }_F|q$ (respectively, $p\stackrel{\alpha }{\Rightarrow }_F|q$)
in this paper.

\quad

{\bf Definition 2.1 (}[43]{\bf )} An LTS $(P,Act_\tau ,\rightarrow ,F)$ is
said to be a LLTS if, for each $p\in P$,

\begin{description}
\item  {\bf (LTS1)} $p\in F$ if $\exists \alpha \in I(p)\forall q\in P(p%
\stackrel{\alpha }{\rightarrow }q\;\text{implies}\;q\in F)$,

\item  {\bf (LTS2)} $p\in F$ if $\neg \exists q\in P.p\stackrel{\varepsilon
}{\Rightarrow }_F|q$.
\end{description}

A LLTS $(P,Act_\tau ,\rightarrow ,F)$ is said to be $\tau -pure$ if, for
each $p\in P$, $p\stackrel{\tau }{\rightarrow }$ implies $\neg \exists a\in
Act.\;p\stackrel{a}{\rightarrow }$. Hence, for any state $p$ in a $\tau $%
-pure LTS, either $I(p)=\{\tau \}$ or $I(p)\subseteq Act$.

\qquad

Here the predicate $F$ is used to denote the set of all inconsistent states.
Compared with usual LTSs, it is one distinguishing feature of LLTS that it
involves consideration of inconsistencies. Roughly speaking, the motivation
behind such consideration lies in dealing with inconsistencies caused by
conjunctive composition. In the sequel, we shall use the phrase ``{\it %
inconsistency} {\it predicate}'' to refer to $F$. The condition (LTS1)
formalizes the backward propagation of inconsistencies, and (LTS2) captures
the intuition that divergence (i.e., infinite sequences of $\tau $%
-transitions) should be viewed as catastrophic. For more intuitive idea
about inconsistency and motivation behind (LTS1) and (LTS2), the reader may
refer to [43, 44].

\subsection{ A variant of ready simulation}

In [43, 44], the notion of ready simulation below is adopted to formalize
the refinement relation, which is a modified version of the usual notion of
ready simulation (see, e.g., [25]).

\qquad

{\bf Definition 2.2} ([43, 44]) Given a LLTS $(P,Act_\tau ,\rightarrow ,F)$,
a relation $R\subseteq P\times P$ is said to be a stable ready simulation
relation if, for any $(t,s)\in R$ and $a\in Act$, the following conditions
hold

\begin{description}
\item  (RS1) Both $t$ and $s$ are stable;

\item  (RS2) $t\notin F$ implies $s\notin F$;

\item  (RS3) $t\stackrel{a}{\Rightarrow }_F|u$ implies $\exists v.s\stackrel{%
a}{\Rightarrow }_F|v\ $and\ $(u,v)\in R$;

\item  (RS4) $t\notin F$ implies $I(t)=I(s)$.
\end{description}

We say that $t$ is stable ready simulated by $s$, in symbols $t\stackunder{%
\sim }{\sqsubset }_{RS}s$, if there exists a stable ready simulation
relation $R$ with $(t,s)\in R$. Further, $t$ is said to be ready simulated
by $s$, written $t\sqsubseteq _{RS}s$, if

\begin{center}
$\forall u(t\stackrel{\epsilon }{\Rightarrow }_F|u$ implies $\exists v.s%
\stackrel{\epsilon }{\Rightarrow }_F|v\ $and$\ u\stackunder{\sim }{\sqsubset
}_{RS}v)$.
\end{center}

It is easy to see that both $\stackunder{\sim }{\sqsubset }_{RS}$ and $%
\sqsubseteq _{RS}$ are pre-order (i.e., reflexive and transitive). The
equivalence relations induced by them are denoted by $\approx _{RS}$ and $%
=_{RS}$, respectively, that is

\begin{center}
$\approx _{RS}$ $\stackrel{\bigtriangleup }{=}$ $\stackunder{\sim }{%
\sqsubset }_{RS}\bigcap $ $(\stackunder{\sim }{\sqsubset }_{RS})^{-1}$ and $%
=_{RS}$ $\stackrel{\bigtriangleup }{=}$ $\sqsubseteq _{RS}\bigcap $ $%
(\sqsubseteq _{RS})^{-1}$.
\end{center}

The notion of ready simulation presented in Def. 2.2 is a central notion in
[43, 44, 60] and this paper. It is natural to wonder why such notion is
adopted to formalize the refinement relation. From our point of view,
whenever we try to mix process-algebraic and logic styles of specification
in a uniform framework, the requirements below should be met by such
framework.

\begin{itemize}
\item  It is well known that parallel composition and conjunction are two
fundamental ways of combining specifications: the former is adopted to
structurally compose two or more subsystems, and the latter is used to
combine specifications expressed by logic formulae. Thus such uniform
framework should include these two constructors.

\item  Since such framework involves specifications in logic style, we
should take account of the consistency of specifications. A trivial and
desired property is that an inconsistent specification can only be refined
by inconsistent ones.

\item  Such uniform framework should support compositional reasoning. Hence
the behavior relation adopted in this framework need to be (pre)congruent
w.r.t all operators within it.
\end{itemize}

Consequently, the result below reveals that it is reasonable to adopt the
notion of ready simulation in Def. 2.2 as behavior relation when we intend
to explore such uniform framework.

\qquad

{\bf Theorem 2.1} ([43]) The ready simulation $\sqsubseteq _{RS}$ exactly is
the largest precongruence $\preceq $ w.r.t parallel composition and
conjunction such that $p\preceq q$ and $q\in F$ implies $p\in F$ .

\TeXButton{Proof}{\proof}See Theorem 21 in [43]. \TeXButton{End Proof}
{\endproof}

\subsection{Transition system specifications}

Structural Operational Semantics (SOS) is a logic method of giving
operational semantics, which provides a syntax oriented view on operational
semantics [52]. Transition System Specifications (TSSs), as presented by
Groote and Vaandrager in [31], are formalizations of SOS. This subsection
recalls basic concepts related to TSS. Further information on this issue may
be found in [9, 13, 31].

Given an infinite set $V$ of variables and a signature $\Sigma $, we assume
that the resulting notions of term, closed (ground) terms, substitution and
closed (ground) substitution are already familiar to the reader. Following
standard usage, the set of all $\Sigma $-terms (or, $\Sigma $-closed terms)
over $V$ is denoted by $T(\Sigma ,V)$ ($T(\Sigma )$, respectively).

A TSS is a quadruple $\Gamma =(\Sigma ,A,\Lambda ,\Xi )$, where $\Sigma $ is
a signature, $A$ is a set of labels, $\Lambda $ is a set of predicate
symbols and $\Xi $ is a set of rules. Positive literals are all expressions
of the form $t\stackrel{\alpha }{\rightarrow }s$ or $tP$, while negative
literals are all expressions of the form $t\stackrel{\alpha }{\not
\rightarrow }$ or $t\neg P$, where $t,s\in T(\Sigma ,V)$, $\alpha \in A$ and
$P\in \Lambda $. A rule $r\in \Xi $ has the form like $\frac{prem(r)}{conc(r)%
}$, where $prem(r)$, the premises of the rule $r$, is a set of (positive or
negative) literals, and $conc(r)$, the conclusion of the rule $r$, is a
positive literal. Given a rule $r$, the set of positive premises (or,
negative premises) of $r$ is denoted by $pprem(r)$ (respectively, $nprem(r)$
), moreover, $r$ is said to be positive if $nprem(r)=\emptyset $. A TSS is
said to be positive if it has only positive rules. Given a substitution $%
\sigma $ and a rule $r\in \Xi $, $r\sigma $ is the rule obtained from $r$ by
replacing each variable in $r$ by its $\sigma $-image, that is, $r\sigma =%
\frac{\{\varphi \sigma |\varphi \in prem(r)\}}{conc(r)\sigma }$. Moreover,
if $\sigma $ is closed then $r\sigma $ is said to be a ground instance of $r$%
.

\qquad

{\bf Definition 2.3 }(Proof in Positive TSS) Let $\Gamma =(\Sigma ,A,\Lambda
,\Xi )$ be a positive TSS. A proof of a closed positive literal $\psi $ from
$\Gamma $ is a well-founded, upwardly branching tree, whose nodes are
labelled by closed literals, such that

\begin{itemize}
\item  the root is labelled with $\psi $,

\item  if $\chi $ is the label of a node $q$ and $\{\chi _i:i\in I\}$ is the
set of labels of the nodes directly above $q$, then there is a rule $%
\{\varphi _i:i\in I\}/\varphi $ in $\Xi $ and a closed substitution $\sigma $
such that $\chi =\varphi \sigma $ and $\chi _i=\varphi _i\sigma $ for each $%
i\in I$.
\end{itemize}

If a proof of $\psi $ from $\Gamma $ exists, then $\psi $ is said to be
provable from $\Gamma $, in symbols $\Gamma \vdash \psi $.

\qquad

Given a TSS $\Gamma =(\Sigma ,A,\Lambda ,\Xi )$, a transition model $M$ is a
subset of $Tr(\Sigma ,A)\cup Pred(\Sigma ,\Lambda )$, where $Tr(\Sigma
,A)=T(\Sigma )\times A\times T(\Sigma )$ and $Pred(\Sigma ,\Lambda
)=T(\Sigma )\times \Lambda $. Following standard usage, elements $(t,a,s)$
and $(t,P)$ in $M$ are written as $t\stackrel{a}{\rightarrow }s$ and $tP$
respectively. A positive closed literal $\psi $ is said to be valid in $M$,
in symbols $M\models \psi $, if $\psi \in M$. A negative closed literal $t%
\stackrel{a}{\not \rightarrow }$ (or, $t\neg P$) holds in $M$, in symbols $%
M\models t\stackrel{a}{\not \rightarrow }$ ($M\models t\neg P$,
respectively), if there is no $s$ such that $t\stackrel{a}{\rightarrow }s\in
M$ ($tP\notin M$, respectively). As usual, for a set $\Psi $ of closed
literals, $M\models \Psi $ iff $M\models \psi $ for each $\psi \in \Psi $.

\qquad

{\bf Definition 2.4} Let $\Gamma =(\Sigma ,A,\Lambda ,\Xi )$ be a TSS and $M$
a transition model. $M$ is said to be a model of $\Gamma $ if, for each $%
r\in \Xi $ and $\sigma :V\rightarrow T(\Sigma )$ such that $M\models
prem(r\sigma )$, we have $M\models conc(r\sigma )$. $M$ is said to be
supported by $\Gamma $ if, for each $\psi \in M$, there exist $r\in \Xi $
and $\sigma :V\rightarrow T(\Sigma )$ such that $M\models prem(r\sigma )$
and $conc(r\sigma )=\psi $. $M$ is said to be a supported model of $\Gamma $
if $M$ is supported by $\Gamma $ and $M$ is a model of $\Gamma $.

\qquad

A natural and simple method of describing the operational nature of
processes is in terms of LTSs. Given a TSS, an important problem is how to
associate LTSs with process terms. For positive TSS, the answer is
straightforward. It is well known that every positive TSS $\Gamma $ has a
least transition model, which exactly consists of provable transitions of $%
\Gamma $ and induces a LTS naturally. However, since it is not immediately
clear what can be considered as a ``proof'' for a negative formula, it is
much less trivial to associate a model with a TSS containing negative
premises [32]. The first generic answer to this question is formulated in
[32, 12], where the above notion of supported model is introduced. However,
this notion doesn't always work well. Several alternatives have been
proposed, and a good overview on this issue is provided in [26]. In the
following, we recall the notions of stratification and stable transition
model, which play an important role in this field.

\quad

{\bf Definition 2.5 }(Stratification [13]) Let $\Gamma =(\Sigma ,A,\Lambda
,\Xi )$ be a TSS and $\zeta $ an ordinal number. A function $S:Tr(\Sigma
,A)\cup Pred(\Sigma ,\Lambda )\rightarrow \zeta $ is said to be a
stratification of $\Gamma $ if, for every rule $r\in \Xi $ and every
substitution $\sigma :V\longrightarrow T(\Sigma )$, the following conditions
hold.

\begin{itemize}
\item  $S(\psi )\leq S(conc(r\sigma ))$ for each $\psi \in pprem(r\sigma )$,

\item  $S(tP)<S(conc(r\sigma ))$ for each $t\neg P\in nprem(r\sigma )$, and

\item  $S(t\stackrel{\alpha }{\rightarrow }s)<S(conc(r\sigma ))$ for each $%
s\in T(\Sigma )$ and $t\stackrel{\alpha }{\not \rightarrow }\in
nprem(r\sigma )$.
\end{itemize}

A TSS is said to be stratified iff there exists a stratification function
for it.

\qquad

{\bf Definition 2.6} (Stable Transition Model [13, 24]) Let $\Gamma =(\Sigma
,A,\Lambda ,\Xi )$ be a TSS and $M$ a transition model. $M$ is said to be a
stable transition model for $\Gamma $ if

\begin{center}
$M=M_{Strip(\Gamma \text{, }M)}$,
\end{center}

where $Strip(\Gamma ,M)$ is the TSS $(\Sigma ,A,\Lambda ,Strip(\Xi ,M))$ with

\begin{center}
$Strip(\Xi ,M)\stackrel{\bigtriangleup }{=}\left\{ \frac{pprem(r)}{conc(r)}%
:\left.
\begin{array}{c}
r\;
\text{is a ground instance of some rule in }\Xi \\ \text{and}\;M\models
nprem(r)
\end{array}
\right. \right\} $,
\end{center}

and $M_{Strip(\Gamma \text{, }M)}$ is the least transition model of the
positive TSS $Strip(\Gamma ,M)$.

\qquad

As is well known, stable models are supported models, and each stratified
TSS $\Gamma $ has a unique stable model [13], moreover, such stable model
does not depend on particular stratification function [32].

\subsection{Process calculus CLL}

For the convenience of the reader this subsection will briefly sketch the
process calculus CLL proposed in [60], thus making our exposition
self-contained. The processes in CLL are given by BNF below, where $\alpha
\in Act_\tau $ and $A\subseteq Act$.

\begin{center}
$p::=0$ $\left| \text{ }\bot \text{ }\right| $ $(\alpha .p)$ $\left| \text{ }%
(p\Box p)\text{ }\right| $ $(p||_Ap)$ $\left| \text{ }(p\vee p)\text{ }%
\right| $ $(p\wedge p)$.
\end{center}

As usual, 0 is a process that can do nothing. The prefix $\alpha .t$ has a
single capability expressed by $\alpha $, and the process $t$ cannot proceed
until $\alpha $ has been exercised. $\Box $ is an external choice operator. $%
\parallel _A$ is a CSP-style parallel operator, $t_1\parallel _At_2$
represents a process that behaves as $t_1$ in parallel with $t_2$ under the
synchronization set $A$. $\bot $ represents an inconsistent process which
cannot engage in any transition. $\vee $ and $\wedge $ are logic operators,
which are intended for describing logic combinations of processes. In
addition to operators over processes, CLL also contains predicate symbols $F$
and $F_\alpha $ for each $\alpha \in Act_\tau $. Intuitively, given a
process $p$, $pF$ says that $p$ is inconsistent, and $pF_\alpha $ says that $%
p$ has a {\em consistent} $\alpha -$derivative, which is useful when
describing (LTS1) (see, Def. 2.1) in terms of SOS rules. The SOS rules of
CLL are divided into two parts: transition rules and predicate rules, which
are given below.

\begin{description}
\item  (Ra$_1$)$\dfrac {}{\alpha .p\stackrel{\alpha }{\rightarrow }p}\qquad
\qquad \quad $(Ra$_2$)$\dfrac{p_1\stackrel{a}{\rightarrow }t\qquad p_2%
\stackrel{\tau }{\not \rightarrow }}{p_1\Box p_2\stackrel{a}{\rightarrow }t}%
\qquad $(Ra$_3$)$\dfrac{p_2\stackrel{a}{\rightarrow }t\qquad p_1\stackrel{%
\tau }{\not \rightarrow }}{p_1\Box p_2\stackrel{a}{\rightarrow }t}$

\item  (Ra$_4$)$\dfrac{p_1\stackrel{\tau }{\rightarrow }t}{p_1\Box p_2%
\stackrel{\tau }{\rightarrow }t\Box p_2}\qquad $(Ra$_5$)$\dfrac{p_2\stackrel{%
\tau }{\rightarrow }t}{p_1\Box p_2\stackrel{\tau }{\rightarrow }p_1\Box t}%
\qquad \quad $(Ra$_6$)$\dfrac{p_1\stackrel{a}{\rightarrow }t_1\qquad p_2%
\stackrel{a}{\rightarrow }t_2}{p_1\wedge p_2\stackrel{a}{\rightarrow }%
t_1\wedge t_2}$

\item  (Ra$_7$) $\dfrac{p_1\stackrel{\tau }{\rightarrow }t}{p_1\wedge p_2%
\stackrel{\tau }{\rightarrow }t\wedge p_2}\quad $(Ra$_8$)$\dfrac{p_2%
\stackrel{\tau }{\rightarrow }t}{p_1\wedge p_2\stackrel{\tau }{\rightarrow }%
p_1\wedge t}\qquad $(Ra$_9$)$\dfrac {}{p_1\vee p_2\stackrel{\tau }{%
\rightarrow }p_1}$

\item  (Ra$_{10}$)$\dfrac {}{p_1\vee p_2\stackrel{\tau }{\rightarrow }%
p_2}\qquad $(Ra$_{11}$)$\dfrac{p_1\stackrel{\tau }{\rightarrow }t}{%
p_1\parallel _Ap_2\stackrel{\tau }{\rightarrow }t\parallel _Ap_2}\quad $(Ra$%
_{12}$)$\dfrac{p_2\stackrel{\tau }{\rightarrow }t}{p_1\parallel _Ap_2%
\stackrel{\tau }{\rightarrow }p_1\parallel _At}$

\item  (Ra$_{13}$)$\dfrac{p_1\stackrel{a}{\rightarrow }t\qquad p_2\stackrel{%
\tau }{\not \rightarrow }\qquad a\notin A}{p_1\parallel _Ap_2\stackrel{a}{%
\rightarrow }t\parallel _Ap_2}\qquad $(Ra $_{14}$)$\dfrac{p_2\stackrel{a}{%
\rightarrow }t\qquad p_1\stackrel{\tau }{\not \rightarrow }\qquad a\notin A}{%
p_1\parallel _Ap_2\stackrel{a}{\rightarrow }p_1\parallel _At}$

\item  (Ra$_{15}$)$\dfrac{p_1\stackrel{a}{\rightarrow }t_1\qquad p_2%
\stackrel{a}{\rightarrow }t_2\qquad a\in A}{p_1\parallel _Ap_2\stackrel{a}{%
\rightarrow }t_1\parallel _At_2}\qquad \qquad $
\end{description}

\begin{center}
{\bf Table 1 The transition rules of CLL}
\end{center}

\begin{description}
\item  (Rp$_1$) $\dfrac{\quad \quad }{\bot F}\qquad \qquad $(Rp$_2$)$\dfrac{%
\quad pF\quad }{\alpha .pF}\qquad \qquad $(Rp$_3$)$\dfrac{pF\text{,}\quad qF%
}{p\vee qF}\qquad $(Rp$_4$)$\dfrac{\quad pF\quad }{p\Box qF}\qquad $

\item  (Rp$_5$)$\dfrac{\quad qF\quad }{p\Box qF}\qquad $(Rp$_6$)$\dfrac{%
\quad pF\quad }{p\parallel _AqF}\qquad \quad \qquad $(Rp$_7$)$\dfrac{\quad
qF\quad }{p\parallel _AqF}\qquad $(Rp$_8$)$\dfrac{\quad pF\quad }{p\wedge qF}
$

\item  (Rp$_9$)$\dfrac{\quad qF\quad }{p\wedge qF}\quad $(Rp$_{10}$)$\dfrac{%
\quad p\stackrel{a}{\rightarrow }p_1,\text{ }q\stackrel{a}{\not \rightarrow }%
\text{, }p\wedge q\stackrel{\tau }{\not \rightarrow }}{p\wedge qF}\quad $(Rp$%
_{11}$)$\dfrac{\quad q\stackrel{a}{\rightarrow }q_1,\text{ }p\stackrel{a}{%
\not \rightarrow }\text{, }p\wedge q\stackrel{\tau }{\not \rightarrow }}{%
p\wedge qF}$
\end{description}

\begin{center}
{\bf Table 2 The predicate rules about }$F$
\end{center}

\begin{description}
\item  (Rp$_{CLL12}$) $\dfrac{p\wedge q\stackrel{\alpha \quad }{\rightarrow r%
}\quad r\neg F}{p\wedge q{}F_\alpha }\qquad $\quad (Rp$_{CLL13}$)$\dfrac{%
p\wedge q\stackrel{\alpha \quad }{\rightarrow r}\quad p\wedge q\neg F_\alpha
\quad }{p\wedge qF}$
\end{description}

\begin{center}
{\bf Table 3 The predicate rules about }$F_\alpha $\qquad
\end{center}

Table~1 consists of transition rules $Ra_i(1\leq i\leq 15)$, where $a\in Act$%
, $\alpha \in Act_\tau $ and $A\subseteq Act$. Negative premises in rules $%
Ra_2$, $Ra_3$, $Ra_{13}$ and $Ra_{14}$ give $\tau $-transition precedence
over transitions labelled by visible actions, which guarantees that the
transition model of CLL is $\tau $-pure. Rules $Ra_9$ and $Ra_{10}$
illustrate that the operational aspect of $t_1\vee t_2$ is same as internal
choice in usual process calculus. The rule $Ra_6$ reflects that the
conjunction operator $\wedge $ is a synchronous product for visible
transitions.

Table~2 contains predicate rules about the inconsistency predicate $F$.
Although both $0$ and $\bot $ have empty behavior, they represent different
processes. The rule $Rp_1$ says that $\bot $ is inconsistent, but $0$ is
consistent as there is no proof of $0F$. The rule $Rp_3$ reflects that if
both two disjunctive parts are inconsistent then so is the disjunction.
Rules $Rp_4-Rp_9$ describe the system design strategy that if one part is
inconsistent, then so is the whole composition. The rules $Rp_{10}$ and $%
Rp_{11}$ reveal that a stable conjunction is inconsistent if its conjuncts
have distinct ready sets.

Table~3 contains predicate rules (Rp$_{CLL12}$) and (Rp$_{CLL13}$) which
formalize (LTS1) in Def.~2.1 for processes with the format $p\wedge q$.\quad

Following [43], the notion of ready simulation (see, Def. 2.2) is adopted to
formalize the refinement relation in [60]. Moreover, a sound and
ground-complete axiomatic system is provided to characterize the operators
within CLL in terms of (in)equational laws in [60].

\quad

\section{Process calculus CLLT}

This section will introduce the process calculus CLLT, which is obtained by
enriching CLL with two temporal operators and two useful auxiliary
operators, but omitting all predicate symbols $F_\alpha $ with $\alpha \in
Act_\tau $. In the following, we will give syntax and SOS rules of CLLT, and
demonstrate that CLLT has a unique stable model. Moreover, a number of
simple but useful properties of such model are given.

\subsection{Syntax and SOS rules of CLLT}

In addition to operators in CLL, new process operators $true$, $\sharp $ and
$\varpi $, and auxiliary operators $\bigtriangleup $ and $\odot $ are added
to CLLT. Before describing their behavior formally in terms of SOS rules, we
give a brief, informal account of the intended interpretation of these
operators. The constant (i.e., 0-ary operator) $true$ represents the
``loosest'' specification: it does not require anything except consistency,
while admitting any possible move. The operators $\sharp $ and $\varpi $ are
intended to capture modal operators $always$ and $unless$ respectively
through providing graphical representations of logic specifications ``$%
always $ $p$'' and ``$p$ $unless$ $q$''. They turn out to be suitable in
describing the ``loosest'' implementations that realize these two logic
specifications respectively. Auxiliary operators $\bigtriangleup $ and $%
\odot $ themselves have little computational (or logic) meaning, but they
are useful stepping-stones when we assign operational semantics to operators
$\sharp $ and $\varpi $ by means of SOS rules. Roughly speaking, the whole
point of using $\bigtriangleup $ (or, $\odot $) is to record the evolving
paths of processes with the format $\sharp p$ ($p\varpi q$, respectively).

\qquad

{\bf Definition 3.1} The processes in CLLT are defined by BNF below

\begin{center}
$p::=q\left| \text{ }true\text{ }\left| \text{ }(\sharp p)\text{ }\right|
\text{ }(\text{ }p\varpi p)\text{ }\right| $ $(p\odot (p\varpi p))$ $\left|
\text{ }(p\bigtriangleup p)\right. $ with $q\in T(\Sigma _{CLL})$.
\end{center}

Here $T(\Sigma _{CLL})$ is the set of all processes in $CLL$. In the
remainder, we shall always use $t_1\equiv t_2$ to mean that the expressions $%
t_1$ and $t_2$ are syntactically identical, and use the notation $%
\stackunder{i<n}{\Box }t_i$ for a generalized external choice, which is
defined formally below.

\quad

{\bf Definition 3.2} Let $<t_0,t_1,\dots ,t_{n-1}>$ be a finite sequence of
process terms with $n\geq 0$. The generalized external choice $\stackunder{%
i<n}{\Box }t_i$ is defined recursively as

\begin{description}
\item  1 $\stackunder{i<0}{\Box }t_i=0,$

\item  2 $\stackunder{i<1}{\Box }t_i=t_0,$

\item  3 $\stackunder{i<k+1}{\Box }t_i=(\stackunder{i<k}{\Box }t_i)\Box t_k$
for $k\geq 1$.
\end{description}

In fact, modulo $=_{RS}$, the order and grouping of terms in $\stackunder{i<n%
}{\Box }t_i$ may be ignored by virtue of the commutative and associative
laws [60]. Therefore we also often use the notation $\stackunder{i\in I}{%
\Box }t_i$ to denote generalized external choice, where $I$ is an arbitrary
finite indexed set.

\begin{description}
\item  (Ra$_{16}$)\footnote{%
In particular, $\dfrac {}{true\stackrel{\tau }{\rightarrow }0}$ by setting $%
A=\emptyset $.}$\dfrac {}{true\stackrel{\tau }{\rightarrow }\stackunder{a\in
A}{\Box }a.true}$\qquad $\qquad \qquad $(Ra$_{17}$) $\dfrac{p\stackrel{\tau
\quad }{\rightarrow q}}{\sharp p\stackrel{\tau }{\rightarrow }%
q\bigtriangleup p}$

\item  (Ra$_{18}$)$\dfrac{p\stackrel{a\quad }{\rightarrow q}}{\sharp p%
\stackrel{a}{\rightarrow }(q\wedge p)\bigtriangleup p}\qquad \qquad \quad
\qquad $(Ra$_{19}$) $\dfrac{p\stackrel{\tau \quad }{\rightarrow q}}{%
p\triangle r\stackrel{\tau }{\rightarrow }q\bigtriangleup r}\qquad \qquad
\qquad \qquad \quad $

\item  (Ra$_{20}$) $\dfrac{p\stackrel{a\quad }{\rightarrow q}}{p\triangle r%
\stackrel{a}{\rightarrow }(q\wedge r)\bigtriangleup r}\qquad \qquad \quad
\quad $(Ra$_{21}$) $\dfrac {}{p\varpi q\stackrel{\tau }{\rightarrow }p\odot
(p\varpi q)}$

\item  (Ra$_{22}$) $\dfrac {}{p\varpi q\stackrel{\tau }{\rightarrow }%
q}\qquad \quad \qquad \qquad \qquad \quad $(Ra$_{23}$) $\dfrac{r\stackrel{%
\tau }{\rightarrow }s}{r\odot (p\varpi q)\stackrel{\tau }{\rightarrow }%
s\odot (p\varpi q)}$

\item  (Ra$_{24}$) $\dfrac{r\stackrel{a}{\rightarrow }s}{r\odot (p\varpi q)%
\stackrel{a}{\rightarrow }s\wedge q}\qquad \qquad \quad \quad $(Ra$_{25}$) $%
\dfrac{r\stackrel{a}{\rightarrow }s}{r\odot (p\varpi q)\stackrel{a}{%
\rightarrow }(s\wedge p)\odot (p\varpi q)}$
\end{description}

\begin{center}
{\bf Table 4 Additional transition rules}
\end{center}

\qquad

\begin{description}
\item  (Rp$_{12}$)$\dfrac{qF\quad p\odot (p\varpi q)F}{p\varpi qF}\qquad
\qquad \qquad \qquad \qquad $(Rp$_{13}$)$\dfrac{\quad pF\quad }{p\triangle qF%
}$

\item  (Rp$_{14}$) $\dfrac{rF}{r\odot (p\varpi q)F}\qquad \qquad \qquad
\qquad \qquad \qquad $(Rp$_{15}$)$\dfrac{\quad pF\quad }{\sharp pF}$

\item  (Rp$_{16}$) $\dfrac{p\stackrel{\alpha }{\rightarrow }s\quad \{rF:p%
\stackrel{\alpha \quad }{\rightarrow r}\}}{pF}$, where the topmost operator
of $p$ is in $\left\{ \wedge ,\sharp ,\bigtriangleup ,\odot \right\}
\footnote{%
That is, $p$ has one of the formats: $r\wedge t$, $\sharp t$, $%
t\bigtriangleup s$ and $t\odot (r\varpi q)$.}$.
\end{description}

\begin{center}
{\bf Table 5 Additional predicate rules}
\end{center}

Similar to CLL, the SOS rules of CLLT are divided into two parts: transition
rules and predicate rules. In effect these rules capture L\"uttgen and
Vogler's constructions\footnote{%
See Definition 9 and 10 in [44].} in process algebraic style.

On the side of transition rules, in addition to all transition rules of CLL
(i.e., rules in Table 1), rules in Table 4 are adopted to describe the
behavior of $true$, $\sharp $, $\varpi $, $\bigtriangleup $ and $\odot $,
where $a\in Act$ and $A$ is any finite subset of $Act$.

On the side of rules concerning inconsistency predicate $F$, rules in Table
2 are preserved, and rules in Table 5 are added to CLLT. Notice that the
rules (Rp$_{CLL12}$) and (Rp$_{CLL13}$) in Table 3 are replaced by the rule
(Rp$_{16}$). The motivation behind this modification may be found in the
next subsection (see, Remark 3.1).

Summarizing, the TSS for CLLT is $\Gamma _{CLLT}=(\Sigma _{CLLT},Act_\tau
,\Lambda _{CLLT},\Xi _{CLLT})$, where

\begin{itemize}
\item  $\Sigma _{CLLT}=\{\Box ,\wedge ,\vee ,0,\bot \}\bigcup \{\alpha
.()|\alpha \in Act_\tau \}\bigcup \{\parallel _A|A\subseteq Act\}\bigcup
\{true,\sharp ,\varpi ,\bigtriangleup ,\odot \}$,

\item  $\Lambda _{CLLT}=\{F\}$, and

\item  $\Xi _{CLLT}=\{Ra_1,\dots ,Ra_{25}\}\bigcup \{Rp_1,\dots ,Rp_{16}\}$%
.\quad
\end{itemize}

\subsection{Stable transition model of CLLT}

This subsection will illustrate that $\Gamma _{CLLT}$ has a unique stable
model. To this end, a few preliminary definitions are needed.

\qquad \qquad

{\bf Definition 3.3} The degree of terms is defined inductively below

$|0|=|\bot |=|true|=1$

$|\sharp t|=|\alpha .t|=|t|+1$

$|t_1\clubsuit t_2|=|t_1|+|t_2|+1$ for $\clubsuit \in \{\varpi ,\wedge ,\vee
,||_A,\Box \}$

$|t_1\bigtriangleup t_2|=|t_1\odot (t_2\varpi t_3)|=|t_1|$

\qquad

{\bf Definition 3.4} The function $S$ from $Tr(\Sigma _{CLLT},Act_\tau
)\;\cup \;Pred(\Sigma _{CLLT},\Lambda _{CLLT})$ to $\omega +1$ is defined
as: $S(t\stackrel{\alpha }{\rightarrow }r)=\left| t\right| $ for any $t%
\stackrel{\alpha }{\rightarrow }r\in Tr(\Sigma _{CLLT},Act_\tau )$, and$%
\;S(tF)=\omega $ for any $tF\in Pred(\Sigma _{CLLT},\Lambda _{CLLT})$, where
$\omega $ is the initial limit ordinal.

\qquad

It is easy to check that this function $S$ is a stratification of $\Gamma
_{CLLT}$. Thus $\Gamma _{CLLT}$ has a unique stable transition model.
Henceforward such model is denoted by $M_{CLLT}$. As usual, the LTS
associated with CLLT is defined below.

\qquad

{\bf Definition 3.5} The LTS associated with CLLT, in symbols $LTS(CLLT)$,
is the quadruple $(T(\Sigma _{CLLT}),Act_\tau ,\rightarrow _{CLLT},F_{CLLT})$
such that for any $t$, $s\in T(\Sigma _{CLLT})$ and $\alpha \in Act_\tau $, $%
t\stackrel{\alpha }{\rightarrow }_{CLLT}s$ iff $t\stackrel{\alpha }{%
\rightarrow }s\in M_{CLLT}$, and $t\in F_{CLLT}$ iff $tF\in M_{CLLT}$.

\qquad

Since $M_{CLLT}$ is a stable transition model, which exactly consists of
provable transitions of the positive TSS $Strip(\Gamma _{CLLT},M_{CLLT})$,
the result below follows immediately.

\qquad

{\bf Theorem 3.1} For any $t,t_1,t_2\in T(\Sigma _{CLLT})$ and $\alpha \in
Act_\tau $, we have

(1) $t_1\stackrel{\alpha }{\rightarrow }_{CLLT}t_2$ iff $Strip(\Gamma
_{CLLT},M_{CLLT})\vdash t_1\stackrel{\alpha }{\rightarrow }t_2$.\qquad

(2) $t\in F_{CLLT}$ iff $Strip(\Gamma _{CLLT},M_{CLLT})\vdash tF$.

\TeXButton{Proof}{\proof} Straightforward. \qquad \TeXButton{End Proof}
{\endproof}

\qquad

This theorem is trivial but useful. It provides a way to establish the
properties of $LTS(CLLT)$. That is, we can demonstrate some conclusions by
proceeding induction on the depth of inferences in the positive TSS $%
Strip(\Gamma _{CLLT},M_{CLLT})$. In the remainder of this paper, we will
apply this theorem without any reference.

\quad

{\bf Remark 3.1} Although the universal quantifier symbol does not occur in
(Rp$_{16}$) explicitly, it is not difficult to see that the premise of (Rp$%
_{16}$) involves universal quantifier in spirit. Analogous to CLL, we may
adopt the method given in [58] to avoid this. In detail, for each $\alpha
\in Act_\tau $, the auxiliary predicates $F_\alpha $ is added to CLLT and
the rule (Rp$_{16}$) is replaced by two rules below, where the topmost
operator of $p$ is in $\left\{ \wedge ,\sharp ,\bigtriangleup ,\odot
\right\} $.

\begin{center}
(Rp$_{16-1}$) $\dfrac{p\stackrel{\alpha \quad }{\rightarrow q}\quad q\neg F}{%
p{}F_\alpha }$\qquad (Rp$_{16-2}$)$\dfrac{p\stackrel{\alpha \quad }{%
\rightarrow q}\quad p\neg F_\alpha \quad }{pF}$
\end{center}

Similar to (Rp$_{16}$), these two rules also capture (LTS1). However, it is
easy to see that, due to two rules above, the stratifying function does not
exist for resulting calculus. By means of technique so-called $positive$ $%
after$ $reduction$ [13, 26], we can also get its stable transition model as
done in [60]. Moreover, such stable transition model coincides with $%
M_{CLLT} $. To avoid cumbersome reduction procedure, our current system
employs (Rp$_{16}$) instead of (Rp$_{16-1}$) and (Rp$_{16-2}$).

\qquad

{\bf Convention 3.1 } For the sake of convenience, in the remainder of this
paper, we shall omit the subscript in labelled transition relations $%
\stackrel{\alpha }{\rightarrow }_{CLLT}$, that is, we shall use $\stackrel{%
\alpha }{\rightarrow }$ to denote transition relations within $LTS(CLLT)$.
Thus, the notation $\stackrel{\alpha }{\rightarrow }$ has double utility:
predicate symbol in the TSS $\Gamma _{CLLT}$ and labelled transition
relation on processes in $LTS(CLLT)$. However, it usually does not lead to
confusion in a given context. Similarly, the notation $F_{CLLT}$ is
abbreviated to $F$. Hence the symbol $F$ is overloaded, predicate symbol in
the TSS $\Gamma _{CLLT}$ and the set of all inconsistent processes in $%
LTS(CLLT)$, in each case the context of use will allow us to make the
distinction.

\subsection{Basic properties of $LTS(CLLT)$}

This subsection will provide a number of simple properties of $LTS(CLLT)$.
In particular, we will show that $LTS(CLLT)$ is indeed a $\tau -pure$ LLTS.
We begin with listing a few simple properties in the next three lemmas,
which will be frequently used in subsequent sections.

\qquad

{\bf Lemma 3.1} Let $t,p,q\in T(\Sigma _{CLLT})$ and $\alpha ,\beta \in
Act_\tau $.

(1) $\alpha .t\stackrel{\beta }{\rightarrow }r$ iff $\alpha =\beta $ and $%
r\equiv t$.

(2) $p\vee q\stackrel{\beta }{\rightarrow }r$ iff $\beta =\tau $ and either $%
p\equiv r$ or $q\equiv r$.

(3) $true\stackrel{\beta }{\rightarrow }r$ iff $\beta =\tau $ and either $%
r\equiv 0$ or $r\equiv \stackunder{a\in A}{\Box }a.true$ for some nonempty
finite set $A\subseteq Act$.

(4) $\sharp p\stackrel{\tau }{\rightarrow }r$ iff $r\equiv p_1\bigtriangleup
p$ for some $p_1$ with $p\stackrel{\tau }{\rightarrow }p_1$.

(5) $p\bigtriangleup q\stackrel{\tau }{\rightarrow }r$ iff $r\equiv
p_1\bigtriangleup q$ for some $p_1$ with $p\stackrel{\tau }{\rightarrow }p_1$%
.

(6) $p\varpi q\stackrel{\beta }{\rightarrow }r$ iff $\beta =\tau $ and
either $r\equiv q$ or $r\equiv p\odot (p\varpi q)$.

(7) $t\odot (p\varpi q)\stackrel{\tau }{\rightarrow }r$ iff $r\equiv
t_1\odot (p\varpi q)$ for some $t_1$ with $t\stackrel{\tau }{\rightarrow }%
t_1 $.

(8) $p\heartsuit q\stackrel{\tau }{\rightarrow }r$ iff either ($r\equiv
s\heartsuit q$ and $p\stackrel{\tau }{\rightarrow }s$) or ($r\equiv
p\heartsuit s$ and $q\stackrel{\tau }{\rightarrow }s$) for some $s$, where $%
\heartsuit \in \{\wedge ,\Box ,\parallel _A\}$.

\TeXButton{Proof}{\proof}For each item, the implication from right to left
is obvious. The proof of converse implication is a routine case analysis on
the last rule applied in the inference. As a sample case, we consider (6),
the remainder may be handled in the similar manner and omitted. It follows
from $p\varpi q\stackrel{\beta }{\rightarrow }r$ that $Strip(\Gamma
_{CLLT},M_{CLLT})\vdash p\varpi q\stackrel{\beta }{\rightarrow }r$. Clearly,
the last rule applied in the inference has the format below

\begin{center}
\qquad either $\quad \dfrac {}{p\varpi q\stackrel{\tau }{\rightarrow }%
q}\quad $or \quad $\dfrac {}{p\varpi q\stackrel{\tau }{\rightarrow }p\odot
(p\varpi q)}$ .
\end{center}

Then $\beta =\tau $ and either $r\equiv q$ or $r\equiv p\odot (p\varpi q)$,
as desired. \TeXButton{End Proof}{\endproof}

\qquad

{\bf Lemma 3.2} Let $t,p,q\in T(\Sigma _{CLLT})$ and $a\in Act$.

(1) $p\Box q\stackrel{a}{\rightarrow }r$ iff either $p\stackrel{a}{%
\rightarrow }r$ and $q\stackrel{\tau }{\not \rightarrow }$, or $q\stackrel{a%
}{\rightarrow }r$ and $p\stackrel{\tau }{\not \rightarrow }$.

(2) $p\wedge q\stackrel{a}{\rightarrow }r$ iff $p\stackrel{a}{\rightarrow }%
r_1$, $q\stackrel{a}{\rightarrow }r_2$ and $r\equiv r_1\wedge r_2$ for some $%
r_1,$ $r_2$.

(3) $\sharp p\stackrel{a}{\rightarrow }r$ iff $r\equiv (q\wedge
p)\bigtriangleup p$ for some $q$ with $p\stackrel{a}{\rightarrow }q$.

(4) $p\bigtriangleup q\stackrel{a}{\rightarrow }r$ iff $r\equiv (s\wedge
q)\bigtriangleup q$ for some $s$ with $p\stackrel{a}{\rightarrow }s$.

(5) $t\odot (p\varpi q)\stackrel{a}{\rightarrow }r$ iff there exists $s$
such that $t\stackrel{a}{\rightarrow }s$ and either $r\equiv $ $s\wedge q$
or $r\equiv (s\wedge p)\odot (p\varpi q)$.

(6) If $a\notin A$ then, $p||_Aq\stackrel{a}{\rightarrow }r$ iff either ($%
r\equiv s||_Aq$ , $q\stackrel{\tau }{\not \rightarrow }$ and $p\stackrel{a}{%
\rightarrow }s$) or ($r\equiv p||_As$, $q\stackrel{\tau }{\not \rightarrow }$
and $q\stackrel{a}{\rightarrow }s$) for some $s$.

(7) If $a\in A$ then, $p||_Aq\stackrel{a}{\rightarrow }r$ iff $r\equiv
s||_At $, $p\stackrel{a}{\rightarrow }s$ and $q\stackrel{a}{\rightarrow }t$
for some $s,t$.

\TeXButton{Proof}{\proof}Analogous to Lemma 3.1, omitted.
\TeXButton{End Proof}{\endproof}

\qquad

{\bf Lemma 3.3} Suppose $p,q,r\in T(\Sigma _{CLLT})$.

(1) $p\vee q\in F$ iff $p,q\in F$.

(2) $\alpha .p\in F$ iff $p\in F$.

(3) $p\heartsuit q\in F$ iff either $p\in F$ or $q\in F$ for $\heartsuit \in
\{\Box ,\parallel _A\}$.

(4) Either $p\in F$ or $q\in F$ implies $p\wedge q\in F$.

(5) $p\varpi q$ $\in F$ iff $q,p\odot (p\varpi q)\in F$.

(6) $r\in F$ implies $r\bigtriangleup p,\sharp r,r\odot (p\varpi q)\in F$.

(7) $0\notin F$, $true\notin F$ and $\perp \in F$.

\TeXButton{Proof}{\proof}Straightforward. \TeXButton{End Proof}{\endproof}

\qquad

{\bf Lemma 3.4} $LTS(CLLT)$ is $\tau -pure$.

\TeXButton{Proof}{\proof}Let $t\stackrel{a}{\rightarrow }s$. It is enough to
show that $t\stackrel{\tau }{\not \rightarrow }$. The proof is done by
induction on the inference $Strip(\Gamma _{CLLT},M_{CLLT})\vdash t\stackrel{a%
}{\rightarrow }s$, which is a long but routine case analysis based on the
last rule applied in the inference, omitted. \TeXButton{End Proof}{\endproof}

\qquad

{\bf Lemma 3.5} $LTS(CLLT)$ satisfies (LTS1).

\TeXButton{Proof}{\proof}Let $t$ be any process and assume that $\forall s(t%
\stackrel{\alpha }{\rightarrow }s\;\text{implies}\;s\in F)$ for some $\alpha
\in I(t)$. We intend to verify $t\in F$ by induction on $t$.

\begin{itemize}
\item  $t\equiv 0,\perp ,true$, $\alpha .p$, $p\vee q$ or $p\varpi q$
\end{itemize}

Follows from Lemma 3.1 and 3.3. In particular, for $t\equiv 0,\perp $ or $%
true$, since the premise does not hold at all, it holds trivially.

\begin{itemize}
\item  $t\equiv p\wedge q$, $\sharp p$, $p\bigtriangleup q$ or $p\odot
(r\varpi q)$.
\end{itemize}

Immediately follows from the rule (Rp$_{16}$).

\begin{itemize}
\item  $t\equiv p\Box q$
\end{itemize}

By Lemma 3.3(3), it suffices to show that either $p\in F$ or $q\in F$.
Conversely, suppose that $p\notin F$ and $q\notin F$. By Lemma 3.2 (1) and
3.1(8), we have \text{either}\ $\alpha \in I(p)$\ \text{or}\ $\alpha \in
I(q) $. W.l.o.g, we consider the first alternative. If $\alpha \in Act$,
then, by Lemma 3.2(1), we get

\begin{center}
$\left\{ s:p\stackrel{\alpha }{\rightarrow }s\right\} \subseteq \left\{
s:p\Box q\stackrel{\alpha }{\rightarrow }s\right\} \subseteq F$ .
\end{center}

Hence, by induction hypothesis (IH, for short), we have $p\in F$, a
contradiction. If $\alpha =\tau $, then, by Lemma 3.1(8), it follows that

\begin{center}
$\left\{ s\Box q:p\stackrel{\tau }{\rightarrow }s\right\} \subseteq \left\{
s:p\Box q\stackrel{\tau }{\rightarrow }s\right\} \subseteq F$ .
\end{center}

Further, by Lemma 3.3(3), it follows from $q\notin F$ that $\left\{ s:p%
\stackrel{\tau }{\rightarrow }s\right\} \subseteq F$. Then, by IH, we also
obtain $p\in F$, a contradiction.

\begin{itemize}
\item  $t\equiv p||_Aq$
\end{itemize}

Again by Lemma 3.3(3), it is sufficient to show that either $p\in F$ or $%
q\in F$. On the contrary, suppose that $p\notin F$ and $q\notin F$. We
distinguish two cases depending on whether $\alpha $ is in $A$.

\begin{description}
\item  Case 1 $\alpha \notin A.$
\end{description}

Then \text{either}\ $\alpha \in I(p)$\ \text{or}\ $\alpha \in I(q)$ by Lemma
3.1(8) and 3.2(6). W.l.o.g, we handle the first alternative. Hence

\begin{center}
$\left\{ s||_Aq:p\stackrel{\alpha }{\rightarrow }s\right\} \subseteq \left\{
s:p||_Aq\stackrel{\alpha }{\rightarrow }s\right\} \subseteq F$ .
\end{center}

Further, by Lemma 3.3(3), it follows from $q\notin F$ that $\left\{ s:p%
\stackrel{\alpha }{\rightarrow }s\right\} \subseteq F$. Then $p\in F$ due to
IH, a contradiction.

\begin{description}
\item  Case 2 $\alpha \in A.$
\end{description}

In such situation, we get $\alpha \in I(p)$ and $\alpha \in I(q)$. Then, by
IH, it follows from $p\notin F$ and $q\notin F$ that there exist $p_1$ and $%
q_1$ such that

\begin{center}
$p\stackrel{\alpha }{\rightarrow }p_1$, $q\stackrel{\alpha }{\rightarrow }%
q_1 $, $p_1\notin F$ and $q_1\notin F$.
\end{center}

Thus $p||_Aq\stackrel{\alpha }{\rightarrow }p_1||_Aq_1$ and $%
p_1||_Aq_1\notin F$ by Lemma 3.2(7) and 3.3(3), a contradiction.
\TeXButton{End Proof}{\endproof}

\qquad

A simple but useful result is given below, which provides a necessary and
sufficient condition for a non-stable process to be inconsistent. An
analogous result have been obtained for CLL in [60].

\qquad

{\bf Lemma 3.6} For any $t\in T(\Sigma _{CLLT}),$ we have

(1) $t\in F$ iff $\forall s(t\stackrel{\tau }{\rightarrow }s$ implies $s\in
F)$ whenever $\tau \in I(t)$.

(2) If $t\stackrel{\varepsilon }{\Rightarrow }|s$ and $s\notin F$ then $t$ $%
\notin F$ and $t\stackrel{\varepsilon }{\Rightarrow }_F|s$.

\TeXButton{Proof}{\proof}Clearly, (2) immediately follows from (1). In the
following, we consider (1). Assume that $\tau \in I(t)$. Then, by Lemma 3.5,
we need only show that the left implies the right. We can prove it by
induction on the inference $Strip(\Gamma _{CLLT},M_{CLLT})\vdash tF$ , which
is a case analysis based on the format of $t$. As an instance, we shall deal
with the case $t\equiv p\bigtriangleup q$, the remainder may be handled in a
similar way and omitted.

Since $t\equiv p\bigtriangleup q$, the last rule applied in the inference
has the format

\begin{center}
either\quad $\dfrac{p\bigtriangleup q\stackrel{\alpha \quad }{\rightarrow r}%
\quad \{rF:p\bigtriangleup q\stackrel{\alpha \quad }{\rightarrow r}\}}{%
p\bigtriangleup qF}\quad $or\quad $\dfrac{\quad pF\quad }{p\triangle qF}.$
\end{center}

For the first alternative, since $\tau \in I(p\bigtriangleup q)$, we get $%
\alpha =\tau $ by Lemma 3.4. Then it immediately follows that $%
\{r:p\bigtriangleup q\stackrel{\tau }{\rightarrow }r\}\subseteq F$. For the
second alternative, we have $p\in F$. Moreover, by Lemma 3.1(5), we get $%
\tau \in I(p)$ because of $\tau \in I(p\bigtriangleup q)$. Hence, by IH, it
follows that $\{r:p\stackrel{\tau }{\rightarrow }r\}\subseteq F$. Further,
since $\{r:p\bigtriangleup q\stackrel{\tau }{\rightarrow }%
r\}=\{r\bigtriangleup q:p\stackrel{\tau }{\rightarrow }r\}$, we obtain $%
\{r:p\bigtriangleup q\stackrel{\tau }{\rightarrow }r\}\subseteq F$ by Lemma
3.3(6), as desired. \TeXButton{End Proof}{\endproof}

\qquad

In order to show that $LTS(CLLT)$ satisfies (LTS2), we introduce the notion
of $\tau -$degree as follows, which measures processes's capability of
executing successive $\tau $ actions.

\quad

{\bf Definition 3.6} The $\tau -$degree of processes is defined inductively
below

$d(true)=1$

$d(0)=d(\bot )=d(a.t)=0$ whenever $a\in Act$

$d(\tau .t)=d(t)+1$

$d(t_1\varpi t_2)=d(t_1\vee t_2)=\max \{d(t_1),d(t_2)\}+1$

$d(t_1\wedge t_2)=d(t_1||_At_2)=d(t_1\Box t_2)=d(t_1)+d(t_2)$

$d(\sharp t)=d(t\bigtriangleup t_1)=d(t\odot (t_1\varpi t_2))=d(t)$

\qquad

{\bf Lemma 3.7} If $t\stackrel{\tau }{\rightarrow }r$ then $d(r)<d(t)$ for
any $t,r$ $\in T(\Sigma _{CLLT})$.

\TeXButton{Proof}{\proof}Proceeding by induction on the inference $%
Strip(\Gamma _{CLLT},M_{CLLT})\vdash t\stackrel{\tau }{\rightarrow }r$,
which is a routine case analysis on the last rule applied in the inference.
\TeXButton{End Proof}{\endproof}

\qquad

This elementary property makes it effective to apply the induction on the $%
\tau -$degree in the next proof.

\qquad

{\bf Lemma 3.8} $LTS(CLLT)$ satisfies (LTS2).

\TeXButton{Proof}{\proof}Let $t\in T(\Sigma _{CLLT})$ with $t\notin F$. It
suffices to find $p$ such that $t\stackrel{\varepsilon }{\Rightarrow }_F|p$.
We prove it by induction on the $\tau -$degree of $t$. Assume that it holds
for all $p$ with $d(p)<d(t)$. If $t$ is stable, then $t\stackrel{\varepsilon
}{\Rightarrow }_F|t$ follows from $t\notin F$. Next we consider another case
where $\tau \in I(t)$. Since $t\notin F$ and $\tau \in I(t)$, by Lemma
3.6(1), we have $t\stackrel{\tau }{\rightarrow }_Fs$ for some $s$. Hence $%
d(s)<d(t)$ by Lemma 3.7. Thus $s$ $\stackrel{\varepsilon }{\Rightarrow }_F|r$
for some $r$ due to IH. Then $t\stackrel{\varepsilon }{\Rightarrow }_F|r$,
as desired.\TeXButton{End Proof}{\endproof}

\qquad

Now we get the main result of this section as follows.

\quad

{\bf Theorem 3.2} $LTS(CLLT)$ is a $\tau -pure$ LLTS.

\TeXButton{Proof}{\proof}Obvious from Lemma 3.4, 3.5 and 3.8.
\TeXButton{End Proof}{\endproof}

\quad

In contrast with usual process calculuses, one of features of LLTS-oriented
process calculuses is that these calculuses take into account consistency of
processes. The inconsistency predicate is central to the description of
behavior. We often need to prove that a given process $p$ is consistent,
which boils down to show that there is no inference for $pF$ in $%
Strip(\Gamma _{CLLT},M_{CLLT})$. To this end, we introduce the notion below,
which is useful for demonstrating the consistency of processes. The
motivation behind this notion is that we intend to establish the consistency
of a given process based on the well-foundedness of proof trees.

\quad

{\bf Definition 3.7 (}$F-${\bf hole)} A set $\Omega $ of processes is said
to be a $F$-hole if, for each $q\in \Omega $, any proof tree of $%
Strip(\Gamma _{CLLT},M_{CLLT})\vdash qF$ has a $proper$ subtree with the
root labelled with $uF$ for some $u\in \Omega $.

\qquad

As the name suggests, each process in a $F-$hole is not in $F$. Formally, we
have the result below.

\qquad

{\bf Lemma 3.9 }If $\Omega $ is a $F-$hole then $\Omega \cap F=\emptyset $.

\TeXButton{Proof}{\proof}Conversely, suppose that $\Omega \cap F\not
=\emptyset $, say, $q\in \Omega \cap F$. Thus there exists a proof tree of $%
Strip(\Gamma _{CLLT},M_{CLLT})\vdash qF$. However, by Definition 3.7, such
proof tree is not well-founded, which contradicts Def. 2.3.
\TeXButton{End Proof}{\endproof}

\quad

Therefore, in order to verify that a given process $p$ is consistent, it
suffices to provide a $F-$hole including $p$. The next lemma has been showed
for CLL in pure process-algebraic style in [60], where the proof essentially
depends on the fact that, for any process $t$ within CLL and $\alpha \in
Act_\tau $, $t$ is of more complex structure than its $\alpha $-derivatives.
Unfortunately, such property does not always hold for CLLT. For instance,
consider processes $true,$ $p\triangle r$ and $r\odot (p\varpi q)$. Here we
give an alternative proof for it and indicate how the notion of $F-$hole may
be used to show the consistency of a given process.

\quad \quad

{\bf Lemma 3.10} If $s\stackunder{\sim }{\sqsubset }_{RS}r,s\stackunder{\sim
}{\sqsubset }_{RS}t$ and $s\notin F$ then $r\wedge t\notin F$.

\TeXButton{Proof}{\proof}Put

\begin{center}
$\Omega =\left\{ p_1\wedge p_2:q\stackunder{\sim }{\sqsubset }_{RS}p_1,q%
\stackunder{\sim }{\sqsubset }_{RS}p_2\text{ and }q\notin F\right\} $.
\end{center}

It is enough to show that $\Omega $ is a $F-$hole. Let $p_1\wedge p_2\in
\Omega $ and $\Im $ be any proof tree of $p_1\wedge p_2F$. Thus $q%
\stackunder{\sim }{\sqsubset }_{RS}p_1,q\stackunder{\sim }{\sqsubset }%
_{RS}p_2$ and $q\notin F$ for some $q$. So, it follows that $%
I(p_1)=I(p_2)=I(q)$, $p_1\notin F$ and $p_2\notin F$. Further, since $%
p_1\wedge p_2\stackrel{\tau }{\not \rightarrow }$, the last rule applied in $%
\Im $ has the format below

\begin{center}
$\dfrac{p_1\wedge p_2\stackrel{a}{\rightarrow }u,\text{ }\left\{
rF:p_1\wedge p_2\stackrel{a}{\rightarrow }r\right\} }{p_1\wedge p_2F}$ for
some $a\in Act$.\qquad (3.10.1)
\end{center}

Hence $a\in I(q)$. Moreover, by Lemma 3.5 and 3.8, it follows from $q\notin
F $ that $q\stackrel{a}{\Rightarrow }_F\left| q_1\right. $ for some $q_1$.
Since $q\stackunder{\sim }{\sqsubset }_{RS}p_1$ and $q\stackunder{\sim }{%
\sqsubset }_{RS}p_2$, there exist $r_i$, $t_j$ with $i\leq n$ and $j\leq m$
such that $p_1\stackrel{a}{\rightarrow }_Fr_1\stackrel{\tau }{\rightarrow }%
_Fr_2...\stackrel{\tau }{\rightarrow }_F\left| r_n\right. $, $p_2\stackrel{a%
}{\rightarrow }_Ft_1\stackrel{\tau }{\rightarrow }_Ft_2...\stackrel{\tau }{%
\rightarrow }_F\left| t_m\right. $ and

\begin{center}
$q_1\stackunder{\sim }{\sqsubset }_{RS}r_n$ and $q_1\stackunder{\sim }{%
\sqsubset }_{RS}t_m$.\qquad (3.10.2)\qquad
\end{center}

Moreover, we also have $p_1\wedge p_2\stackrel{a}{\rightarrow }r_1\wedge t_1$
and

\begin{center}
$\left. r_1\wedge t_1\stackrel{\tau }{\rightarrow }...\stackrel{\tau }{%
\rightarrow }r_n\wedge t_1\stackrel{\tau }{\rightarrow }...\stackrel{\tau }{%
\rightarrow }r_n\wedge t_m.\qquad \text{(3.10.3)}\right. $
\end{center}

Then, by (3.10.1), $\Im $ contains a proper subtree $\Im _1$ with the root
labelled with $r_1\wedge t_1F$. On the other hand, it follows from (3.10.2)
that $r_n\wedge t_m\in \Omega $. Thus, to complete the proof, it suffices to
show that $\Im $ contains a proper subtree with the root labelled with $%
r_n\wedge t_mF$. If $m=n=1$, this holds obviously due to $r_1\wedge
t_1\equiv r_n\wedge t_m$. Otherwise, w.l.o.g, we assume $n>1$. Hence $%
r_1\wedge t_1$ is not stable. Moreover, since $r_1,t_1\notin F$, the last
rule applied in $\Im _1$ is

\begin{center}
$\dfrac{r_1\wedge t_1\stackrel{\tau }{\rightarrow }s,\text{ }\left\{
rF:r_1\wedge t_1\stackrel{\tau }{\rightarrow }r\right\} }{r_1\wedge t_1F}.$
\end{center}

Thus $\Im _1$ contains a node labelled with $r_2\wedge t_1F$. By repeating
this procedure along (3.10.3), it is easily seen that $\Im $ contains a
proper subtree with the root labelled with $r_n\wedge t_mF$, as desired.
\TeXButton{End Proof}{\endproof}

\quad

We end this section with recalling some useful properties of the operator $%
\wedge $, which has been obtained in [43] and [60] in different style.

\qquad

{\bf Lemma 3.11 }For any process $p_1$, $p_{2\text{ }}$and $q$, we have

(1) $p_1\wedge $ $p_{2\text{ }}\stackunder{\sim }{\sqsubset }_{RS}p_i$ for $%
i=1,2$ whenever $p_1\stackrel{\tau }{\not \rightarrow }$ and $p_2\stackrel{%
\tau }{\not \rightarrow }$,

(2) if $q\stackunder{\sim }{\sqsubset }_{RS}p_1$ and $q\stackunder{\sim }{%
\sqsubset }_{RS}p_2$ then $q\stackunder{\sim }{\sqsubset }_{RS}p_1\wedge p_2$%
,

(3) $p_1\wedge $ $p_{2\text{ }}\sqsubseteq _{RS}p_i$ for $i=1,2$, and

(4) if $q\sqsubseteq _{RS}p_1$ and $q\sqsubseteq _{RS}p_2$ then $%
q\sqsubseteq _{RS}p_1\wedge p_2.$

\TeXButton{Proof}{\proof}A proof in pure process-algebraic style has been
given in [60]. Here we only draw the outline of its proof. Item (3) and (4)
follow from (1) and (2), respectively. For (1) and (2), we set

\begin{center}
$R_1=\left\{ \left\langle s\wedge t,s\right\rangle :s\wedge t\text{ }%
\stackrel{\tau }{\not \rightarrow }\right\} $ and $R_2=\left\{ \left\langle
s,r\wedge t\right\rangle :s\stackunder{\sim }{\sqsubset }_{RS}r\text{ and }s%
\stackunder{\sim }{\sqsubset }_{RS}t\right\} .$
\end{center}

It suffices to show that these two relations are stable ready simulation
relations. Notice that Lemma 3.10 is used to prove that $R_2$ satisfies
(RS2) and (RS3). For more details we refer the reader to [60].%
\TeXButton{End Proof}{\endproof}

\quad

It is an immediate consequence of the above lemma that, modulo $=_{RS}$ (or,
$\approx _{RS}$), the operator $\wedge $ satisfies the idempotent,
commutative and associative laws.

\section{The operator $\sharp $\qquad}

This section aims to explore properties of the operator $\sharp $. In
particular, we shall characterize processes that refine processes with the
format $\sharp p$. This result supports the claim that the operator $\sharp $
captures the modal operator $always$. Since the behavior of $\sharp $ is
described in terms of $\bigtriangleup $, we will study the latter firstly.

\quad

{\bf Lemma 4.1} If $p\stackunder{\sim }{\sqsubset }_{RS}r\triangle t$ and $p%
\stackrel{a}{\Rightarrow }_F\left| p_1\right. $then $p_1\stackunder{\sim }{%
\sqsubset }_{RS}(s\wedge u)\bigtriangleup t$ for some $s$ and $u$ such that $%
r\stackrel{a}{\Rightarrow }_F\left| s\right. $ and $t\stackrel{\varepsilon }{%
\Rightarrow }_F\left| u\right. $.

\TeXButton{Proof}{\proof}Since $p\stackunder{\sim }{\sqsubset }%
_{RS}r\triangle t$ and $p\stackrel{a}{\Rightarrow }_F\left| p_1\right. $, $%
p_1\stackunder{\sim }{\sqsubset }_{RS}q$ for some $q$ with $r\triangle t%
\stackrel{a}{\Rightarrow }_F\left| q\right. $. Then it follows from $%
r\triangle t\stackrel{\tau }{\not \rightarrow }$ that $r\triangle t\stackrel{%
\qquad \quad }{\stackrel{a}{\rightarrow }_Fq_1}\stackrel{\varepsilon }{%
\Rightarrow }_F\left| q\right. $ for some $q_1$. Further, by Lemma 3.1(5)(8)
and 3.2 (4), there exist $r_1,r_2$ and $t_1$ such that $q_1\equiv (r_1\wedge
t)\triangle t$ with $r\stackrel{a}{\rightarrow }_Fr_1$, and $q\equiv
(r_2\wedge t_1)\triangle t$ with $t\stackrel{\varepsilon }{\Rightarrow }%
_F\left| t_1\right. $ and $r_1\stackrel{\varepsilon }{\Rightarrow }_F\left|
r_2\right. $. Hence $p_1\stackunder{\sim }{\sqsubset }_{RS}(r_2\wedge
t_1)\triangle t$ with $t\stackrel{\varepsilon }{\Rightarrow }_F\left|
t_1\right. $ and $r\stackrel{a}{\Rightarrow }_F\left| r_2\right. $, as
desired.\TeXButton{End Proof}{\endproof}

\qquad

A simple method for showing that one process simulates another one is to
find a stable ready simulation relating them. It is well known that up-to
technique is a tractable way for such coinduction proof. Here we introduce
the notion of a stable ready simulation up to $\stackunder{\sim }{\sqsubset }%
_{RS}$ as follows.

\quad

{\bf Definition 4.1} (stable ready simulation up to $\stackunder{\sim }{%
\sqsubset }_{RS}$) A binary relation $R\subseteq T(\Sigma _{CLLT})\times
T(\Sigma _{CLLT})$ is said to be a stable ready simulation relation up to $%
\stackunder{\sim }{\sqsubset }_{RS}$ if for any $\left\langle
t,s\right\rangle \in R$, it satisfies (RS1), (RS2), (RS4) in Def. 2.2 and

\begin{description}
\item  {\bf (RS3-up to)} $t\stackrel{a}{\Rightarrow }_F\left| t_1\right. $%
implies $\exists s_1(s\stackrel{a}{\Rightarrow }_F\left| s_1\right. $and $%
\left\langle t_1,s_1\right\rangle \in R\circ \stackunder{\sim }{\sqsubset }%
_{RS})$ for any $a\in Act$.
\end{description}

\qquad

As usual, given a relation $R$ satisfying the above conditions, $R$ itself
is not in general a stable ready simulation relation. But the simple result
below ensures that up-to technique based on the above notion is sound.

\quad

{\bf Lemma 4.2} If $R$ is a stable ready simulation relation up to $%
\stackunder{\sim }{\sqsubset }_{RS}$ then $R\subseteq $ $\stackunder{\sim }{%
\sqsubset }_{RS}$.

\TeXButton{Proof}{\proof}Due to the reflexivity of $\stackunder{\sim }{%
\sqsubset }_{RS}$, we have $R\subseteq $ $R\circ \stackunder{\sim }{%
\sqsubset }_{RS}$. Thus it suffices to show that $R\circ \stackunder{\sim }{%
\sqsubset }_{RS}$is a stable ready simulation. For any pair $\left\langle
s,t\right\rangle \in R\circ \stackunder{\sim }{\sqsubset }_{RS}$, based on
Def. 4.1 and the transitivity of $\stackunder{\sim }{\sqsubset }_{RS}$, it
is straightforward to check that $\left\langle s,t\right\rangle $ satisfies
four conditions in Def. 2.2. \TeXButton{End Proof}{\endproof}

\qquad

{\bf Lemma 4.3} If $p$ $\stackunder{\sim }{\sqsubset }_{RS}u\triangle t$
then $p$ $\stackunder{\sim }{\sqsubset }_{RS}u$. Hence $u\triangle
t\sqsubseteq _{RS}u$ for any $u$ and $t$.

\TeXButton{Proof}{\proof} Set

\begin{center}
$R=\left\{ \left\langle q,s\right\rangle :q\stackunder{\sim }{\sqsubset }%
_{RS}s\triangle r\text{ for some }r\right\} $.
\end{center}

We wish to prove that $R$ is a stable ready simulation relation up to $%
\stackunder{\sim }{\sqsubset }_{RS}$. Let $\left\langle q,s\right\rangle \in
R$. Then $q\stackunder{\sim }{\sqsubset }_{RS}s\triangle r$ for some $r$.
Thus both $q$ and $s\triangle r$ are stable. By item (5) in Lemma 3.1, so is
$s$. Hence (RS1) holds.

(RS2) Suppose $q\notin F$. Due to $q\stackunder{\sim }{\sqsubset }%
_{RS}s\triangle r$, we get $s\triangle r\notin F$, which implies $s\notin F$
by Lemma 3.3 (6).

(RS3-upto) Let $q\stackrel{a}{\Rightarrow }_F\left| q_1\right. $. Since $q%
\stackunder{\sim }{\sqsubset }_{RS}s\triangle r$, by Lemma 4.1, $q_1$ $%
\stackunder{\sim }{\sqsubset }_{RS}(s_1\wedge r_1)\bigtriangleup r$ for some
$r_1$ and $s_1$ such that $r\stackrel{\varepsilon }{\Rightarrow }_F\left|
r_1\right. $ and $s\stackrel{a}{\Rightarrow }_F\left| s_1\right. $. Thus $%
\left\langle q_1,s_1\wedge r_1\right\rangle \in R$. On the other hand, by
item (1) in Lemma 3.11, we get $s_1\wedge r_1\stackunder{\sim }{\sqsubset }%
_{RS}s_1$. Hence $\left\langle q_1,s_1\right\rangle \in R\circ \stackunder{%
\sim }{\sqsubset }_{RS}$ and $s\stackrel{a}{\Rightarrow }_F\left| s_1\right.
$, as desired.

(RS4) If $q\notin F$ then it follows from $q\stackunder{\sim }{\sqsubset }%
_{RS}s\triangle r$ that $I(q)=I(s\triangle r)$, and hence $I(q)=I(s)$ by
Lemma 3.2 (4).\TeXButton{End Proof}{\endproof}

\qquad

{\bf Notation 4.1 }For a more convenient notation, we introduce the
notations below.

(1) Following{\bf \ }[44]{\bf , }the notation $\stackrel{Act}{\Rightarrow }%
_F $ is used to stand for $\stackunder{a\in Act}{\bigcup }\stackrel{a}{%
\Rightarrow }_F$.

(2) The notation $p\sqsubseteq _{RS}^\forall t$ means that $\forall n\in
\omega \forall p_0,$ $p_1,...$ $p_n$ $(p\stackrel{\varepsilon }{\Rightarrow }%
_F\left| p_0\right. \stackrel{Act}{\Rightarrow }_F\left| p_1\right. $ $...$ $%
\stackrel{Act}{\Rightarrow }_F\left| p_n\right. $ implies $p_n\sqsubseteq
_{RS}t)$.

(3) The notation $p\stackunder{\sim }{\sqsubset }_{RS}^\forall t$ means that
$\forall n\in \omega \forall p_0,$ $p_1,...$ $p_n(p\stackrel{Act}{%
\Rightarrow }_F\left| p_0\right. \stackrel{Act}{\Rightarrow }_F\left|
p_1\right. ...\stackrel{Act}{\Rightarrow }_F\left| p_n\right. $ implies $p_n%
\stackunder{\sim }{\sqsubset }_{RS}t)$.

\qquad

The next two results provide a necessary condition for a process to refine $%
\sharp t$ , where the refinement relation is captured by $\stackunder{\sim }{%
\sqsubset }_{RS}$ and $\sqsubseteq _{RS}$ respectively.

\qquad

{\bf Lemma 4.4} If $p\stackunder{\sim }{\sqsubset }_{RS}\sharp t$ then $p%
\stackunder{\sim }{\sqsubset }_{RS}^\forall t$.

\TeXButton{Proof}{\proof}Assume that $p\stackrel{Act}{\Rightarrow }_F\left|
p_0\right. \stackrel{Act}{\Rightarrow }_F\left| p_1\right. ...\stackrel{Act}{%
\Rightarrow }_F\left| p_n\right. $. If it were true that

\begin{center}
$p_n\stackunder{\sim }{\sqsubset }_{RS}(r\wedge t)\bigtriangleup t$ for some
$r$\qquad (4.4.1)
\end{center}

we would have $p_n\stackunder{\sim }{\sqsubset }_{RS}t$ by Lemma 4.3 and
3.11 (1), and hence the proof would be complete. In the following, we thus
intend to prove (4.4.1) by induction on $n$.

For the induction basis $n=0$, we have $p\stackrel{a}{\Rightarrow }_F\left|
p_0\right. $for some $a\in Act$. It follows from $p\stackunder{\sim }{%
\sqsubset }_{RS}\sharp t$ that $p_0\stackunder{\sim }{\sqsubset }_{RS}t_1$
for some $t_1$with $\sharp t\stackrel{a}{\Rightarrow }_F\left| t_1\right. $.
Due to the stableness of $\sharp t$, we get $\sharp t\stackrel{a}{%
\rightarrow }_Ft_2\stackrel{\varepsilon }{\Rightarrow }_F\left| t_1\right. $%
for some $t_2$. Then, by Lemma 3.2 (3), $t_2\equiv (s\wedge t)\bigtriangleup
t$ for some $s$. Further, by Lemma 3.1 (5)(8) and 3.3 (6), it follows from $%
(s\wedge t)\bigtriangleup t\stackrel{\varepsilon }{\Rightarrow }_F\left|
t_1\right. $ that there exist $t_3$ and $s_1$ such that $t\stackrel{%
\varepsilon }{\Rightarrow }_F\left| t_3\right. $, $s\stackrel{\varepsilon }{%
\Rightarrow }_F\left| s_1\right. $ and $t_1\equiv (s_1\wedge
t_3)\bigtriangleup t$. Since $t$ is stable, we get $t\equiv t_3$. Thus $p_0%
\stackunder{\sim }{\sqsubset }_{RS}(s_1\wedge t)\bigtriangleup t$, as
desired.

For the induction step $n=k+1$, suppose that $p\stackrel{Act}{\Rightarrow }%
_F\left| p_0\right. \stackrel{Act}{\Rightarrow }_F\left| p_1\right. ...%
\stackrel{Act}{\Rightarrow }_F\left| p_k\right. \stackrel{a}{\Rightarrow }%
_F\left| p_{k+1}\right. $. By IH, $p_k\stackunder{\sim }{\sqsubset }%
_{RS}(s\wedge t)\bigtriangleup t$ for some $s$. Then, by Lemma 4.1, it
follows from $p_k\stackrel{a}{\Rightarrow }_F\left| p_{k+1}\right. $ and $t%
\stackrel{\tau }{\not \rightarrow }$ that $p_{k+1}\stackunder{\sim }{%
\sqsubset }_{RS}(r\wedge t)\bigtriangleup t$ for some $r$.
\TeXButton{End Proof}{\endproof}

\qquad

This result is of independent interest, but its principal significance is
that it will serve as a stepping stone in demonstrating the next lemma.

\qquad

{\bf Lemma 4.5} $p\sqsubseteq _{RS}\sharp t$ implies $p\sqsubseteq
_{RS}^\forall t$.

\TeXButton{Proof}{\proof}Assume that $p\stackrel{\varepsilon }{\Rightarrow }%
_F\left| p_0\right. \stackrel{Act}{\Rightarrow }_F\left| p_1\right.
\stackrel{Act}{\Rightarrow }_F\left| p_2\right. ...\stackrel{Act}{%
\Rightarrow }_F\left| p_n\right. $. We intend to prove that $p_n\sqsubseteq
_{RS}t$. The argument splits into two cases depending on whether $t$ is
stable.

\begin{description}
\item  Case 1 $t\stackrel{\tau }{\not \rightarrow }$.
\end{description}

So, $\sharp t\stackrel{\tau }{\not \rightarrow }$. Then it follows from $%
p\sqsubseteq _{RS}\sharp t$ and $p\stackrel{\varepsilon }{\Rightarrow }%
_F\left| p_0\right. $ that $p_0\stackunder{\sim }{\sqsubset }_{RS}\sharp t$.
Hence $p_n\stackunder{\sim }{\sqsubset }_{RS}t$ by Lemma 4.4. Consequently, $%
p_n\sqsubseteq _{RS}t$ holds due to $t\stackrel{\tau }{\not \rightarrow }$
and $p_n\stackrel{\tau }{\not \rightarrow }$.

\begin{description}
\item  Case 2 $t\stackrel{\tau }{\rightarrow }$.
\end{description}

Since $p\stackrel{\varepsilon }{\Rightarrow }_F\left| p_0\right. $ and $%
p\sqsubseteq _{RS}\sharp t$, it follows that $p_0\stackunder{\sim }{%
\sqsubset }_{RS}t_0$ for some $t_0$ with $\sharp t\stackrel{\tau }{%
\Rightarrow }_F\left| t_0\right. $. By Lemma 3.1(4) and 3.3(6), there exists
$r$ such that $t_0\equiv r\triangle t$ and $t\stackrel{\tau }{\Rightarrow }%
_F\left| r\right. $. Thus, by Lemma 4.3, we have $p_0\stackunder{\sim }{%
\sqsubset }_{RS}r$. If $n=0$ then $p_0\sqsubseteq _{RS}t$ comes from $p_0%
\stackrel{\tau }{\not \rightarrow }$ and $t\stackrel{\tau }{\Rightarrow }%
_F\left| r\right. $. We now turn to the case $n\geq 1$. Due to $p_0%
\stackunder{\sim }{\sqsubset }_{RS}r\triangle t$, applying Lemma 4.1
repeatedly, it may be proved without any difficulty that $p_n\stackunder{%
\sim }{\sqsubset }_{RS}(s\wedge t_1)\bigtriangleup t$ for some $s$ and $t_1$
such that $t\stackrel{\varepsilon }{\Rightarrow }_F\left| t_1\right. $.
Further, by Lemma 4.3 and 3.11(1), we have $p_n\stackunder{\sim }{\sqsubset }%
_{RS}s\wedge t_1\stackunder{\sim }{\sqsubset }_{RS}t_1$. Then it follows
from $p_n\stackrel{\tau }{\not \rightarrow }$ and $t\stackrel{\varepsilon }{%
\Rightarrow }_F\left| t_1\right. $ that $p_n\sqsubseteq _{RS}t$.
\TeXButton{End Proof}{\endproof}

\qquad

The converse of the above lemma also holds. However, its proof is far from
straightforward. A few of preliminary results are needed. Two results
concerning consistency are given firstly.

\quad

{\bf Lemma 4.6} Let $p$ and $t$ be any process such that $p\sqsubseteq
_{RS}^\forall t$, and put

\begin{center}
$\Omega =\left\{ r\bigtriangleup t:\exists q_0,q_1,...,q_n\left( p\stackrel{%
\varepsilon }{\Rightarrow }_F\left| q_0\right. \stackrel{Act}{\Rightarrow }%
_F\left| q_1\right. ...\stackrel{Act}{\Rightarrow }_F\left| q_n\right.
\stackunder{\sim }{\sqsubset }_{RS}r\text{ }\right) \right\} $.
\end{center}

Then $\Omega $ is a $F-$hole.

\TeXButton{Proof}{\proof}Let $r\bigtriangleup t\in \Omega $. Then there
exist $p_0,p_1,p_2...p_n$ such that $p\stackrel{\varepsilon }{\Rightarrow }%
_F\left| p_0\right. \stackrel{Act}{\Rightarrow }_F\left| p_1\right. ...%
\stackrel{Act}{\Rightarrow }_F\left| p_n\right. $ and $p_n\stackunder{\sim }{%
\sqsubset }_{RS}r$. Let $\Im $ be any proof tree of $Strip(\Gamma
_{CLLT},M_{CLLT})\vdash r\bigtriangleup tF$. Since $p_n\stackunder{\sim }{%
\sqsubset }_{RS}r$ and $p_n\notin F$, we get $r\notin F$. Moreover, $%
r\bigtriangleup t$ is stable due to $r\stackrel{\tau }{\not \rightarrow }$
and Lemma 3.1(5). Thus the last rule applied in $\Im $ is of the format below

\begin{center}
$\dfrac{r\bigtriangleup t\stackrel{a}{\rightarrow }u,\text{ }\left\{
qF:r\bigtriangleup t\stackrel{a}{\rightarrow }q\right\} }{r\bigtriangleup tF}
$ for some $a\in Act$.\qquad (4.6.1)
\end{center}

Due to $p_n\stackunder{\sim }{\sqsubset }_{RS}r$ and $p_n\notin F$, we have $%
I(p_n)=I(r)=I(r\bigtriangleup t)$ by Lemma 3.2(4). Hence $a\in I(p_n)$.
Moreover, by Lemma 3.5 and 3.8, it follows from $p_n\notin F$ that $p_n%
\stackrel{a}{\Rightarrow }_F\left| p_{n+1}\right. $ for some $p_{n+1}$.
Then, due to $p_n\stackunder{\sim }{\sqsubset }_{RS}r$, there exist $r_1$
and $r_2$ with

\begin{center}
$r\stackrel{a}{\rightarrow }_Fr_1\stackrel{\varepsilon }{\Rightarrow }%
_F\left| r_2\right. $ and $p_{n+1}\stackunder{\sim }{\sqsubset }_{RS}r_2$.
\end{center}

Moreover, $p_{n+1}\sqsubseteq _{RS}t$ because of $p\sqsubseteq _{RS}^\forall
t$. Hence $p_{n+1}\stackunder{\sim }{\sqsubset }_{RS}t_1$ for some $t_1$
with $t\stackrel{\varepsilon }{\Rightarrow }_F\left| t_1\right. $. Thus, by
Lemma 3.11 (2), it follows that $p_{n+1}\stackunder{\sim }{\sqsubset }%
_{RS}r_2\wedge t_1$. Hence $(r_2\wedge t_1)\bigtriangleup t\in \Omega $.
Consequently, in order to complete the proof, it is enough to show that $\Im
$ contains a proper subtree with the root labelled with $(r_2\wedge
t_1)\bigtriangleup tF$. Next we shall prove this.

By Lemma 3.2(4) and 3.1(5), it follows from $r\stackrel{a}{\rightarrow }_Fr_1%
\stackrel{\varepsilon }{\Rightarrow }_F\left| r_2\right. $ and $t\stackrel{%
\varepsilon }{\Rightarrow }_F\left| t_1\right. $ that

\begin{center}
$r\bigtriangleup t\stackrel{a}{\rightarrow }(r_1\wedge t)\bigtriangleup t%
\stackrel{\varepsilon }{\Rightarrow }\left| (r_2\wedge t_1)\bigtriangleup
t\right. $.\qquad
\end{center}

Hence, by (4.6.1), $\Im $ contains a proper subtree with the root labelled
with $(r_1\wedge t)\bigtriangleup tF$. Obviously, if $(r_1\wedge
t)\bigtriangleup t$ is stable then $(r_1\wedge t)\bigtriangleup t\equiv
(r_2\wedge t_1)\bigtriangleup t$, and hence $\Im $ contains a node labelled
with $(r_2\wedge t_1)\bigtriangleup tF$, as desired. In the following, we
handle the nontrivial case $(r_1\wedge t)\bigtriangleup t\stackrel{\tau }{%
\rightarrow }$. In such situation, there exist $s_1,s_2,...,s_m$ such that

\begin{description}
\item  $r_1\wedge t\stackrel{\tau }{\rightarrow }s_1\stackrel{\tau }{%
\rightarrow }s_2\stackrel{\tau }{\rightarrow }...\stackrel{\tau }{%
\rightarrow }s_m\stackrel{\tau }{\rightarrow }\left| r_2\wedge t_1\right. $,
and\qquad (4.6.2)

\item  $(r_1\wedge t)\bigtriangleup t\stackrel{\tau }{\rightarrow }%
s_1\bigtriangleup t\stackrel{\tau }{\rightarrow }s_2\bigtriangleup t%
\stackrel{\tau }{\rightarrow }...\stackrel{\tau }{\rightarrow }%
s_m\bigtriangleup t\stackrel{\tau }{\rightarrow }\left| (r_2\wedge
t_1)\bigtriangleup t\right. $.\qquad (4.6.3)
\end{description}

On the other hand, due to $p_{n+1}\notin F$ and $p_{n+1}\stackunder{\sim }{%
\sqsubset }_{RS}r_2\wedge t_1$, we get $r_2\wedge t_1\notin F$. Then, by
Lemma 3.6(1) and (4.6.2), it is easy to see that

\begin{center}
$r_1\wedge t\notin F$ and $s_i\notin F$ with $1\leq i\leq m$.\qquad
\end{center}

Thus, for each $u\in \left\{ r_1\wedge t\right\} \bigcup \left\{ s_i:1\leq
i\leq m\right\} $, the last rule applied in any proof tree of $Strip(\Gamma
_{CLLT},M_{CLLT})\vdash u\bigtriangleup tF$ must be of the format below

\begin{center}
$\dfrac{u\bigtriangleup t\stackrel{\tau }{\rightarrow }r,\text{ }\left\{
qF:u\bigtriangleup t\stackrel{\tau }{\rightarrow }q\right\} }{%
u\bigtriangleup tF}$.
\end{center}

Therefore, by (4.6.3), it follows that $\Im $ contains a proper subtree with
the root labelled with $(r_2\wedge t_1)\bigtriangleup tF$, as desired.
\TeXButton{End Proof}{\endproof}

\qquad

With the helping of this result, we shall prove the assertion below, which
is a crucial part of the proof for the converse of Lemma 4.5.

\qquad

{\bf Lemma 4.7} If $p\sqsubseteq _{RS}^\forall t$ and $p\notin F$ then $%
\sharp t\notin F$.

\TeXButton{Proof}{\proof}Since $p\notin F$, by Lemma 3.8, there exists $q_0$
such that $p\stackrel{\varepsilon }{\Rightarrow }_F\left| q_0\right. $. Then
$q_0\sqsubseteq _{RS}t$ due to $p\sqsubseteq _{RS}^\forall t$. We
distinguish two cases depending on whether $t$ is stable.

\begin{description}
\item  Case 1 $t\stackrel{\tau }{\rightarrow }$.
\end{description}

In such situation, since $q_0\sqsubseteq _{RS}t$, there exists $t_1$ such
that $q_0\stackunder{\sim }{\sqsubset }_{RS}t_1$ and $t\stackrel{\tau }{%
\Rightarrow }_F\left| t_1\right. $. Then $\sharp t\stackrel{\tau }{%
\Rightarrow }\left| t_1\bigtriangleup t\right. $ by Lemma 3.1(4)(5). On the
other hand, by Lemma 3.9 and 4.6, it follows from $p\stackrel{\varepsilon }{%
\Rightarrow }_F\left| q_0\right. \stackunder{\sim }{\sqsubset }_{RS}t_1$
that $t_1\bigtriangleup t\notin F$. Thus $\sharp t\notin F$ by Lemma 3.6(2).

\begin{description}
\item  Case 2 $t\stackrel{\tau }{\not \rightarrow }$.
\end{description}

Assume that $\sharp t\in F$ and let $\Im $ be any proof tree of $%
Strip(\Gamma _{CLLT},M_{CLLT})\vdash \sharp tF$. Since $t$ is stable, so is $%
\sharp t$ by Lemma 3.1(4). Moreover, it follows from $t\stackrel{\tau }{\not
\rightarrow },$ $p\stackrel{\varepsilon }{\Rightarrow }_F\left| q_0\right. $
and $q_0\sqsubseteq _{RS}t$ that $q_0\stackunder{\sim }{\sqsubset }_{RS}t$
and $t\notin F$. Thus the last rule applied in $\Im $ is of the format below

\begin{center}
$\dfrac{\sharp t\stackrel{a}{\rightarrow }u,\text{ }\left\{ qF:\sharp t%
\stackrel{a}{\rightarrow }q\right\} }{\sharp tF}$ for some $a\in Act$.\qquad
(4.7.1)
\end{center}

Since $q_0\stackunder{\sim }{\sqsubset }_{RS}t$ and $q_0\notin F$, by Lemma
3.2(3), we have $I(q_0)=I(t)=I(\sharp t)$. Hence $a\in I(q_0)$. Further, by
Lemma 3.5 and 3.8, it follows from $q_0\notin F$ that $q_0\stackrel{a}{%
\Rightarrow }_F\left| q_1\right. $ for some $q_1$. Thus there exist $t_1$
and $t_2$ such that

\begin{center}
$t\stackrel{a}{\rightarrow }_Ft_1\stackrel{\varepsilon }{\Rightarrow }%
_F\left| t_2\right. $ and $q_1\stackunder{\sim }{\sqsubset }_{RS}t_2.\qquad $%
(4.7.2)
\end{center}

Clearly, we also have

\begin{center}
$\sharp t\stackrel{a}{\rightarrow }(t_1\wedge t)\bigtriangleup t\stackrel{%
\varepsilon }{\Rightarrow }\left| (t_2\wedge t)\bigtriangleup t\right. $%
.\qquad (4.7.3)\qquad
\end{center}

Since $p\stackrel{\varepsilon }{\Rightarrow }_F\left| q_0\right. \stackrel{a%
}{\Rightarrow }_F\left| q_1\right. $ and $p\sqsubseteq _{RS}^\forall t$, we
obtain $q_1\sqsubseteq _{RS}t$. Then $q_1\stackunder{\sim }{\sqsubset }%
_{RS}t $ because of $t\stackrel{\tau }{\not \rightarrow }$, which, together
with (4.7.2), implies that $q_1\stackunder{\sim }{\sqsubset }_{RS}t_2\wedge t
$ by Lemma 3.11(2). From this and $p\stackrel{\varepsilon }{\Rightarrow }%
_F\left| q_0\right. \stackrel{a}{\Rightarrow }_F\left| q_1\right. $, we
conclude $(t_2\wedge t)\bigtriangleup t\notin F$ by Lemma 3.9 and 4.6. Then,
by Lemma 3.6(2), it follows from (4.7.3) that $(t_1\wedge t)\bigtriangleup
t\notin F$. But we also have $(t_1\wedge t)\bigtriangleup t\in F$ due to
(4.7.1) and (4.7.3), a contradiction.\qquad \TeXButton{End Proof}{\endproof}

\qquad

In addition to preceding two lemmas, the next result will be applied in
demonstrating the converse of Lemma 4.5.

\quad

{\bf Lemma 4.8} Let $p$ and $t$ be any process such that $p\sqsubseteq
_{RS}^\forall t$. For any process $u$ and $v$, if $\exists
u_0,u_1,u_2...u_{n-1}$ $(p\stackrel{\varepsilon }{\Rightarrow }_F\left|
u_0\right. \stackrel{Act}{\Rightarrow }_F\left| u_1\right. ...\stackrel{Act}{%
\Rightarrow }_F\left| u_{n-1}\right. \stackrel{Act}{\Rightarrow }_F\left|
u\right. )$ \footnote{%
It means $p\stackrel{\varepsilon }{\Rightarrow }_F\left| u\right. $ whenever
$n=0$.} and $u\stackunder{\sim }{\sqsubset }_{RS}v$ then $u\stackunder{\sim
}{\sqsubset }_{RS}v\bigtriangleup t$.

\TeXButton{Proof}{\proof} Set

\begin{center}
$R=\left\{ \left\langle q,r\bigtriangleup t\right\rangle :\exists
q_0,q_1,...,q_{n-1}\left(
\begin{array}{c}
q
\stackunder{\sim }{\sqsubset }_{RS}r\text{ and} \\ p\stackrel{\varepsilon }{%
\Rightarrow }_F\left| q_0\right. \stackrel{Act}{\Rightarrow }_F\left|
q_1\right. ...\stackrel{Act}{\Rightarrow }_F\left| q_{n-1}\right. \stackrel{%
Act}{\Rightarrow }_F\left| q\right.
\end{array}
\right) \right\} $.
\end{center}

Obviously, it suffices to show that $R$ is a stable ready simulation
relation. Suppose $\left\langle q,r\bigtriangleup t\right\rangle \in R$.
Then it is easy to see that both $q$ and $r\bigtriangleup t$ are stable, and
$r\bigtriangleup t\notin F$ by Lemma 3.9 and 4.6. Moreover, by Lemma 3.2(4),
since $q\notin F$ and $q\stackunder{\sim }{\sqsubset }_{RS}r,$ we also have $%
I(q)=I(r)=I(r\bigtriangleup t)$. Thus it remains only to prove that the pair
$\left\langle q,r\bigtriangleup t\right\rangle $ satisfies (RS3).

Let $q\stackrel{a}{\Rightarrow }_F\left| s\right. $. Then $s\sqsubseteq
_{RS}t$ due to $p\sqsubseteq _{RS}^\forall t$. Moreover, it follows from $q%
\stackunder{\sim }{\sqsubset }_{RS}r$ that $s\stackunder{\sim }{\sqsubset }%
_{RS}r_1$ for some $r_1$ with $r\stackrel{a}{\Rightarrow }_F\left|
r_1\right. $. On the other hand, since $s\sqsubseteq _{RS}t$ and $s\stackrel{%
\varepsilon }{\Rightarrow }_F\left| s\right. $, we have $s\stackunder{\sim }{%
\sqsubset }_{RS}t_1$ for some $t_1$ such that $t\stackrel{\varepsilon }{%
\Rightarrow }_F\left| t_1\right. $. Hence $s\stackunder{\sim }{\sqsubset }%
_{RS}r_1\wedge t_1$ by Lemma 3.11 (2). Thus $\left\langle s,(r_1\wedge
t_1)\bigtriangleup t\right\rangle \in R$.

Next we shall show that $r\bigtriangleup t\stackrel{a}{\Rightarrow }_F\left|
(r_1\wedge t_1)\bigtriangleup t\right. $. Since $r\stackrel{a}{\Rightarrow }%
_F\left| r_1\right. $ and $r\stackrel{\tau }{\not \rightarrow }$, we have $r%
\stackrel{a}{\rightarrow }_Fv\stackrel{\varepsilon }{\Rightarrow }_F\left|
r_1\right. $ for some $v$. Then, by Lemma 3.2(4), it follows that $%
r\bigtriangleup t\stackrel{a}{\rightarrow }(v\wedge t)\bigtriangleup t$.
Further, by Lemma 3.1(5), it follows from $t\stackrel{\varepsilon }{%
\Rightarrow }_F\left| t_1\right. $ and $v\stackrel{\varepsilon }{\Rightarrow
}_F\left| r_1\right. $ that

\begin{center}
$r\bigtriangleup t\stackrel{a}{\rightarrow }(v\wedge t)\bigtriangleup t%
\stackrel{\varepsilon }{\Rightarrow }\left| (r_1\wedge t_1)\bigtriangleup
t\right. $.$\qquad $(4.8.1)
\end{center}

Moreover, since $q\stackrel{a}{\Rightarrow }_F\left| s\right. \stackunder{%
\sim }{\sqsubset }_{RS}r_1\wedge t_1$, by Lemma 3.9 and 4.6, we get $%
(r_1\wedge t_1)\bigtriangleup t\notin F$. Then, by Lemma 3.6(2), it follows
from $r\bigtriangleup t\notin F$ and (4.8.1) that $r\bigtriangleup t%
\stackrel{a}{\Rightarrow }_F\left| (r_1\wedge t_1)\bigtriangleup t\right. $,
as desired. \TeXButton{End Proof}{\endproof}

\qquad

We are now ready to prove the converse of Lemma 4.5.

\quad

{\bf Lemma 4.9} $p\sqsubseteq _{RS}^\forall t$ implies $p\sqsubseteq
_{RS}\sharp t$.

\TeXButton{Proof}{\proof}Let $p\stackrel{\varepsilon }{\Rightarrow }_F\left|
s\right. $. It is enough to find $q$ such that $\sharp t\stackrel{%
\varepsilon }{\Rightarrow }_F\left| q\right. $ and $s\stackunder{\sim }{%
\sqsubset }_{RS}q$. We consider two cases below.

\begin{description}
\item  Case 1 $t\stackrel{\tau }{\rightarrow }$.
\end{description}

Since $s\sqsubseteq _{RS}t$ and $s\stackrel{\varepsilon }{\Rightarrow }%
_F\left| s\right. $, there exists $r$ such that $s\stackunder{\sim }{%
\sqsubset }_{RS}r$ and $t\stackrel{\tau }{\Rightarrow }_F\left| r\right. $.
Then $s\stackunder{\sim }{\sqsubset }_{RS}r\bigtriangleup t$ by Lemma 4.8.
On the other hand, by Lemma 3.1(4)(5), it follows from $t\stackrel{\tau }{%
\Rightarrow }_F\left| r\right. $ that $\sharp t\stackrel{\tau }{\Rightarrow }%
\left| r\bigtriangleup t\right. $. Moreover, $r\bigtriangleup t\notin F$ due
to $s\stackunder{\sim }{\sqsubset }_{RS}r\bigtriangleup t$ and $s\notin F$.
Hence $\sharp t\stackrel{\tau }{\Rightarrow }_F\left| r\bigtriangleup
t\right. $ by Lemma 3.6(2). Consequently, $r\bigtriangleup t$ is exactly one
that we seek.

\begin{description}
\item  Case 2 $t\stackrel{\tau }{\not \rightarrow }$.
\end{description}

In such situation, since $s\sqsubseteq _{RS}t$ and $s\stackrel{\varepsilon }{%
\Rightarrow }_F\left| s\right. $, we have $s\stackunder{\sim }{\sqsubset }%
_{RS}t$. Moreover, $\sharp t$ is stable because of $t\stackrel{\tau }{\not
\rightarrow }$. To complete the proof, it suffices to prove that $s%
\stackunder{\sim }{\sqsubset }_{RS}\sharp t$. Put

\begin{center}
$R=\left\{ \left\langle s,\sharp t\right\rangle \right\} \bigcup \stackunder{%
\sim }{\sqsubset }_{RS}$.
\end{center}

We intend to show that $R$ is a stable ready simulation relation. Clearly,
since both $s$ and $\sharp t$ are stable, it is enough to prove that the
pair $\left\langle s,\sharp t\right\rangle $ satisfies (RS2)-(RS4). By Lemma
4.7, we have $\sharp t\notin F$. So, $\left\langle s,\sharp t\right\rangle $
satisfies (RS2). Moreover, by Lemma 3.2(3), it follows from $s\notin F$ and $%
s\stackunder{\sim }{\sqsubset }_{RS}t$ that $I(s)=I(t)=I(\sharp t)$, that
is, such pair satisfies (RS4). The remaining work has then to be spent on
checking (RS3).

Let $s\stackrel{a}{\Rightarrow }_F\left| q\right. $. Clearly, it suffices to
find a process $w$ such that $\sharp t\stackrel{a}{\Rightarrow }_F\left|
w\right. $ and $q\stackunder{\sim }{\sqsubset }_{RS}w$. It follows from $s%
\stackunder{\sim }{\sqsubset }_{RS}t$ that $q\stackunder{\sim }{\sqsubset }%
_{RS}t_1$ for some $t_1$ such that $t\stackrel{a}{\Rightarrow }_F\left|
t_1\right. $. Then, due to $t\stackrel{\tau }{\not \rightarrow }$, we have $t%
\stackrel{a}{\rightarrow }_Fv\stackrel{\varepsilon }{\Rightarrow }_F\left|
t_1\right. $ for some $v$. Hence, by Lemma 3.2(3) and 3.1(5), it follows that

\begin{center}
$\sharp t\stackrel{a}{\rightarrow }(v\wedge t)\bigtriangleup t\stackrel{%
\varepsilon }{\Rightarrow }\left| (t_1\wedge t)\bigtriangleup t\right. $.
\quad (4.9.1)
\end{center}

On the other hand, since $p\stackrel{\varepsilon }{\Rightarrow }_F\left|
s\right. \stackrel{a}{\Rightarrow }_F\left| q\right. $, $p\sqsubseteq
_{RS}^\forall t$ and $t\stackrel{\tau }{\not \rightarrow }$, we get $q%
\stackunder{\sim }{\sqsubset }_{RS}t$. Thus $q\stackunder{\sim }{\sqsubset }%
_{RS}t_1\wedge t$ by Lemma 3.11(2). Further, by Lemma 4.8, it follows that $q%
\stackunder{\sim }{\sqsubset }_{RS}(t_1\wedge t)\bigtriangleup t$ , and
hence $(t_1\wedge t)\bigtriangleup t\notin F$. Then, by Lemma 3.6(2), it
comes from $\sharp t\notin F$ and (4.9.1) that $\sharp t\stackrel{a}{%
\Rightarrow }_F\left| (t_1\wedge t)\bigtriangleup t\right. $. Consequently,
the process $(t_1\wedge t)\bigtriangleup t$ is one that we need.
\TeXButton{End Proof}{\endproof}

\qquad

The development so far can be summarized in the following theorem, which
provides a natural and intrinsic characterization of processes that refine
ones with the format $\sharp t$.

\qquad

{\bf Theorem 4.1} For any process $p$ and $t$, we have

(1) $p\sqsubseteq _{RS}\sharp t$ iff $p\sqsubseteq _{RS}^\forall t$.

(2) $p\stackunder{\sim }{\sqsubset }_{RS}\sharp t$ iff $p\stackunder{\sim }{%
\sqsubset }_{RS}^\forall t$ whenever $p$ and $t$ are stable.

\TeXButton{Proof}{\proof}Immediately follows from Lemma 4.9, 4.5 and 4.4. In
particular, by Lemma 4.9, it is a simple matter to verify the implication
from right to left in item (2).\TeXButton{End Proof}{\endproof}

\qquad

As an immediate consequence of the above theorem, we have the result below,
which reveals that both $\stackunder{\sim }{\sqsubset }_{RS}$ and $%
\sqsubseteq _{RS}$ are precongruent w.r.t the operator $\sharp $.

\quad

{\bf Corollary 4.1 (}Monotonicity Law for $\sharp ${\bf )} $t\sqsubseteq
_{RS}s$ implies $\sharp t\sqsubseteq _{RS}\sharp s$. Hence $\sharp t%
\stackunder{\sim }{\sqsubset }_{RS}\sharp s$ whenever $t\stackunder{\sim }{%
\sqsubset }_{RS}s$.

\TeXButton{Proof}{\proof}Suppose that $t\sqsubseteq _{RS}s$. Then it follows
from Theorem 4.1 and the transitivity of $\sqsubseteq _{RS}$ that, for any
process $p$, $p$ $\sqsubseteq _{RS}\sharp t$ implies $p$ $\sqsubseteq
_{RS}\sharp s$. Further, due to the reflexivity of $\sqsubseteq _{RS}$, we
have $\sharp t\sqsubseteq _{RS}\sharp s$. \TeXButton{End Proof}{\endproof}

\quad

We conclude this section with proving that $\sqsubseteq _{RS}$ is also
precongruent w.r.t the operator $\bigtriangleup $. To this end, a
preliminary result concerning inconsistency predicate is given below.
Although it can be proved by an analogous argument of Lemma 4.6, for the
sake of integrality, we still show it in detail.

\quad

{\bf Lemma 4.10} The set $\Omega $ is a $F-$hole, where $\Omega $ is given as

\begin{center}
$\Omega =\left\{ r\bigtriangleup t:\exists p,u\left( p\stackunder{\sim }{%
\sqsubset }_{RS}r\text{ , }u\sqsubseteq _{RS}t\text{ and }p\bigtriangleup
u\notin F\right) \right\} $.
\end{center}

\TeXButton{Proof}{\proof}Suppose $r\bigtriangleup t\in \Omega $. Then there
exist $p$ and $u$ such that $p\bigtriangleup u\notin F$, $u\sqsubseteq
_{RS}t $ and $p\stackunder{\sim }{\sqsubset }_{RS}r$. Let $\Im $ be any
proof tree of $Strip(\Gamma _{CLLT},M_{CLLT})\vdash r\bigtriangleup tF$.
Since $p\bigtriangleup u\notin F$, we have $p\notin F$ by Lemma 3.3(6).
Hence $r\notin F$ due to $p\stackunder{\sim }{\sqsubset }_{RS}r$. Moreover,
by Lemma 3.1(5), $r\bigtriangleup t$ is stable because of $r\stackrel{\tau }{%
\not \rightarrow }$. Thus the last rule applied in $\Im $ has the format
below

\begin{center}
$\dfrac{r\bigtriangleup t\stackrel{a}{\rightarrow }w,\text{ }\left\{
qF:r\bigtriangleup t\stackrel{a}{\rightarrow }q\right\} }{r\bigtriangleup tF}
$ for some $a\in Act$.\qquad (4.10.1)
\end{center}

By Lemma 3.2(4), since $p\stackunder{\sim }{\sqsubset }_{RS}r$ and $p\notin
F $, we get $I(p)=I(r)=I(r\bigtriangleup t)$. Hence $a\in
I(p)=I(p\bigtriangleup u)$. Moreover, by Lemma 3.5 and 3.8, it follows from $%
p\bigtriangleup u\notin F$ that $p\bigtriangleup u\stackrel{a}{\rightarrow }%
_Fs\stackrel{\varepsilon }{\Rightarrow }_F\left| v\right. $ for some $s$ and
$v$. Further, by Lemma 3.2(4), we obtain $s\equiv (p_1\wedge
u)\bigtriangleup u$ and $v\equiv (p_2\wedge u_1)\bigtriangleup u$ for some $%
p_1,$ $p_2$ and $u_1$ with

\begin{center}
$p\stackrel{a}{\rightarrow }_Fp_1\stackrel{\varepsilon }{\Rightarrow }%
_F\left| p_2\right. $ and $u\stackrel{\varepsilon }{\Rightarrow }_F\left|
u_1\right. $.
\end{center}

Then it follows from $p\stackunder{\sim }{\sqsubset }_{RS}r$ and $%
u\sqsubseteq _{RS}t$ that there exist $t_1,r_1$ and $r_2$ such that $p_2%
\stackunder{\sim }{\sqsubset }_{RS}r_2$ with $r\stackrel{a}{\rightarrow }%
_Fr_1\stackrel{\varepsilon }{\Rightarrow }_F\left| r_2\right. $, and $u_1%
\stackunder{\sim }{\sqsubset }_{RS}t_1$ with $t\stackrel{\varepsilon }{%
\Rightarrow }_F\left| t_1\right. $. By Lemma 3.11(1)(2), 3.2(4) and 3.1(5),
this clearly forces $p_2\wedge u_1\stackunder{\sim }{\sqsubset }%
_{RS}r_2\wedge t_1$ and

\begin{center}
$r\bigtriangleup t\stackrel{a}{\rightarrow }(r_1\wedge t)\bigtriangleup t%
\stackrel{\varepsilon }{\Rightarrow }\left| (r_2\wedge t_1)\bigtriangleup
t\right. $.\qquad (4.10.2)
\end{center}

Further, due to $v\equiv (p_2\wedge u_1)\bigtriangleup u\notin F$ and $%
u\sqsubseteq _{RS}t$, we have

\begin{center}
$(r_2\wedge t_1)\bigtriangleup t\in \Omega $. \qquad
\end{center}

Then it remains to show that $\Im $ contains a proper subtree with the root
labelled with $(r_2\wedge t_1)\bigtriangleup tF$. By (4.10.1) and (4.10.2), $%
\Im $ contains a proper subtree with the root labelled with $(r_1\wedge
t)\bigtriangleup tF$. If $(r_1\wedge t)\bigtriangleup t$ is stable then $\Im
$ contains a node labelled with $(r_2\wedge t_1)\bigtriangleup tF$ because
of $(r_1\wedge t)\bigtriangleup t\equiv (r_2\wedge t_1)\bigtriangleup t$. In
the following, we consider another case $(r_1\wedge t)\bigtriangleup t%
\stackrel{\tau }{\rightarrow }$. In such situation, there exist $%
s_1,s_2,...,s_m$ such that

\begin{description}
\item  $r_1\wedge t\stackrel{\tau }{\rightarrow }s_1\stackrel{\tau }{%
\rightarrow }s_2\stackrel{\tau }{\rightarrow }...\stackrel{\tau }{%
\rightarrow }s_m\stackrel{\tau }{\rightarrow }\left| r_2\wedge t_1\right. $,
and\qquad \qquad (4.10.3)

\item  $(r_1\wedge t)\bigtriangleup t\stackrel{\tau }{\rightarrow }%
s_1\bigtriangleup t\stackrel{\tau }{\rightarrow }...\stackrel{\tau }{%
\rightarrow }s_m\bigtriangleup t\stackrel{\tau }{\rightarrow }\left|
(r_2\wedge t_1)\bigtriangleup t\right. .$\qquad (4.10.4)
\end{description}

Since $v\equiv (p_2\wedge u_1)\bigtriangleup u\notin F$, by Lemma 3.3(6), we
get $p_2\wedge u_1\notin F$. Then $r_2\wedge t_1\notin F$ due to $p_2\wedge
u_1\stackunder{\sim }{\sqsubset }_{RS}r_2\wedge t_1$. Hence, by Lemma 3.6
(1) and (4.10.3), it is evident that

\begin{center}
$r_1\wedge t\notin F$ and $s_i\notin F$ for each $i$ with 1$\leq i\leq m$%
.\qquad
\end{center}

Thus, for each $w\in \left\{ r_1\wedge t\right\} \bigcup \left\{ s_i:1\leq
i\leq m\right\} $, the last rule applied in any proof tree of $Strip(\Gamma
_{CLLT},M_{CLLT})\vdash w\bigtriangleup tF$ must be of the format below

\begin{center}
$\dfrac{w\bigtriangleup t\stackrel{\tau }{\rightarrow }u,\text{ }\left\{
qF:w\bigtriangleup t\stackrel{\tau }{\rightarrow }q\right\} }{%
w\bigtriangleup tF}$.
\end{center}

Consequently, by (4.10.4), it is not difficult to see that $\Im $ contains a
proper subtree with the root labelled with $(r_2\wedge t_1)\bigtriangleup tF$%
, as desired. \TeXButton{End Proof}{\endproof}

\qquad

{\bf Theorem 4.2 (}Monotonicity Law for $\bigtriangleup ${\bf )} For any
process $t_i$, $p_i$ ($i=1,2$), we have

(1) If $p_1\stackunder{\sim }{\sqsubset }_{RS}p_2$ and $t_1\sqsubseteq
_{RS}t_2$ then $p_1\bigtriangleup t_1\stackunder{\sim }{\sqsubset }%
_{RS}p_2\bigtriangleup t_2$.

(2) If $p_1\sqsubseteq _{RS}p_2$ and $t_1\sqsubseteq _{RS}t_2$ then $%
p_1\bigtriangleup t_1\sqsubseteq _{RS}p_2\bigtriangleup t_2$. Hence $%
\sqsubseteq _{RS}$ is a precongruence w.r.t the operator $\bigtriangleup $.

\TeXButton{Proof}{\proof}(1) Set

\begin{center}
$R=\left\{ \left\langle p\bigtriangleup t,\text{ }q\bigtriangleup
w\right\rangle :p\stackunder{\sim }{\sqsubset }_{RS}q\text{ and }%
t\sqsubseteq _{RS}w\right\} $.
\end{center}

It suffices to show that $R$ is a stable ready simulation relation. Let $%
\left\langle p\bigtriangleup t,\text{ }q\bigtriangleup w\right\rangle $ $\in
R$. Hence $p\stackunder{\sim }{\sqsubset }_{RS}q$ and $t\sqsubseteq _{RS}w$.
Then, by item (5) in Lemma 3.1, both $p\bigtriangleup t$ and $%
q\bigtriangleup w$ are stable, that is, (RS1) holds. Moreover, it
immediately follows from Lemma 3.9 and 4.10 that (RS2) holds.

(RS3) Let $p\bigtriangleup t\stackrel{a}{\Rightarrow }_F\left| u\right. $.
Hence $p\bigtriangleup t\stackrel{a}{\rightarrow }_Fs\stackrel{\varepsilon }{%
\Rightarrow }_F\left| u\right. $ for some $s$ due to (RS1). Further, by
Lemma 3.2(4), we get $s\equiv (p_1\wedge t)\bigtriangleup t$ and $u\equiv
(p_2\wedge t_1)\bigtriangleup t$ for some $p_1$, $p_2$ and $t_1$ such that

\begin{center}
$p\stackrel{a}{\rightarrow }_Fp_1\stackrel{\varepsilon }{\Rightarrow }%
_F\left| p_2\right. $ and $t\stackrel{\varepsilon }{\Rightarrow }_F\left|
t_1\right. $.
\end{center}

Then it follows from $p\stackunder{\sim }{\sqsubset }_{RS}q$ and $%
t\sqsubseteq _{RS}w$ that there exist $w_1,$ $q_1$ and $q_2$ such that $t_1%
\stackunder{\sim }{\sqsubset }_{RS}w_1$ with $w\stackrel{\varepsilon }{%
\Rightarrow }_F\left| w_1\right. $, and $p_2\stackunder{\sim }{\sqsubset }%
_{RS}q_2$ with $q\stackrel{a}{\rightarrow }_Fq_1\stackrel{\varepsilon }{%
\Rightarrow }_F\left| q_2\right. $. Hence, by Lemma 3.2(4) and 3.1(5), we
obtain

\begin{center}
$q\bigtriangleup w\stackrel{a}{\rightarrow }(q_1\wedge w)\bigtriangleup w%
\stackrel{\varepsilon }{\Rightarrow }\left| (q_2\wedge w_1)\bigtriangleup
w\right. $. \quad (4.2.1)
\end{center}

Moreover, by Lemma 3.11 (1)(2), we have $p_2\wedge t_1\stackunder{\sim }{%
\sqsubset }_{RS}q_2\wedge w_1$. Combining this with $t\sqsubseteq _{RS}w$ we
conclude that

\begin{center}
$\left\langle (p_2\wedge t_1)\bigtriangleup t\text{, }(q_2\wedge
w_1)\bigtriangleup w\right\rangle \in R$.
\end{center}

On the other hand, by Lemma 3.9 and 4.10, it follows from $p\bigtriangleup
t\notin F$ and $\left\langle p\bigtriangleup t,\text{ }q\bigtriangleup
w\right\rangle $ $\in R$ that $q\bigtriangleup w\notin F$. Similarly, we
also have $(q_2\wedge w_1)\bigtriangleup w\notin F$. Further, by (RS2),
(4.2.1) and Lemma 3.6 (2), it follows that

\begin{center}
$q\bigtriangleup w\stackrel{a}{\rightarrow }_F(q_1\wedge w)\bigtriangleup w%
\stackrel{\varepsilon }{\Rightarrow }_F\left| (q_2\wedge w_1)\bigtriangleup
w\right. $.
\end{center}

(RS4) Assume that $p\bigtriangleup t\notin F$. Then $p\notin F$ by Lemma
3.3(6). Thus it follows from $p\stackunder{\sim }{\sqsubset }_{RS}q$ that $%
I(p)=I(q)$. Further, by Lemma 3.2 (4), we get $I(p\bigtriangleup
t)=I(p)=I(q)=I(q\bigtriangleup w)$, as desired.

(2) Suppose $p_1\bigtriangleup t_1\stackrel{\varepsilon }{\Rightarrow }%
_F\left| u\right. $. The task is now to seek $t$ such that $%
p_2\bigtriangleup t_2\stackrel{\varepsilon }{\Rightarrow }_F\left| t\right. $
and $u\stackunder{\sim }{\sqsubset }_{RS}t$. By Lemma 3.1(5), we get $%
u\equiv s\bigtriangleup t_1$ for some $s$ with $p_1\stackrel{\varepsilon }{%
\Rightarrow }_F\left| s\right. $. Moreover, since $p_1\sqsubseteq _{RS}p_2$,
we have $s\stackunder{\sim }{\sqsubset }_{RS}w$ for some $w$ with $p_2%
\stackrel{\varepsilon }{\Rightarrow }_F\left| w\right. $. Then $%
p_2\bigtriangleup t_2\stackrel{\varepsilon }{\Rightarrow }\left|
w\bigtriangleup t_2\right. $ by Lemma 3.1(5). On the other hand, by Lemma
3.9 and 4.10, it follows from $s\stackunder{\sim }{\sqsubset }_{RS}w$, $%
t_1\sqsubseteq _{RS}t_2$ and $u\equiv s\bigtriangleup t_1\notin F$ that $%
w\bigtriangleup t_2\notin F$. So, by Lemma 3.6, we have $p_2\bigtriangleup
t_2\stackrel{\varepsilon }{\Rightarrow }_F\left| w\bigtriangleup t_2\right. $%
. Moreover, by item (1) in this lemma, it follows from $s\stackunder{\sim }{%
\sqsubset }_{RS}w$ and $t_1\sqsubseteq _{RS}t_2$ that $u\equiv
s\bigtriangleup t_1\stackunder{\sim }{\sqsubset }_{RS}w\bigtriangleup t_2$.
Therefore $w\bigtriangleup t_2$ indeed is one that we need.
\TeXButton{End Proof}{\endproof}

\quad

By the way, according to item (1) in the above theorem, it is obvious that $%
\stackunder{\sim }{\sqsubset }_{RS}$ is also a precongruece w.r.t the
operator $\bigtriangleup $, that is, $p_1\bigtriangleup t_1\stackunder{\sim
}{\sqsubset }_{RS}p_2\bigtriangleup t_2$ holds whenever $p_1\stackunder{\sim
}{\sqsubset }_{RS}p_2$ and $t_1\stackunder{\sim }{\sqsubset }_{RS}t_2$.

\section{The operator $\varpi $ \qquad}

This section will focus on the temporal operator $\varpi $, and characterize
processes that refine processes with the topmost operator $\varpi $. Since
the auxiliary operator $\odot $ plays an important role in describing the
behavior of $\varpi $, we begin with exploring the properties of it. We
first want to indicate some simple properties.

\quad

{\bf Lemma 5.1} For any process $s,t,p$ and $q$, we have

(1) If $s\stackunder{\sim }{\sqsubset }_{RS}t\odot (p\varpi q)$ then $s%
\stackunder{\sim }{\sqsubset }_{RS}t$.

(2) $t\odot (p\varpi q)\stackunder{\sim }{\sqsubset }_{RS}t$ whenever $t%
\stackrel{\tau }{\not \rightarrow }$.

(3) $t\odot (p\varpi q)\sqsubseteq _{RS}t$.

\TeXButton{Proof}{\proof} (1) Set

\begin{center}
$R=\left\{ \left\langle u,v\right\rangle :u\stackunder{\sim }{\sqsubset }%
_{RS}v\odot (r\varpi w)\right\} \bigcup \stackunder{\sim }{\sqsubset }_{RS}$.
\end{center}

We intend to show that $R$ is a stable ready simulation up to $\stackunder{%
\sim }{\sqsubset }_{RS}$. Suppose that $u\stackunder{\sim }{\sqsubset }%
_{RS}v\odot (r\varpi w)$. It is straightforward to verify that the pair $%
\left\langle u,v\right\rangle $ satisfies (RS1), (RS2) and (RS4). To deal
with (RS3-upto), we suppose $u\stackrel{a}{\Rightarrow }_F\left| u_1\right. $%
. It suffices to find $v_1$ such that $v\stackrel{a}{\Rightarrow }_F\left|
v_1\right. $ and $\left\langle u_1,v_1\right\rangle \in R\circ $ $%
\stackunder{\sim }{\sqsubset }_{RS}$.

Clearly, it follows from $u\stackunder{\sim }{\sqsubset }_{RS}v\odot
(r\varpi w)$ that $u_1\stackunder{\sim }{\sqsubset }_{RS}t$ for some $t$
with $v\odot (r\varpi w)\stackrel{a}{\Rightarrow }_F\left| t\right. $. Since
$v\odot (r\varpi w)$ is stable, there exists $t_1$ such that $v\odot
(r\varpi w)\stackrel{a}{\rightarrow }_Ft_1\stackrel{\varepsilon }{%
\Rightarrow }_F\left| t\right. $. We proceed by considering two cases
depending on the last rule applied in the proof tree of $Strip(\Gamma
_{CLLT},M_{CLLT})\vdash v\odot (r\varpi w)\stackrel{a}{\rightarrow }t_1$.

\begin{description}
\item  Case 1 $\dfrac{v\stackrel{a}{\rightarrow }s}{v\odot (r\varpi w)%
\stackrel{a}{\rightarrow }s\wedge w}$
\end{description}

Then $t_1\equiv s\wedge w$. Moreover, by Lemma 3.1(8), $t\equiv s_1\wedge
w_1 $ for some $s_1$, $w_1$ such that $s\stackrel{\varepsilon }{\Rightarrow }%
_F\left| s_1\right. $ and $w\stackrel{\varepsilon }{\Rightarrow }_F\left|
w_1\right. $. Thus $v\stackrel{a}{\Rightarrow }_F\left| s_1\right. $. On the
other hand, by Lemma 3.11(1), it follows that $u_1\stackunder{\sim }{%
\sqsubset }_{RS}t$ $\equiv s_1\wedge w_1$ $\stackunder{\sim }{\sqsubset }%
_{RS}s_1$. Then $\left\langle u_1,s_1\right\rangle \in R\circ $ $\stackunder{%
\sim }{\sqsubset }_{RS}$due to $\stackunder{\sim }{\sqsubset }_{RS}\subseteq
R$.

\begin{description}
\item  Case 2 $\dfrac{v\stackrel{a}{\rightarrow }s}{v\odot (r\varpi w)%
\stackrel{a}{\rightarrow }(s\wedge r)\odot (r\varpi w)}$
\end{description}

Hence $t_1\equiv (s\wedge r)\odot (r\varpi w)$. By Lemma 3.1(7) and (8), $%
t\equiv (s_1\wedge r_1)\odot (r\varpi w)$ for some $s_1$, $r_1$ such that $s%
\stackrel{\varepsilon }{\Rightarrow }_F\left| s_1\right. $ and $r\stackrel{%
\varepsilon }{\Rightarrow }_F\left| r_1\right. $. Thus it follows from $u_1%
\stackunder{\sim }{\sqsubset }_{RS}t\equiv (s_1\wedge r_1)\odot (r\varpi w)$
that $\left\langle u_1,(s_1\wedge r_1)\right\rangle \in R$. Moreover, by
Lemma 3.11(1), we also have $s_1\wedge r_1\stackunder{\sim }{\sqsubset }%
_{RS}s_1$. Hence $\left\langle u_1,s_1\right\rangle \in R\stackunder{\sim }{%
\circ \sqsubset }_{RS}$and $v\stackrel{a}{\Rightarrow }_F\left| s_1\right. $%
, as desired.

(2) Immediately follows from the item (1) and $t\odot (p\varpi q)\stackunder{%
\sim }{\sqsubset }_{RS}t\odot (p\varpi q)$.

(3) Let $t\odot (p\varpi q)\stackrel{\varepsilon }{\Rightarrow }_F\left|
s\right. $. By Lemma 3.1(7) and 3.3(6), $s\equiv r\odot (p\varpi q)$ for
some $r$ such that $t\stackrel{\varepsilon }{\Rightarrow }_F\left| r\right. $%
. Moreover, by item (2) in this lemma, we have $s\equiv r\odot (p\varpi q)%
\stackunder{\sim }{\sqsubset }_{RS}r$. \TeXButton{End Proof}{\endproof}

\qquad

The next result provides a necessary condition for a process to refine $%
t_1\varpi t_2$. Before giving it, for the sake of convenience, we introduce
the notation below.

\qquad

{\bf Notation 5.1} For any process $p$, $t_1$ and $t_2$, the notation $%
p\sqsubseteq _{RS}^\forall t_1\uparrow t_2$ is used to stand for $\forall
n\in \omega \forall p_0,$ $p_1,...$ $p_n$ $(p\stackrel{\varepsilon }{%
\Rightarrow }_F\left| p_0\right. \stackrel{Act}{\Rightarrow }_F\left|
p_1\right. $ $...$ $\stackrel{Act}{\Rightarrow }_F\left| p_n\right. $
implies $p_n\sqsubseteq _{RS}t_1$ or $\exists i\leq n(p_i\sqsubseteq
_{RS}t_2))$.

\quad

{\bf Lemma 5.2} If $p\sqsubseteq _{RS}t_1\varpi t_2$ then $p\sqsubseteq
_{RS}^\forall t_1\uparrow t_2$.

\TeXButton{Proof}{\proof}Assume that $p\stackrel{\varepsilon }{\Rightarrow }%
_F\left| p_0\right. \stackrel{Act}{\Rightarrow }_F\left| p_1\right.
\stackrel{Act}{\Rightarrow }_F\left| p_2\right. ...\stackrel{Act}{%
\Rightarrow }_F\left| p_n\right. $. We intend to prove that either $%
p_n\sqsubseteq _{RS}t_1$ or $\exists i\leq n(p_i\sqsubseteq _{RS}t_2)$ by
induction on $n$.

For the induction basis $n=0$, since $p\sqsubseteq _{RS}t_1\varpi t_2$ and $%
t_1\varpi t_2$ is not stable, there exist $s$ and $s_1$ such that $t_1\varpi
t_2\stackrel{\tau }{\rightarrow }_Fs\stackrel{\varepsilon }{\Rightarrow }%
_F\left| s_1\right. $ and $p_0\stackunder{\sim }{\sqsubset }_{RS}s_1$. The
argument splits into two cases based on the last rule applied in the
inference $Strip(\Gamma _{CLLT},M_{CLLT})\vdash t_1\varpi t_2\stackrel{\tau
}{\rightarrow }s$.

\begin{description}
\item  Case 1 $\dfrac {}{t_1\varpi t_2\stackrel{\tau }{\rightarrow }t_2}$
\end{description}

Thus $s\equiv t_2$ and $t_2\stackrel{\varepsilon }{\Rightarrow }_F\left|
s_1\right. $. Then it follows from $p_0\stackunder{\sim }{\sqsubset }%
_{RS}s_1 $ that $p_0\sqsubseteq _{RS}t_2$.

\begin{description}
\item  Case 2 $\dfrac {}{t_1\varpi t_2\stackrel{\tau }{\rightarrow }t_1\odot
(t_1\varpi t_2)}$
\end{description}

Then $s\equiv t_1\odot (t_1\varpi t_2)$. Moreover, by Lemma 3.1(7), $%
s_1\equiv u\odot (t_1\varpi t_2)$ for some $u$ with $t_1\stackrel{%
\varepsilon }{\Rightarrow }_F\left| u\right. $. By Lemma 5.1(1), it follows
from $p_0\stackunder{\sim }{\sqsubset }_{RS}s_1\equiv u\odot (t_1\varpi t_2)$
that $p_0\stackunder{\sim }{\sqsubset }_{RS}u$. Hence $p_0\sqsubseteq
_{RS}t_1$.

For the induction step $n=k+1$, by IH, we have either $\exists i\leq
k(p_i\sqsubseteq _{RS}t_2)$ or $p_k\sqsubseteq _{RS}t_1$. If the former
holds, then we get $\exists i\leq k+1(p_i\sqsubseteq _{RS}t_2)$ immediately.
In the following, we consider another case where $\neg \exists i\leq
k(p_i\sqsubseteq _{RS}t_2)$.

Since $p\sqsubseteq _{RS}t_1\varpi t_2$ and $p_i\stackrel{a_{i+1}}{%
\Rightarrow }_F\left| p_{i+1}\right. $ for any $i\leq k+1$, there exist $r_0$%
, $r_1$, ... $r_{k+1}$ such that $t_1\varpi t_2\stackrel{\varepsilon }{%
\Rightarrow }_F\left| r_0\right. ,$ $r_i\stackrel{a_{i+1}}{\Rightarrow }%
_F\left| r_{i+1}\right. $and $p_i\stackunder{\sim }{\sqsubset }_{RS}r_i$ for
each $i\leq k+1$. To conclude the proof, we need the claim below.

\begin{description}
\item  {\bf Claim 1} For each $j\leq k$, $r_j\equiv v\odot (t_1\varpi t_2)$
for some $v$.
\end{description}

We proceed by induction on $j$. For the induction basis $j=0$, due to $%
t_1\varpi t_2\stackrel{\tau }{\rightarrow }$, we obtain $t_1\varpi t_2%
\stackrel{\tau }{\rightarrow }_Fs\stackrel{\varepsilon }{\Rightarrow }%
_F\left| r_0\right. $ for some $s$. It is easy to see that either $s\equiv
t_2$ or $s\equiv t_1\odot (t_1\varpi t_2)$. If the first alternative holds,
then $p_0\sqsubseteq _{RS}t_2$ by the similar argument applied to Case 1 in
the above. This contradicts the assumption that $\neg \exists j\leq
k(p_j\sqsubseteq _{RS}t_2)$. Hence $s\equiv t_1\odot (t_1\varpi t_2)$. Then
it immediately follows that $r_0\equiv v_1\odot (t_1\varpi t_2)$ for some $%
v_1$ with $t_1\stackrel{\varepsilon }{\Rightarrow }_F\left| v_1\right. $.

For the induction step $j=i+1\leq k$, we assume that $r_i\equiv v_i\odot
(t_1\varpi t_2)$ for some $v_i$. Since $r_i\stackrel{a_{i+1}}{\Rightarrow }%
_F\left| r_{i+1}\right. $ and $r_i$ is stable, we obtain $r_i\stackrel{%
a_{i+1}}{\rightarrow }_Fs\stackrel{\varepsilon }{\Rightarrow }_F\left|
r_{i+1}\right. $ for some $s$. Clearly, the last rule applied in the
inference $Strip(\Gamma _{CLLT},M_{CLLT})\vdash r_i\stackrel{a_{i+1}}{%
\rightarrow }s$ is

\begin{center}
either$\quad \dfrac{v_i\stackrel{a_{i+1}}{\rightarrow }u}{v_i\odot
(t_1\varpi t_2)\stackrel{a_{i+1}}{\rightarrow }u\wedge t_2}\quad $or $\dfrac{%
v_i\stackrel{a_{i+1}}{\rightarrow }u}{v_i\odot (t_1\varpi t_2)\stackrel{%
a_{i+1}}{\rightarrow }(u\wedge t_1)\odot (t_1\varpi t_2)}.$
\end{center}

For the first alternative, we get $s\equiv u\wedge t_2$ and $r_{i+1}\equiv
q\wedge w$ for some $q$ and $w$ such that $u\stackrel{\varepsilon }{%
\Rightarrow }_F\left| q\right. $ and $t_2\stackrel{\varepsilon }{\Rightarrow
}_F\left| w\right. $. On the other hand, since $p_{i+1}\stackunder{\sim }{%
\sqsubset }_{RS}r_{i+1}\equiv q\wedge w$, by Lemma 3.11(1), we have $p_{i+1}%
\stackunder{\sim }{\sqsubset }_{RS}w$. Further, it follows from $t_2%
\stackrel{\varepsilon }{\Rightarrow }_F\left| w\right. $ that $%
p_{i+1}\sqsubseteq _{RS}t_2$, which, due to $i<k$, contradicts the
assumption $\neg \exists j\leq k(p_j\sqsubseteq _{RS}t_2)$. Thus we can
conclude that the last rule applied in the inference is the second
alternative. Then it is clear that $r_{_{i+1}}\equiv v_{i+1}\odot (t_1\varpi
t_2)$ for some $v_{i+1}$ as the operator $\odot $ is static w.r.t the $\tau $%
-labelled transition relation$\footnote{%
That is, the structure that $\odot $ represents is preserved under $\tau $%
-transitions.}$.

\qquad

Returning now to the proof of the lemma, by the above claim, we may assume
that $r_k\equiv t\odot (t_1\varpi t_2)$ for some $t$. Since $r_k\stackrel{%
a_{k+1}}{\Rightarrow }_F\left| r_{k+1}\right. $ and $r_k$ is stable, we
obtain $r_k\stackrel{a_{k+1}}{\rightarrow }_Fs\stackrel{\varepsilon }{%
\Rightarrow }_F\left| r_{k+1}\right. $ for some $s$. The last rule applied
in the inference $Strip(\Gamma _{CLLT},M_{CLLT})\vdash r_k\stackrel{a_{k+1}}{%
\rightarrow }s$ is

\begin{center}
either$\quad \dfrac{t\stackrel{a_{k+1}}{\rightarrow }u}{t\odot (t_1\varpi
t_2)\stackrel{a_{k+1}}{\rightarrow }u\wedge t_2}\quad $or $\dfrac{t\stackrel{%
a_{k+1}}{\rightarrow }u}{t\odot (t_1\varpi t_2)\stackrel{a_{k+1}}{%
\rightarrow }(u\wedge t_1)\odot (t_1\varpi t_2)}.$
\end{center}

For the former, $p_{k+1}\sqsubseteq _{RS}t_2$ follows by the argument
similar to that in the proof of the induction step in Claim 1. For the
latter, we have $s\equiv (u\wedge t_1)\odot (t_1\varpi t_2)$, and $%
r_{k+1}\equiv (w\wedge q)\odot (t_1\varpi t_2)$ for some $w$, $q$ such that $%
u\stackrel{\varepsilon }{\Rightarrow }_F\left| w\right. $ and $t_1\stackrel{%
\varepsilon }{\Rightarrow }_F\left| q\right. $. Moreover, by Lemma 5.1, it
follows from $p_{k+1}\stackunder{\sim }{\sqsubset }_{RS}r_{k+1}$ $\equiv
(w\wedge q)\odot (t_1\varpi t_2)$ that $p_{k+1}\stackunder{\sim }{\sqsubset }%
_{RS}w\wedge q$. Then $p_{k+1}\stackunder{\sim }{\sqsubset }_{RS}q$ by Lemma
3.11(1). Further, due to $t_1\stackrel{\varepsilon }{\Rightarrow }_F\left|
q\right. $, we get $p_{k+1}\sqsubseteq _{RS}t_1$, as desired.
\TeXButton{End Proof}{\endproof}

\qquad

In order to establish the converse of the above lemma, we need the following
two results which concern themselves with inconsistency predicate.

\quad

{\bf Lemma 5.3} If $u\stackrel{\varepsilon }{\Rightarrow }\left| u_1\right. $%
, $p\stackunder{\sim }{\sqsubset }_{RS}u_1$, $p\notin F$ and $p\sqsubseteq
_{RS}t$ then $u\wedge t\ \notin F$.

\TeXButton{Proof}{\proof}Since $p\sqsubseteq _{RS}t$, it follows from $%
p\notin F$ and $p$ $\stackrel{\tau }{\not \rightarrow }$ that $p\stackunder{%
\sim }{\sqsubset }_{RS}t_1$ for some $t_1$ with $t\stackrel{\varepsilon }{%
\Rightarrow }_F\left| t_1\right. $. Moreover, we have $p\stackunder{\sim }{%
\sqsubset }_{RS}u_1\wedge t_1$ due to $p\stackunder{\sim }{\sqsubset }%
_{RS}u_1$ and Lemma 3.11(2). Then $u_1\wedge t_1\notin F$ because of $%
p\notin F$. Thus, by Lemma 3.6(2), it follows from $u\wedge t\ \stackrel{%
\varepsilon }{\Rightarrow }\left| u_1\wedge t_1\right. $ that $u\wedge t\
\notin F$. \TeXButton{End Proof}{\endproof}

\qquad

{\bf Lemma 5.4} Let $p$, $t_1$ and $t_2$ be any process such that $%
p\sqsubseteq _{RS}^\forall t_1\uparrow t_2$, and set

\begin{center}
$\Omega =\left\{ t\odot (t_1\varpi t_2):\exists p_0,p_1,...p_l\left(
\begin{array}{c}
p
\stackrel{\varepsilon }{\Rightarrow }_F\left| p_0\right. \stackrel{Act}{%
\Rightarrow }_F\left| p_1\right. ...\stackrel{Act}{\Rightarrow }_F\left|
p_l\right. \text{ ,} \\ \text{ }p_l\stackunder{\sim }{\sqsubset }_{RS}t\text{
and }\neg \exists j\leq l(p_j\sqsubseteq _{RS}t_2)
\end{array}
\right) \right\} $.
\end{center}

Then $\Omega $ is a $F-$hole.

\TeXButton{Proof}{\proof}Let $t\odot (t_1\varpi t_2)\in \Omega $ and $\Im $
be any proof tree of $Strip(\Gamma _{CLLT},M_{CLLT})\vdash t\odot (t_1\varpi
t_2)F$. Hence there exist $p_0,p_1,...p_n$, $a_1,a_2,...a_n$ such that

\begin{description}
\item  (a) $p\stackrel{\varepsilon }{\Rightarrow }_F\left| p_0\right.
\stackrel{a_1}{\Rightarrow }_F\left| p_1\right. ...\stackrel{a_n}{%
\Rightarrow }_F\left| p_n\right. $,

\item  (b) $p_n\stackunder{\sim }{\sqsubset }_{RS}t$, and

\item  (c) $\neg \exists j\leq n(p_j\sqsubseteq _{RS}t_2)$.
\end{description}

Since $p_n\stackunder{\sim }{\sqsubset }_{RS}t$ and $p_n\notin F$, we get $%
t\notin F$. Moreover, it follows from $t$ $\stackrel{\tau }{\not \rightarrow
}$ that the last rule applied in $\Im $ is

\begin{center}
$\dfrac{t\odot (t_1\varpi t_2)\stackrel{a}{\rightarrow }w,\text{ }\left\{
qF:t\odot (t_1\varpi t_2)\stackrel{a}{\rightarrow }q\right\} }{t\odot
(t_1\varpi t_2)F}$ for some $a\in Act$.\qquad (5.4.1)
\end{center}

Since $p_n\stackunder{\sim }{\sqsubset }_{RS}t$ and $p_n\notin F$, we have $%
I(p_n)=I(t)=I(t\odot (t_1\varpi t_2))$. Hence $a\in I(p_n)$. Moreover, by
Lemma 3.5 and 3.8, it follows from $p_n\notin F$ that $p_n\stackrel{a}{%
\Rightarrow }_F\left| p_{n+1}\right. $ for some $p_{n+1}$. Due to $p_n%
\stackunder{\sim }{\sqsubset }_{RS}t$, there exist $s_0,s_1$ such that

\begin{center}
$t\stackrel{a}{\rightarrow }_Fs_0\stackrel{\varepsilon }{\Rightarrow }%
_F\left| s_1\right. $ and $p_{n+1}\stackunder{\sim }{\sqsubset }_{RS}s_1$.$\
$
\end{center}

Then $t\odot (t_1\varpi t_2)\stackrel{a}{\rightarrow }s_0\wedge t_2$ and $%
t\odot (t_1\varpi t_2)\stackrel{a}{\rightarrow }(s_0\wedge t_1)\odot
(t_1\varpi t_2)$ are two $a$-labelled transitions from $t\odot (t_1\varpi
t_2)$. Thus, by (5.4.1), we get

\begin{center}
$s_0\wedge t_2\in F$ and $(s_0\wedge t_1)\odot (t_1\varpi t_2)\in F$.
\end{center}

In particular, $\Im $ contains a proper subtree with the root labelled with $%
(s_0\wedge t_1)\odot (t_1\varpi t_2)F$. Clearly, to complete the proof, it
is enough to show that either $(s_0\wedge t_1)\odot (t_1\varpi t_2)\in
\Omega $ or any proof tree of $(s_0\wedge t_1)\odot (t_1\varpi t_2)F$ must
contain a proper subtree with the root labelled with $uF$ for some $u\in
\Omega $. In the following, we intend to prove this.$\qquad $

Since $p\stackrel{\varepsilon }{\Rightarrow }_F\left| p_0\right. \stackrel{%
a_1}{\Rightarrow }_F\left| p_1\right. ...\stackrel{a_n}{\Rightarrow }%
_F\left| p_n\right. \stackrel{a}{\Rightarrow }_F\left| p_{n+1}\right. $ and $%
p\sqsubseteq _{RS}^\forall t_1\uparrow t_2$, we get

\begin{center}
either $p_{n+1}\sqsubseteq _{RS}t_1$ or $\exists i\leq n+1(p_i\sqsubseteq
_{RS}t_2)$.\qquad (5.4.2)
\end{center}

On the other hand, by Lemma 5.3, it follows from $s_0\stackrel{\varepsilon }{%
\Rightarrow }_F\left| s_1\right. $, $p_{n+1}\stackunder{\sim }{\sqsubset }%
_{RS}s_1$ and $s_0\wedge t_2\in F$ that

\begin{center}
$p_{n+1}\not \sqsubseteq _{RS}t_2$.
\end{center}

Further, due to (5.4.2) and (c) (i.e., $\neg \exists j\leq n(p_j\sqsubseteq
_{RS}t_2)$), we have

\begin{center}
$p_{n+1}\sqsubseteq _{RS}t_1$ and $\neg \exists i\leq n+1(p_i\sqsubseteq
_{RS}t_2)$.\qquad (5.4.3)
\end{center}

Since $p_{n+1}\stackrel{\tau }{\not \rightarrow }$, $p_{n+1}\notin F$ and $%
p_{n+1}\sqsubseteq _{RS}t_1$, there exists $v$ such that $p_{n+1}\stackunder{%
\sim }{\sqsubset }_{RS}v$ and $t_1\stackrel{\varepsilon }{\Rightarrow }%
_F\left| v\right. $. Then, by Lemma 3.11(2) and 3.1(8), it follows from $%
p_{n+1}\stackunder{\sim }{\sqsubset }_{RS}s_1$ and $s_0\stackrel{\varepsilon
}{\Rightarrow }_F\left| s_1\right. $ that

\begin{center}
$p_{n+1}\stackunder{\sim }{\sqsubset }_{RS}s_1\wedge v$ and $s_0\wedge t_1%
\stackrel{\varepsilon }{\Rightarrow }\left| s_1\wedge v\right. .\qquad $%
(5.4.4)
\end{center}

If $(s_0\wedge t_1)\odot (t_1\varpi t_2)$ is stable, then so are $s_0$ and $%
t_1$. Thus $s_0\equiv s_1$ and $t_1\equiv v$. Further, by (5.4.4) and
(5.4.3), we get $(s_0\wedge t_1)\odot (t_1\varpi t_2)\in \Omega $, as
desired. In the following, we deal with another case $(s_0\wedge t_1)\odot
(t_1\varpi t_2)\stackrel{\tau }{\rightarrow }$.

Due to (5.4.4) and $p_{n+1}\notin F$, we obtain $s_1\wedge v\notin F$,
moreover, by Lemma 3.6 (2), it follows that $s_0\wedge t_1\stackrel{%
\varepsilon }{\Rightarrow }_F\left| s_1\wedge v\right. $. Since $s_0\wedge
t_1$ $\notin F$ and $(s_0\wedge t_1)\odot (t_1\varpi t_2)\stackrel{\tau }{%
\rightarrow }$, the last rule applied in the proof tree of $(s_0\wedge
t_1)\odot (t_1\varpi t_2)F$ is

\begin{center}
$\dfrac{(s_0\wedge t_1)\odot (t_1\varpi t_2)\stackrel{\tau }{\rightarrow }w,%
\text{ }\left\{ qF:(s_0\wedge t_1)\odot (t_1\varpi t_2)\stackrel{\tau }{%
\rightarrow }q\right\} }{(s_0\wedge t_1)\odot (t_1\varpi t_2)F}$.\quad
(5.4.5)
\end{center}

On the other hand, since $s_0\wedge t_1\stackrel{\varepsilon }{\Rightarrow }%
_F\left| s_1\wedge v\right. $, there exists $r_i$ ($1\leq i\leq m$) such
that $s_0\wedge t_1\stackrel{\tau }{\rightarrow }_Fr_1\stackrel{\tau }{%
\rightarrow }_F...\stackrel{\tau }{\rightarrow }_Fr_m\stackrel{\tau }{%
\rightarrow }_F\left| s_1\wedge v\right. $. Thus

\begin{center}
$(s_0\wedge t_1)\odot (t_1\varpi t_2)\stackrel{\tau }{\rightarrow }r_1\odot
(t_1\varpi t_2)....\stackrel{\tau }{\rightarrow }r_m\odot (t_1\varpi t_2)%
\stackrel{\tau }{\rightarrow }\left| (s_1\wedge v)\odot (t_1\varpi
t_2)\right. $.$\quad $(5.4.6)
\end{center}

For each $i\leq m$, due to $r_i$ $\notin F$, the last rule applied in any
proof tree of $r_i\odot (t_1\varpi t_2)F$ has the format below

\begin{center}
$\dfrac{r_i\odot (t_1\varpi t_2)\stackrel{\tau }{\rightarrow }w,\text{ }%
\left\{ qF:r_i\odot (t_1\varpi t_2)\stackrel{\tau }{\rightarrow }q\right\} }{%
r_i\odot (t_1\varpi t_2)F}.$
\end{center}

Then, by (5.4.6) and (5.4.5), it is obvious that any proof tree of $%
(s_0\wedge t_1)\odot (t_1\varpi t_2)F$ must contain a proper subtree with
the root labelled with $(s_1\wedge v)\odot (t_1\varpi t_2)F$. Moreover, by
(5.4.4) and (5.4.3), we also have $(s_1\wedge v)\odot (t_1\varpi t_2)\in
\Omega $, as desired. \TeXButton{End Proof}{\endproof}

\qquad

We are now in a position to show the converse of Lemma 5.2.

\quad

{\bf Lemma 5.5} If $p\sqsubseteq _{RS}^\forall t_1\uparrow t_2$ then $%
p\sqsubseteq _{RS}t_1\varpi t_2$.

\TeXButton{Proof}{\proof}Let $p\stackrel{\varepsilon }{\Rightarrow }_F\left|
q_0\right. $. The task is to find $r$ such that $q_0\stackunder{\sim }{%
\sqsubset }_{RS}r$ and $t_1\varpi t_2\stackrel{\varepsilon }{\Rightarrow }%
_F\left| r\right. $.

If $q_0\sqsubseteq _{RS}t_2$, then, due to $q_0\stackrel{\tau }{\not
\rightarrow }$ and $q_0\notin F$, we get $q_0\stackunder{\sim }{\sqsubset }%
_{RS}v$ for some $v$ with $t_2\stackrel{\varepsilon }{\Rightarrow }_F\left|
v\right. $. Moreover, by Lemma 3.1(6) and 3.6, it follows from $t_2\notin F$
that $t_1\varpi t_2\stackrel{\tau }{\rightarrow }_Ft_2\stackrel{\varepsilon
}{\Rightarrow }_F\left| v\right. $, as desired.

We now turn to another case $q_0\not \sqsubseteq _{RS}t_2$. Set $%
R=R_0\bigcup \stackunder{\sim }{\sqsubset }_{RS}$with

\begin{center}
\qquad

$R_0=\left\{ \left\langle q,t\odot (t_1\varpi t_2)\right\rangle :\exists
p_0,p_1,...p_l\left(
\begin{array}{c}
p
\stackrel{\varepsilon }{\Rightarrow }_F\left| p_0\right. \stackrel{Act}{%
\Rightarrow }_F\left| p_1\right. ...\stackrel{Act}{\Rightarrow }_F\left|
p_l\right. \equiv q\text{, } \\ \text{ }q\stackunder{\sim }{\sqsubset }_{RS}t%
\text{ and }\neg \exists i\leq l(p_i\sqsubseteq _{RS}t_2)
\end{array}
\right) \right\} $.
\end{center}

The rest of the proof is based on the following claim.

\begin{description}
\item  {\bf Claim 1} $R$ is a stable ready simulation relation.
\end{description}

Clearly, it suffices to prove that each pair in $R_0$ satisfies (RS1)-(RS4).
Let $\left\langle q,t\odot (t_1\varpi t_2)\right\rangle \in R_0$. Thus there
exist $p_0,p_1,...p_n$, $a_1,a_2,...a_n$ such that

\begin{description}
\item  ($a$) $p\stackrel{\varepsilon }{\Rightarrow }_F\left| p_0\right.
\stackrel{a_1}{\Rightarrow }_F\left| p_1\right. ...\stackrel{a_{n-1}}{%
\Rightarrow }_F\left| p_{n-1}\right. \stackrel{a_n}{\Rightarrow }_F\left|
p_n\equiv q\right. $,

\item  ($b$) $p_n\stackunder{\sim }{\sqsubset }_{RS}t$, and

\item  ($c$) $\neg \exists i\leq n(p_i\sqsubseteq _{RS}t_2)$.
\end{description}

Then (RS1) immediately follows from ($b$), and (RS2) is guaranteed by Lemma
3.9 and 5.4. By Lemma 3.2(5) and ($b$), it follows from $p_n\notin F$ that $%
I(p_n)=I(t)=I(t\odot (t_1\varpi t_2))$, and hence (RS4) holds. We next
verify (RS3).

Suppose $q\equiv p_n\stackrel{a}{\Rightarrow }_F\left| p_{n+1}\right. $.
Then, due to ($b$), there exist $w$ and $u$ such that $p_{n+1}\stackunder{%
\sim }{\sqsubset }_{RS}u$ and $t\stackrel{a}{\rightarrow }_Fw\stackrel{%
\varepsilon }{\Rightarrow }_F\left| u\right. $. Moreover, it follows from $%
(a)$, $(c)$, $p_n\stackrel{a}{\Rightarrow }_F\left| p_{n+1}\right. $ and $%
p\sqsubseteq _{RS}^\forall t_1\uparrow t_2$ that

\begin{center}
either $p_{n+1}\sqsubseteq _{RS}t_1$ or $p_{n+1}\sqsubseteq _{RS}t_2$\qquad
(5.5.1)
\end{center}

The argument splits into two cases depending on whether it holds that $%
p_{n+1}\sqsubseteq _{RS}t_2$.

\begin{description}
\item  Case 1 $p_{n+1}\sqsubseteq _{RS}t_2$.
\end{description}

Due to $p_{n+1}\stackrel{\tau }{\not \rightarrow }$ and $p_{n+1}\notin F$,
we get $p_{n+1}\stackunder{\sim }{\sqsubset }_{RS}v$ for some $v$ with $t_2%
\stackrel{\varepsilon }{\Rightarrow }_F\left| v\right. $. By Lemma 3.11(2),
it follows from $p_{n+1}\stackunder{\sim }{\sqsubset }_{RS}v$ and $p_{n+1}%
\stackunder{\sim }{\sqsubset }_{RS}u$ that

\begin{center}
$p_{n+1}\stackunder{\sim }{\sqsubset }_{RS}u\wedge v$. \qquad (5.5.2)
\end{center}

On the other hand, by Lemma 3.2(5) and 3.1(8), we have

\begin{center}
$t\odot (t_1\varpi t_2)\stackrel{a}{\rightarrow }w\wedge t_2\stackrel{%
\varepsilon }{\Rightarrow }\left| u\wedge v\right. $.
\end{center}

Moreover, by Lemma 3.9 and 5.4, $t\odot (t_1\varpi t_2)\notin F$. By (5.5.2)
and $p_{n+1}\notin F$, we also have $u\wedge v\notin F$. Then, by Lemma 3.6
(2), it follows that

\begin{center}
$t\odot (t_1\varpi t_2)\stackrel{a}{\rightarrow }_Fw\wedge t_2\stackrel{%
\varepsilon }{\Rightarrow }_F\left| u\wedge v\right. $. \quad (5.5.3)
\end{center}

On account of (5.5.2) and (5.5.3), we have the diagram below, as desired.

\begin{center}
$\qquad q\equiv p_n\qquad R_0\qquad t\odot (t_1\varpi t_2)$

$\left.
\begin{array}{c}
\\
a \\
{}
\end{array}
\right\Downarrow _F\qquad \qquad \left.
\begin{array}{c}
\\
a \\
\left. {}\right.
\end{array}
\right\Downarrow _F$

\qquad $p_{n+1}\quad \stackunder{\sim }{\sqsubset }_{RS}\quad u\wedge v$
\end{center}

\begin{description}
\item  Case 2 $p_{n+1}\not \sqsubseteq _{RS}t_2$.
\end{description}

Hence $p_{n+1}\sqsubseteq _{RS}t_1$ by (5.5.1). Then it follows from $p_{n+1}%
\stackrel{\tau }{\not \rightarrow }$ and $p_{n+1}\notin F$ that $p_{n+1}%
\stackunder{\sim }{\sqsubset }_{RS}v$ for some $v$ with $t_1\stackrel{%
\varepsilon }{\Rightarrow }_F\left| v\right. $. Moreover, by Lemma 3.11(2)
and $p_{n+1}\stackunder{\sim }{\sqsubset }_{RS}u$, we have

\begin{center}
$p_{n+1}\stackunder{\sim }{\sqsubset }_{RS}u\wedge v$. \qquad
\end{center}

Further, due to $p_{n+1}\not \sqsubseteq _{RS}t_2$ and $\neg \exists i\leq
n(p_i\sqsubseteq _{RS}t_2)$ (i.e., (c)), we get

\begin{center}
$\left\langle p_{n+1},(u\wedge v)\odot (t_1\varpi t_2)\right\rangle \in R_0$%
.\qquad (5.5.4)
\end{center}

By Lemma 3.2(5) and 3.1(7)(8), it follows that

\begin{center}
$t\odot (t_1\varpi t_2)\stackrel{a}{\rightarrow }(w\wedge t_1)\odot
(t_1\varpi t_2)\stackrel{\varepsilon }{\Rightarrow }\left| (u\wedge v)\odot
(t_1\varpi t_2)\right. $.
\end{center}

Moreover, by Lemma 3.9 and 5.4, $t\odot (t_1\varpi t_2)\notin F$ and $%
(u\wedge v)\odot (t_1\varpi t_2)\notin F$. Then, by Lemma 3.6(2), we obtain

\begin{center}
$t\odot (t_1\varpi t_2)\stackrel{a}{\Rightarrow }_F\left| (u\wedge v)\odot
(t_1\varpi t_2)\right. $.\qquad (5.5.5)
\end{center}

According to (5.5.4) and (5.5.5), we get the diagram below, as desired.

\begin{center}
$\qquad q\equiv p_n\qquad R_0\qquad t\odot (t_1\varpi t_2)$

$\left.
\begin{array}{c}
\\
a \\
{}
\end{array}
\right\Downarrow _F\qquad \qquad \left.
\begin{array}{c}
\\
a \\
\left. {}\right.
\end{array}
\right\Downarrow _F$

\qquad $\quad \qquad \quad p_{n+1}\quad R_0\quad (u\wedge v)\odot (t_1\varpi
t_2)$
\end{center}

From the arguments applied to two cases above, it may be concluded that $%
\left\langle q,t\odot (t_1\varpi t_2)\right\rangle $ satisfies (RS3).
Therefore, the binary relation $R$ is indeed a stable ready simulation
relation.

\qquad

We now return to the proof of the lemma itself. Since $p\sqsubseteq
_{RS}^\forall t_1\uparrow t_2$ and $p\stackrel{\varepsilon }{\Rightarrow }%
_F\left| q_0\right. $, it follows from $q_0\not \sqsubseteq _{RS}t_2$ that $%
q_0\sqsubseteq _{RS}t_1$. Then $q_0\stackunder{\sim }{\sqsubset }_{RS}u$ for
some $u$ with $t_1\stackrel{\varepsilon }{\Rightarrow }_F\left| u\right. $.
Thus $\left\langle q_0,u\odot (t_1\varpi t_2)\right\rangle \in R_0$. By
Claim 1, this clearly forces

\begin{center}
$q_0\stackunder{\sim }{\sqsubset }_{RS}u\odot (t_1\varpi t_2)$.\qquad (5.5.6)
\end{center}

On the other hand, by Lemma 3.1 (6) and (7), it holds that

\begin{center}
$t_1\varpi t_2\stackrel{\tau }{\rightarrow }t_1\odot (t_1\varpi t_2)%
\stackrel{\varepsilon }{\Rightarrow }\left| u\odot (t_1\varpi t_2)\right. $.
\end{center}

Moreover, it follows from $q_0\notin F$ and (5.5.6) that $u\odot (t_1\varpi
t_2)\notin F$. Hence, by Lemma 3.6(2), we have

\begin{center}
$t_1\varpi t_2\stackrel{\varepsilon }{\Rightarrow }_F\left| u\odot
(t_1\varpi t_2)\right. $.\qquad (5.5.7)
\end{center}

Consequently, by (5.5.6) and (5.5.7), the process $u\odot (t_1\varpi t_2)$
is indeed the one that we seek. \qquad \TeXButton{End Proof}{\endproof}

\qquad

Now the main theorem of this section is stated below, which, together with
Theorem 5.1, gives a bridge from CLLT to the action-based CTL that will be
considered in Section 8.

\quad

{\bf Theorem 5.1} For any process $p$, $t_1$ and $t_2$, $p\sqsubseteq
_{RS}t_1\varpi t_2$ iff $p\sqsubseteq _{RS}^\forall t_1\uparrow t_2$.

\TeXButton{Proof}{\proof}Immediately follows from Lemma 5.2 and 5.5.
\TeXButton{End Proof}{\endproof}

\qquad

Let us mention two important consequences of the above theorem:

\quad

{\bf Corollary 5.1 }Suppose $p\sqsubseteq _{RS}t_1\varpi t_2$ and $p%
\stackrel{\varepsilon }{\Rightarrow }_F\left| p_0\right. \stackrel{Act}{%
\Rightarrow }_F\left| p_1\right. ...\stackrel{Act}{\Rightarrow }_F\left|
p_k\right. $. If $\neg \exists i\leq k(p_i\sqsubseteq _{RS}t_2)$ then $%
p_k\sqsubseteq _{RS}t_1\varpi t_2$.

\TeXButton{Proof}{\proof}Straightforward. \TeXButton{End Proof}{\endproof}

\qquad

{\bf Corollary 5.2 (}Monotonicity Law of $\varpi ${\bf )} If $t_1\sqsubseteq
_{RS}s_1$ and $t_2\sqsubseteq _{RS}s_2$ then $t_1\varpi t_2\sqsubseteq
_{RS}s_1\varpi s_2$. Hence $\sqsubseteq _{RS}$ is a precongruence w.r.t the
operator $\varpi $.

\TeXButton{Proof}{\proof}Since $\sqsubseteq _{RS}$ is reflexive, it is
enough to prove that, for any $p$, $p$ $\sqsubseteq _{RS}t_1\varpi t_2$
implies $p$ $\sqsubseteq _{RS}s_1\varpi s_2$. This immediately follows from
Theorem 5.1 and the transitivity of $\sqsubseteq _{RS}$.
\TeXButton{End Proof}{\endproof}

\quad

The remainder of this section will be devoted to the proof of that $%
\sqsubseteq _{RS}$ is also precongruent w.r.t the operator $\odot $. To this
end, the following preliminary result concerning inconsistency predicate is
needed.

\qquad

{\bf Lemma 5.6} The set $\Omega $ given below is a $F-$hole.

\begin{center}
$\Omega =\left\{ u\odot (p\varpi q):\exists u_1,p_1,q_1\left(
\begin{array}{c}
u_1
\stackunder{\sim }{\sqsubset }_{RS}u,p_1\sqsubseteq _{RS}p,q_1\sqsubseteq
_{RS}q\text{ and} \\ \text{ }u_1\odot (p_1\varpi q_1)\notin F
\end{array}
\right) \right\} $.
\end{center}

\TeXButton{Proof}{\proof}Suppose $t_1\odot (p_1\varpi q_1)\in \Omega $. That
is, there exist $t_2,$ $p_2$ and $q_2$ such that

\begin{center}
$t_2\stackunder{\sim }{\sqsubset }_{RS}t_1,p_2\sqsubseteq _{RS}p_1$, $%
q_2\sqsubseteq _{RS}q_1$ and $t_2\odot (p_2\varpi q_2)\notin F$.
\end{center}

Let $\Im $ be any proof tree of $Strip(\Gamma _{CLLT},M_{CLLT})\vdash $ $%
t_1\odot (p_1\varpi q_1)F$. Since $t_2\odot (p_2\varpi q_2)\notin F$, $%
t_2\notin F$ by Lemma 3.3(6). Then, due to $t_2\stackunder{\sim }{\sqsubset }%
_{RS}t_1$, we also get $t_1\notin F$. Moreover, by Lemma 3.1(7), $t_1\odot
(p_1\varpi q_1)$ is stable because of $t_1\stackrel{\tau }{\not \rightarrow }
$. Thus the last rule applied in $\Im $ is

\begin{center}
$\dfrac{t_1\odot (p_1\varpi q_1)\stackrel{a}{\rightarrow }w,\text{ }\left\{
rF:t_1\odot (p_1\varpi q_1)\stackrel{a}{\rightarrow }r\right\} }{t_1\odot
(p_1\varpi q_1)F}$ for some $a\in Act$.\qquad (5.6.1)
\end{center}

Then $a\in I(t_1)$ by Lemma 3.2(5). Since $t_2\stackunder{\sim }{\sqsubset }%
_{RS}t_1$ and $t_2\notin F$, we get $I(t_1)=I(t_2)=I(t_2\odot (p_2\varpi
q_2))$ by Lemma 3.2(5). Hence $a\in I(t_2\odot (p_2\varpi q_2))$. Moreover,
by Lemma 3.5, it follows from $t_2\odot (p_2\varpi q_2)\notin F$ that $%
t_2\odot (p_2\varpi q_2)\stackrel{a}{\rightarrow }_Fs$ for some $s$. The
remaining proof depends on the claim below, which yields information about
the format of $s$.

\begin{description}
\item  {\bf Claim 1} $s\equiv (s_1\wedge p_2)\odot (p_2\varpi q_2)$ for some
$s_1$ with $t_2\stackrel{a}{\rightarrow }_Fs_1$.
\end{description}

Since $t_2\odot (p_2\varpi q_2)\stackrel{a}{\rightarrow }_Fs$, by Lemma
3.2(5) and 3.3(4)(6), there exists $s_1$ such that $t_2\stackrel{a}{%
\rightarrow }_Fs_1$ and

\begin{center}
either $s\equiv (s_1\wedge p_2)\odot (p_2\varpi q_2)$ or $s\equiv s_1\wedge
q_2$.
\end{center}

Thus it is enough to show that $s\not \equiv s_1\wedge q_2$. Conversely,
suppose that $s\equiv s_1\wedge q_2$. Due to $s_1\wedge q_2\notin F$, by
Lemma 3.1(8) and 3.8, $s_1\wedge q_2\stackrel{\varepsilon }{\Rightarrow }%
_F\left| s_2\wedge q_3\right. $ for some $s_2,q_3$ with $s_1\stackrel{%
\varepsilon }{\Rightarrow }_F\left| s_2\right. $ and $q_2\stackrel{%
\varepsilon }{\Rightarrow }_F\left| q_3\right. $. Since $t_2\stackunder{\sim
}{\sqsubset }_{RS}t_1$ and $t_2\stackrel{a}{\rightarrow }_Fs_1\stackrel{%
\varepsilon }{\Rightarrow }_F\left| s_2\right. $, $s_2\stackunder{\sim }{%
\sqsubset }_{RS}u$ for some $u$, $v$ with $t_1\stackrel{a}{\rightarrow }_Fv%
\stackrel{\varepsilon }{\Rightarrow }_F\left| u\right. $. Moreover, it
follows from $q_2\sqsubseteq _{RS}q_1$ and $q_2\stackrel{\varepsilon }{%
\Rightarrow }_F\left| q_3\right. $ that $q_3\stackunder{\sim }{\sqsubset }%
_{RS}q_4$ for some $q_4$ with $q_1\stackrel{\varepsilon }{\Rightarrow }%
_F\left| q_4\right. $. Hence $s_2\wedge q_3\stackunder{\sim }{\sqsubset }%
_{RS}u\wedge q_4$ by Lemma 3.11(1)(2). Then $u\wedge q_4\notin F$ because of
$s_2\wedge q_3\notin F$. Further, by Lemma 3.6(2), it follows from $v\wedge
q_1\stackrel{\varepsilon }{\Rightarrow }\left| u\wedge q_4\right. $ that $%
v\wedge q_1\notin F$. However, due to $t_1\odot (p_1\varpi q_1)\stackrel{a}{%
\rightarrow }v\wedge q_1$ and (5.6.1), we have $v\wedge q_1\in F$. Thus a
contradiction arises, as desired.

\qquad

Now we return to the proof of the lemma. Since $s\equiv (s_1\wedge p_2)\odot
(p_2\varpi q_2)\notin F$, by Lemma 3.8 and 3.1(7)(8), $(s_1\wedge p_2)\odot
(p_2\varpi q_2)\stackrel{\varepsilon }{\Rightarrow }_F\left| (s_3\wedge
p_3)\odot (p_2\varpi q_2)\right. $ for some $s_3,p_3$ such that $s_1%
\stackrel{\varepsilon }{\Rightarrow }_F\left| s_3\right. $ and $p_2\stackrel{%
\varepsilon }{\Rightarrow }_F\left| p_3\right. $. Moreover, it follows from $%
t_2\stackunder{\sim }{\sqsubset }_{RS}t_1$ and $t_2\stackrel{a}{\rightarrow }%
_Fs_1\stackrel{\varepsilon }{\Rightarrow }_F\left| s_3\right. $ that $s_3%
\stackunder{\sim }{\sqsubset }_{RS}u_1$ for some $u_1$, $v_1$ with $t_1%
\stackrel{a}{\rightarrow }_Fv_1\stackrel{\varepsilon }{\Rightarrow }_F\left|
u_1\right. $. Due to $p_2\sqsubseteq _{RS}p_1$ and $p_2\stackrel{\varepsilon
}{\Rightarrow }_F\left| p_3\right. $, we also have $p_3\stackunder{\sim }{%
\sqsubset }_{RS}p_4$ for some $p_4$ with $p_1\stackrel{\varepsilon }{%
\Rightarrow }_F\left| p_4\right. $. Then $s_3\wedge p_3\stackunder{\sim }{%
\sqsubset }_{RS}u_1\wedge p_4$ by Lemma 3.11(1)(2). Combining this with $%
(s_3\wedge p_3)\odot (p_2\varpi q_2)\notin F$, we get

\begin{center}
$(u_1\wedge p_4)\odot (p_1\varpi q_1)\in \Omega $.
\end{center}

Clearly, in order to complete the proof, it suffices to prove that $\Im $
contains a proper subtree with the root labelled with $(u_1\wedge p_4)\odot
(p_1\varpi q_1)F$. Since $t_1\stackrel{a}{\rightarrow }v_1$, we have $%
t_1\odot (p_1\varpi q_1)\stackrel{a}{\rightarrow }(v_1\wedge p_1)\odot
(p_1\varpi q_1)$. Thus, by (5.6.1), $\Im $ contains a proper subtree with
the root labelled with $(v_1\wedge p_1)\odot (p_1\varpi q_1)F$. If $%
v_1\wedge p_1$ is stable then $\Im $ contains a node labelled with $%
(u_1\wedge p_4)\odot (p_1\varpi q_1)F$ because of $(v_1\wedge p_1)\odot
(p_1\varpi q_1)\equiv (u_1\wedge p_4)\odot (p_1\varpi q_1)$, as desired. We
next manage another case $v_1\wedge p_1\stackrel{\tau }{\rightarrow }$.

Since $v_1\wedge p_1\stackrel{\tau }{\Rightarrow }\left| u_1\wedge
p_4\right. $, $s_3\wedge p_3\stackunder{\sim }{\sqsubset }_{RS}u_1\wedge p_4$
and $s_3\wedge p_3\notin F$, we get $v_1\wedge p_1\stackrel{\tau }{%
\Rightarrow }_F\left| u_1\wedge p_4\right. $ by Lemma 3.6(2). Hence there
exist $r_1,r_2...r_m$ ($m\geq 1$) such that

\begin{center}
$v_1\wedge p_1\stackrel{\tau }{\rightarrow }_Fr_1\stackrel{\tau }{%
\rightarrow }_Fr_2\stackrel{\tau }{\rightarrow }_F...\stackrel{\tau }{%
\rightarrow }_Fr_m\stackrel{\tau }{\rightarrow }_F\left| u_1\wedge
p_4\right. $.\qquad (5.6.3)\qquad
\end{center}

By Lemma 3.1(7), we also have

\begin{center}
$(v_1\wedge p_1)\odot (p_1\varpi q_1)\stackrel{\tau }{\rightarrow }r_1\odot
(p_1\varpi q_1)...\stackrel{\tau }{\rightarrow }r_m\odot (p_1\varpi q_1)%
\stackrel{\tau }{\rightarrow }\left| (u_1\wedge p_4)\odot (p_1\varpi
q_1)\right. $(5.6.4)
\end{center}

Then, by (5.6.3), the last rule applied in any proof tree of $w\odot
(p_1\varpi q_1)F$ with $w\in \{v_1\wedge p_1$, $r_i:$ $1\leq i\leq m\}$ must
be of the format below

\begin{center}
$\dfrac{w\odot (p_1\varpi q_1)\stackrel{\tau }{\rightarrow }u,\text{ }%
\left\{ rF:w\odot (p_1\varpi q_1)\stackrel{\tau }{\rightarrow }r\right\} }{%
w\odot (p_1\varpi q_1)F}$.
\end{center}

Consequently, by (5.6.4), it is immediate that $\Im $ contains a proper
subtree with the root labelled with $(u_1\wedge p_4)\odot (p_1\varpi q_1)F$,
as desired. \TeXButton{End Proof}{\endproof}

\quad

Having disposed of this preliminary step, we can now establish the
monotonicity laws of the operator $\odot $, which will be useful in the
sequel.

\qquad

{\bf Theorem 5.2 }For any process $u_i,r_i$ and $s_i$ ($1\leq i\leq 2$), we
have

(1) If $u_2\stackunder{\sim }{\sqsubset }_{RS}u_1$, $r_2\sqsubseteq _{RS}r_1$%
, $s_2\sqsubseteq _{RS}s_1$ then $u_2\odot (r_2\varpi s_2)\stackunder{\sim }{%
\sqsubset }_{RS}u_1\odot (r_1\varpi s_1)$.

(2) If $u_2\sqsubseteq _{RS}u_1$, $r_2\sqsubseteq _{RS}r_1$, $s_2\sqsubseteq
_{RS}s_1$ then $u_2\odot (r_2\varpi s_2)\sqsubseteq _{RS}u_1\odot (r_1\varpi
s_1)$.

\TeXButton{Proof}{\proof}Clearly, (2) immediately follows from (1). In the
following, we shall prove (1). Put

\begin{center}
$R=\left\{ \left\langle t_2\odot (p_2\varpi q_2),t_1\odot (p_1\varpi
q_1)\right\rangle :t_2\stackunder{\sim }{\sqsubset }_{RS}t_1,p_2\sqsubseteq
_{RS}p_1,q_2\sqsubseteq _{RS}q_1\right\} \bigcup \stackunder{\sim }{%
\sqsubset }_{RS}$.
\end{center}

We wish to demonstrate that $R$ is a stable ready simulation. Suppose that $%
t_2\stackunder{\sim }{\sqsubset }_{RS}t_1,$ $p_2\sqsubseteq _{RS}p_1$ and $%
q_2\sqsubseteq _{RS}q_1$. By Lemma 3.1(7), 3.9, 5.6 and 3.2(5), it is easy
to verify that the pair $\left\langle t_2\odot (p_2\varpi q_2),\text{ }%
t_1\odot (p_1\varpi q_1)\right\rangle $ satisfies (RS1), (RS2) and (RS4). It
remains to prove that such pair satisfies (RS3). Suppose $t_2\odot
(p_2\varpi q_2)\stackrel{a}{\Rightarrow }_F\left| u\right. $. It is enough
to find $s$ such that

\begin{center}
$t_1\odot (p_1\varpi q_1)\stackrel{a}{\Rightarrow }_F\left| s\right. $ and $%
\left\langle u,\text{ }s\right\rangle \in R$.
\end{center}

Since $t_2\odot (p_2\varpi q_2)\notin F$, by Lemma 3.9 and 5.6, we have

\begin{center}
$t_1\odot (p_1\varpi q_1)\notin F$. \qquad (5.2.1)
\end{center}

Moreover, due to $t_2\odot (p_2\varpi q_2)\stackrel{\tau }{\not \rightarrow }
$ , $t_2\odot (p_2\varpi q_2)\stackrel{a}{\rightarrow }_Fv\stackrel{%
\varepsilon }{\Rightarrow }_F\left| u\right. $ for some $v$. The argument
splits into two cases based on the last rule applied in the inference $%
Strip(\Gamma _{CLLT},M_{CLLT})\vdash t_2\odot (p_2\varpi q_2)\stackrel{a}{%
\rightarrow }v$. Clearly, the last rule is

\begin{center}
either $\dfrac{t_2\stackrel{a}{\rightarrow }s}{t_2\odot (p_2\varpi q_2)%
\stackrel{a}{\rightarrow }s\wedge q_2}$ \quad or \quad $\dfrac{t_2\stackrel{a%
}{\rightarrow }s}{t_2\odot (p_2\varpi q_2)\stackrel{a}{\rightarrow }(s\wedge
p_2)\odot (p_2\varpi q_2)}.$
\end{center}

These two cases may be handled in a similar way. Here we consider only the
second alternative. In such situation, we get $v\equiv (s\wedge p_2)\odot
(p_2\varpi q_2)$ with $t_2\stackrel{a}{\rightarrow }_Fs$, and $u\equiv
(s_1\wedge p_3)\odot (p_2\varpi q_2)$ for some $s_1$ and $p_3$ with $s%
\stackrel{\varepsilon }{\Rightarrow }_F\left| s_1\right. $ and $p_2\stackrel{%
\varepsilon }{\Rightarrow }_F\left| p_3\right. $. Then it follows from $t_2%
\stackunder{\sim }{\sqsubset }_{RS}t_1$ and $p_2\sqsubseteq _{RS}p_1$ that
there exist $t_3,t_4$ and $p_4$ such that $t_1\stackrel{a}{\rightarrow }_Ft_3%
\stackrel{\varepsilon }{\Rightarrow }_F\left| t_4\right. ,$ $p_1\stackrel{%
\varepsilon }{\Rightarrow }_F\left| p_4\right. ,$ $s_1\stackunder{\sim }{%
\sqsubset }_{RS}t_4$ and $p_3\stackunder{\sim }{\sqsubset }_{RS}p_4$. Thus

\begin{center}
$t_1\odot (p_1\varpi q_1)\stackrel{a}{\rightarrow }(t_3\wedge p_1)\odot
(p_1\varpi q_1)\stackrel{\varepsilon }{\Rightarrow }\left| (t_4\wedge
p_4)\odot (p_1\varpi q_1)\right. $. \quad (5.2.2)
\end{center}

By Lemma 3.11, we also have $s_1\wedge p_3\stackunder{\sim }{\sqsubset }%
_{RS}t_4\wedge p_4$. Hence $\left\langle u,(t_4\wedge p_4)\odot (p_1\varpi
q_1)\right\rangle \in R$. Moreover, by Lemma 3.9 and 5.6, it follows from $%
u\equiv (s_1\wedge p_3)\odot (p_2\varpi q_2)\notin F$ that $(t_4\wedge
p_4)\odot (p_1\varpi q_1)\notin F$. Then $t_1\odot (p_1\varpi q_1)\stackrel{a%
}{\Rightarrow }_F\left| (t_4\wedge p_4)\odot (p_1\varpi q_1)\right. $ due to
(5.2.1), (5.2.2) and Lemma 3.6(2). Therefore, the process $(t_4\wedge
p_4)\odot (p_1\varpi q_1)$ is exactly one that we seek. \TeXButton{End Proof}
{\endproof}

\quad

Hitherto we have showed that $\sqsubseteq _{RS}$ is precongruent w.r.t the
operators $\varpi ,\sharp ,\odot $ and $\triangle $. For the remainder
operators (i.e., operators in CLL), such property has been established in
[60]. Consequently, $\sqsubseteq _{RS}$ is precongruent w.r.t all operators
involved in CLLT.

\section{Fixed-point characterization of the operator $\varpi $}

From now on we make the assumption: the set $Act$ is finite. The motivation
behind this assumption will be given in Remark 6.1. This section is devoted
to a few further properties of the operator $\varpi $ including fixed point
characterization and approximation. These properties will serve as a
stepping stone in giving a graphical representation of the temporal operator
$unless$ in a recursive manner. We begin with introducing some preliminary
notions.

\qquad

{\bf Definition 6.1 }Given a finite sequence of processes $<t_0,t_1,\dots
,t_{n-1}>$ with $n>0$, the generalized disjunction $\stackunder{i<n}{\bigvee
}t_i$ is defined inductively as

\begin{description}
\item  (1) $\stackunder{i<1}{\bigvee }t_i=t_0,$

\item  (2) $\stackunder{i<k+1}{\bigvee }t_i=(\stackunder{i<k}{\bigvee }%
t_i)\vee t_k$ for $k\geq 1$.
\end{description}

Moreover, for any nonempty subset $S\subseteq \{t_0,\dots ,t_{n-1}\}$, the
generalized disjunction $\bigvee S$ is defined as $\stackunder{i<\left|
S\right| }{\bigvee }t_i^{^{\prime }}$, where the sequence $<t_0^{^{\prime
}},\dots ,t_{|S|-1}^{^{\prime }}>$ is the restriction of $<t_0,\dots
,t_{n-1}>$ to $S$. Similar to generalized external choice, modulo $=_{RS}$,
the order and grouping of processes in $\bigvee S$ may be ignored due to the
commutative and associative laws [43, 60].

\quad

Analogously, the notion of a generalized conjunction $\bigwedge S$ is
defined in the same manner, and the order and grouping of processes in $%
\bigwedge S$ may also be ignored by the same reason. It should be pointed
out that such generalized conjunction $\bigwedge S$ preserves usual logic
laws of the connective conjunction only if $S$ is finite (see, Remark 6.1).
For the sake of simplicity we also introduce the notions below.

\qquad

{\bf Definition 6.2} Given any process $p$ and $t,$ $\delta _{p,t}$ is a
function assigning to each visible action a process, which is given by $%
\quad $

\begin{center}
for any $a\in Act$, $\delta _{p,t}(a)\equiv \left\{
\begin{array}{c}
\quad \quad
\stackunder{\beta \in I(p)}{\Box }\beta .true\qquad \quad \qquad \text{if }%
a\notin I(p) \\ \left( \stackunder{b\in I(p)-\{a\}}{\Box }b.true\right) \Box
a.t\qquad \text{otherwise}
\end{array}
\right. $ .
\end{center}

\quad

Given an action $a\in Act$, the auxiliary operator $\left\lceil
a\right\rceil $ is introduced below. This operator will be used to explore
the fixed point characterization of $\varpi $. Moreover, itself is also of
logic meaning, that is, it captures the modal operator ``$along$ $a-labelled$
$transitions$, $it$ $is$ $necessary$ $that\ldots $'' in a sense.

\qquad

{\bf Definition 6.3} For any $a\in Act$, the operator $\left\lceil
a\right\rceil $ over processes is defined by

\begin{center}
$\left\lceil a\right\rceil =\lambda X.\left( \stackunder{a\in A\subseteq Act%
}{\bigvee }\left( (\stackunder{b\in A-\{a\}}{\Box }b.true)\Box a.X\right)
\right) \bigvee \left( \stackunder{a\notin A\subseteq Act}{\bigvee }(%
\stackunder{b\in A}{\Box }b.true)\right) $.

\qquad
\end{center}

By the way, since $\left\lceil a\right\rceil p\stackrel{\tau }{\rightarrow }%
\stackunder{a\notin A\subseteq Act}{\bigvee }(\stackunder{b\in A}{\Box }%
b.true)\notin F$, it is easy to see that $\left\lceil a\right\rceil p\notin
F $ for any $a$ and $p$. A simple but useful result is given below.

\qquad

{\bf Lemma 6.1} $p$ $\stackunder{\sim }{\sqsubset }_{RS}\stackunder{a\in I(p)%
}{\Box }a.true$ whenever $p\stackrel{\tau }{\not \rightarrow }$.

\TeXButton{Proof}{\proof}Put

\begin{center}
$R=\left\{ \left\langle q,\stackunder{a\in I(q)}{\Box }a.true\right\rangle :q%
\stackrel{\tau }{\not \rightarrow }\right\} \footnote{%
Notice that, if $I(q)=\emptyset $ then $\stackunder{a\in I(q)}{\Box }a.true$
is defined as $0$, see Def. 3.2.}$.
\end{center}

We only need to show that $R$ is a stable ready simulation relation, which
is routine and is left to the reader. \TeXButton{End Proof}{\endproof}

\qquad

In the following, we shall give some basic properties of the operator $%
\left\lceil a\right\rceil $. The theorem below characterizes processes that
refine ones with the format $\left\lceil a\right\rceil t$.

\qquad

{\bf Theorem 6.1} $p\sqsubseteq _{RS}\left\lceil a\right\rceil t$ iff $%
\forall p_0,p_1\left( p\stackrel{\varepsilon }{\Rightarrow }_F\left|
p_0\right. \stackrel{a}{\Rightarrow }_F\left| p_1\right. \text{ implies }p_1%
\text{ }\sqsubseteq _{RS}t\right) $.

\TeXButton{Proof}{\proof}(Left implies Right) Assume that $p\stackrel{%
\varepsilon }{\Rightarrow }_F\left| p_0\right. \stackrel{a}{\Rightarrow }%
_F\left| p_1\right. $. Then it follows from $p\sqsubseteq _{RS}\left\lceil
a\right\rceil t$ that there exists $r$ such that $p_0$ $\stackunder{\sim }{%
\sqsubset }_{RS}r$ and $\left\lceil a\right\rceil t\stackrel{\varepsilon }{%
\Rightarrow }_F\left| r\right. $. Moreover, since $a\in I(p_0)=I(r)$, we get
$r\equiv \delta _{p_0,t}(a)$. On the other hand, due to $p_0$ $\stackunder{%
\sim }{\sqsubset }_{RS}r$ and $p_0\stackrel{a}{\Rightarrow }_F\left|
p_1\right. $, we have $p_1$ $\stackunder{\sim }{\sqsubset }_{RS}q$ for some $%
q$ with $r\equiv \delta _{p_0,t}(a)\stackrel{a}{\Rightarrow }_F\left|
q\right. $. Further, by Lemma 3.1 (8) and 3.2(1), since $\delta _{p_0,t}(a)%
\stackrel{\tau }{\not \rightarrow }$ and $(\stackunder{b\in I(p_{_0})-\{a\}}{%
\Box }b.true)\stackrel{a}{\not \rightarrow }$, we obtain $a.t\stackrel{a}{%
\Rightarrow }_F\left| q\right. $. Hence $t\stackrel{\varepsilon }{%
\Rightarrow }_F\left| q\right. $. Then $p_1$ $\sqsubseteq _{RS}t$ follows
from $p_1$ $\stackunder{\sim }{\sqsubset }_{RS}q$, as desired.

(Right implies Left) Let $p\stackrel{\varepsilon }{\Rightarrow }_F\left|
p_0\right. $. It suffices to prove that $p_0$ $\stackunder{\sim }{\sqsubset }%
_{RS}q$ for some $q$ with $\left\lceil a\right\rceil t\stackrel{\varepsilon
}{\Rightarrow }_F\left| q\right. $. If $a\notin I(p_0)$, then $\left\lceil
a\right\rceil t\stackrel{\varepsilon }{\Rightarrow }_F|\stackunder{b\in
I(p_0)}{\Box }b.true$, and $p_0$ $\stackunder{\sim }{\sqsubset }_{RS}%
\stackunder{b\in I(p_0)}{\Box }b.true$ due to Lemma 6.1. In the following,
we consider another case $a\in I(p_0)$.

In such situation, by Lemma 3.5 and 3.8, it follows from $p_0\notin F$ that
there exists $p_1$ such that $p\stackrel{\varepsilon }{\Rightarrow }_F\left|
p_0\right. \stackrel{a}{\Rightarrow }_F\left| p_1\right. $. Hence $p_1$ $%
\sqsubseteq _{RS}t$. Then $t\notin F$ because of $p_1\notin F$. Further, by
Lemma 3.3 (2)(3)(7), it follows that $\delta _{p_0,t}(a)\notin F$. Thus, by
Def. 6.3 and Lemma 3.6(2), we obtain $\left\lceil a\right\rceil t\stackrel{%
\varepsilon }{\Rightarrow }_F|\delta _{p_0,t}(a)$. Clearly, in order to
complete the proof, it is enough to show that $p_0$ $\stackunder{\sim }{%
\sqsubset }_{RS}\delta _{p_0,t}(a)$. To do this, we intend to prove that $R$
given below is a stable ready simulation relation.

\begin{center}
$R=\left\{ \left\langle p_0,\delta _{p_0,t}(a)\right\rangle \right\} \bigcup
\stackunder{\sim }{\sqsubset }_{RS}$.
\end{center}

It is straightforward to verify that $R$ satisfies (RS1), (RS2) and (RS4).
For (RS3), suppose $p_0\stackrel{c}{\Rightarrow }_F\left| p_1\right. $. If $%
c\neq a$, we have $p_1\stackunder{\sim }{\sqsubset }_{RS}\stackunder{b\in
I(p_1)}{\Box }b.true$ by Lemma 6.1, and $\delta _{p_0,t}(a)\stackrel{c}{%
\rightarrow }_Ftrue\stackrel{\tau }{\rightarrow }_F\stackunder{b\in I(p_1)}{%
\Box }b.true$. If $c=a$, then $\delta _{p_0,t}(a)\stackrel{c}{\rightarrow }%
_Ft$, and it follows from $p_1$ $\sqsubseteq _{RS}t$ and $p_1\stackrel{%
\varepsilon }{\Rightarrow }_F\left| p_1\right. $ that $p_1\stackunder{\sim }{%
\sqsubset }_{RS}t_1$ for some $t_1$ with $t\stackrel{\varepsilon }{%
\Rightarrow }_F\left| t_1\right. $. Summarizing, we can conclude that there
exists $r$ such that $p_1\stackunder{\sim }{\sqsubset }_{RS}r$ and $\delta
_{p_0,t}(a)\stackrel{c}{\Rightarrow }_F\left| r\right. $. Hence (RS3) holds,
as desired. \TeXButton{End Proof}{\endproof}

\quad

{\bf Corollary 6.1 (}Monotonicity Law of $\left\lceil a\right\rceil ${\bf )}
If $t\sqsubseteq _{RS}s$ then $\left\lceil a\right\rceil t\sqsubseteq
_{RS}\left\lceil a\right\rceil s$ for each $a\in Act.$ Hence $\sqsubseteq
_{RS}$ is a precongruence w.r.t the operator $\left\lceil a\right\rceil $.

\TeXButton{Proof}{\proof}Analogous to that of Corollary 5.2, but using
Theorem 6.1 instead of Theorem 5.1.\TeXButton{End Proof}{\endproof}

\quad

Now we are ready to discuss the fixed-point characterization of $\varpi $.
For this purpose, a series of functions $\eta _{p,q}$ is introduced below.

\quad

{\bf Definition 6.4} For any process{\bf \ }$p$ and $q$, the function $\eta
_{p,q}$ over processes is defined by

\begin{center}
$\eta _{p,q}=\lambda _X.$ $q\vee \left( p\wedge \left( \stackunder{a\in Act}{%
\bigwedge }\left\lceil a\right\rceil X\right) \right) $.\qquad
\end{center}

\quad

Obviously, as all operators involved in $\eta _{p,q}$ are monotonic w.r.t $%
\sqsubseteq _{RS}$, the function $\eta _{p,q}$ itself is also monotonic. In
the following, we intend to show that $p\varpi q$ is the largest fixed point
of $\eta _{p,q}$. We begin with arguing that $p\varpi q$ is a post-fixed
point of $\eta _{p,q}$.

\qquad

{\bf Lemma 6.2} For{\bf \ }any process $p$ and $q$, $p\varpi q\sqsubseteq
_{RS}\eta _{p,q}\left( p\varpi q\right) $.

\TeXButton{Proof}{\proof}If $p\varpi q\in F$ then it holds trivially. In the
following, we consider the nontrivial case $p\varpi q\notin F$. For
simplicity of notation, we shall omit the subscript in $\eta _{p,q}$.
Clearly, it is enough to show that, for any process $v$,

\begin{center}
$v\sqsubseteq _{RS}$ $p\varpi q$ implies $v\sqsubseteq _{RS}\eta \left(
p\varpi q\right) $.
\end{center}

Assume that $t$ is any process such that $t\sqsubseteq _{RS}$ $p\varpi q$.
Let $t\stackrel{\varepsilon }{\Rightarrow }_F\left| t_0\right. $. We want to
find $s$ such that $\eta \left( p\varpi q\right) \stackrel{\varepsilon }{%
\Rightarrow }_F\left| s\right. $ and $t_0\stackunder{\sim }{\sqsubset }%
_{RS}s.$ It proceeds by distinguishing two cases below.

\begin{description}
\item  Case 1 $t_0\sqsubseteq _{RS}q$.
\end{description}

Thus $t_0\stackunder{\sim }{\sqsubset }_{RS}q_0$ for some $q_0$ with $q%
\stackrel{\varepsilon }{\Rightarrow }_F\left| q_0\right. $. Easily, $\eta
\left( p\varpi q\right) \stackrel{\tau }{\rightarrow }q\stackrel{\varepsilon
}{\Rightarrow }\left| q_0\right. $. Then, by $q_0\notin F$ and Lemma 3.6
(1)(2), it follows that $\eta \left( p\varpi q\right) \stackrel{\tau }{%
\rightarrow }_Fq\stackrel{\varepsilon }{\Rightarrow }_F\left| q_0\right. $.
Hence $q_0$ is indeed the one that we seek.

\begin{description}
\item  Case 2 $t_0\not \sqsubseteq _{RS}q$.
\end{description}

In this case, by Theorem 5.1, it follows from $t\sqsubseteq _{RS}$ $p\varpi
q $ and $t\stackrel{\varepsilon }{\Rightarrow }_F\left| t_0\right. $ that $%
t_0\sqsubseteq _{RS}$ $p$. Then $t_0\stackunder{\sim }{\sqsubset }_{RS}p_0$
for some $p_0$ with $p\stackrel{\varepsilon }{\Rightarrow }_F\left|
p_0\right. $. The claim below is needed to complete the proof.

\begin{description}
\item  {\bf Claim 1 }$\left\lceil a\right\rceil p\varpi q\stackrel{%
\varepsilon }{\Rightarrow }_F|$ $\delta _{t_0,p\varpi q}(a)$ and $t_0%
\stackunder{\sim }{\sqsubset }_{RS}\delta _{t_0,p\varpi q}(a)$ for each $%
a\in Act$.
\end{description}

For any $a\in Act$, due to $p\varpi q\notin F$, it is easy to see that $%
\delta _{t_0,p\varpi q}(a)\notin F$. Further, by Lemma 3.6 and Def. 6.3, it
follows that $\left\lceil a\right\rceil p\varpi q\stackrel{\varepsilon }{%
\Rightarrow }_F|$ $\delta _{t_0,p\varpi q}(a)$. We next prove that $t_0%
\stackunder{\sim }{\sqsubset }_{RS}\delta _{t_0,p\varpi q}(a)$. By Lemma
6.1, this is immediate whenever $a\notin I(t_0)$. In the following, we
consider the case $a\in I(t_0)$. Put

\begin{center}
$R=\left\{ \left\langle t_0,\delta _{t_0,p\varpi q}(a)\right\rangle \right\}
\bigcup \stackunder{\sim }{\sqsubset }_{RS}.$
\end{center}

We want to show that $R$ is a stable ready simulation relation. Since it can
be checked without any difficulty that the pair $\left\langle t_0,\delta
_{t_0,p\varpi q}(a)\right\rangle $ satisfies (RS1), (RS2) and (RS4), we put
attention to verify that such pair satisfies (RS3).

Assume $t_0\stackrel{b}{\Rightarrow }_F\left| t_1\right. $. Then $b\in
I(t_0) $ because of $t_0\stackrel{\tau }{\not \rightarrow }$. If $b\neq a$
then $\delta _{t_0,p\varpi q}(a)\stackrel{b}{\rightarrow }_Ftrue\stackrel{%
\tau }{\rightarrow }_F\left| \stackunder{c\in I(t_1)}{\Box }c.true\right. $,
and $t_1$ $\stackunder{\sim }{\sqsubset }_{RS}\stackunder{c\in I(t_1)}{\Box }%
c.true$ by Lemma 6.1, as desired. We next handle another case $a=b$. In such
situation, we get $\delta _{t_0,p\varpi q}(a)\stackrel{b}{\rightarrow }%
_Fp\varpi q$. If $t_1\sqsubseteq _{RS}q$ then $t_1\stackunder{\sim }{%
\sqsubset }_{RS}q_1$ for some $q_1$ with $\delta _{t_0,p\varpi q}(a)%
\stackrel{b}{\rightarrow }_Fp\varpi q\stackrel{\tau }{\rightarrow }_Fq%
\stackrel{\varepsilon }{\Rightarrow }_F|$ $q_1$. If $t_1\not \sqsubseteq
_{RS}q$ then, by Corollary 5.1, it follows from $t_0\not \sqsubseteq _{RS}q$%
, $t\sqsubseteq _{RS}$ $p\varpi q$ and $t\stackrel{\varepsilon }{\Rightarrow
}_F\left| t_0\right. \stackrel{b}{\Rightarrow }_F\left| t_1\right. $ that $%
t_1\sqsubseteq _{RS}p\varpi q$, moreover, due to $t_1\stackrel{\varepsilon }{%
\Rightarrow }_F|$ $t_1$, we have $t_1\stackunder{\sim }{\sqsubset }_{RS}v$
for some $v$ with $\delta _{t_0,p\varpi q}(a)\stackrel{b}{\rightarrow }%
_Fp\varpi q\stackrel{\varepsilon }{\Rightarrow }_F|$ $v$, as desired.

\qquad

Now we return to the proof of the lemma. From Claim 1 and $t_0\stackunder{%
\sim }{\sqsubset }_{RS}p_0$, by Lemma 3.11(2), it follows that

\begin{center}
$t_0\stackunder{\sim }{\sqsubset }_{RS}p_0\wedge \left( \stackunder{a\in Act%
}{\bigwedge }\delta _{t_0,p\varpi q}(a)\right) $.\quad (6.2.1)
\end{center}

Moreover, it is obvious that

\begin{center}
$\eta \left( p\varpi q\right) \stackrel{\varepsilon }{\Rightarrow }|$ $%
p_0\wedge \left( \stackunder{a\in Act}{\bigwedge }\delta _{t_0,p\varpi
q}(a)\right) .$\quad (6.2.2)
\end{center}

Further, by Lemma 3.6(2), it follows from $t_0\notin F$, (6.2.1) and (6.2.2)
that

\begin{center}
$\eta \left( p\varpi q\right) \stackrel{\varepsilon }{\Rightarrow }_F|$ $%
p_0\wedge \left( \stackunder{a\in Act}{\bigwedge }\delta _{t_0,p\varpi
q}(a)\right) .$
\end{center}

Hence the process $p_0\wedge \left( \stackunder{a\in Act}{\bigwedge }\delta
_{t_0,p\varpi q}(a)\right) $ is exactly one that we seek.
\TeXButton{End Proof}{\endproof}

\qquad

We are almost ready now to establish the fixed point characterization of the
operator $\varpi $. The following lemma is instrumental in doing this.

\qquad

{\bf Lemma 6.3} For any $k<\omega $, if $t\sqsubseteq _{RS}\eta
_{p,q}^{k+1}(u)$, $t\stackrel{\varepsilon }{\Rightarrow }_F\left| t_0\right.
\stackrel{a_1}{\Rightarrow }_F\left| t_1\right. ...\stackrel{a_k}{%
\Rightarrow }_F\left| t_k\right. $ and $\neg \exists i\leq k(t_i\sqsubseteq
_{RS}q)$ then $t_k\stackunder{\sim }{\sqsubset }_{RS}w\wedge \left(
\stackunder{a\in Act}{\bigwedge }\delta _{w,u}(a)\right) $ for some $w$ with
$p\stackrel{\varepsilon }{\Rightarrow }_F\left| w\right. $, and hence $%
t_k\sqsubseteq _{RS}p$.

\TeXButton{Proof}{\proof}Prove it by induction on $k$. For the induction
basis $k=0$, since $t\sqsubseteq _{RS}\eta _{p,q}(u)$, we get $t_0%
\stackunder{\sim }{\sqsubset }_{RS}r$ for some $r$ with $\eta _{p,q}(u)%
\stackrel{\varepsilon }{\Rightarrow }_F\left| r\right. $. Then it
immediately follows from $t_0\not \sqsubseteq _{RS}q$ that

\begin{center}
$\eta _{p,q}(u)\stackrel{\tau }{\rightarrow }_Fp\wedge \left( \stackunder{%
a\in Act}{\bigwedge }\left\lceil a\right\rceil u\right) \stackrel{%
\varepsilon }{\Rightarrow }_F\left| r\right. $ .
\end{center}

Thus $r\equiv w\wedge s$ for some $w$ and $s$ with $p\stackrel{\varepsilon }{%
\Rightarrow }_F\left| w\right. $ and $\stackunder{a\in Act}{\bigwedge }%
\left\lceil a\right\rceil u\stackrel{\varepsilon }{\Rightarrow }_F\left|
s\right. $. Due to $w\wedge s\notin F$ and $w\wedge s\stackrel{\tau }{\not
\rightarrow }$, we get $I(w)=I(s).$ Further, by Def. 6.3, it is easy to see
that $s\equiv \stackunder{a\in Act}{\bigwedge }\delta _{w,u}(a)$, as desired.

For the induction step $k=n+1$, since $t\sqsubseteq _{RS}\eta
_{p,q}^{k+1}(u)=\eta _{p,q}^{n+1}(\eta _{p,q}(u))$, by IH, there exists $w$
such that $p\stackrel{\varepsilon }{\Rightarrow }_F\left| w\right. $ and

\begin{center}
$t_n\stackunder{\sim }{\sqsubset }_{RS}w\wedge \left( \stackunder{a\in Act}{%
\bigwedge }\delta _{w,\eta _{p,q}(u)}(a)\right) $. \qquad (6.3.1)
\end{center}

Since $t_n\stackrel{a_{n+1}}{\Rightarrow }_F\left| t_{n+1}\right. $ and $t_n%
\stackrel{\tau }{\not \rightarrow }$, we have $a_{n+1}\in I(t_n)$. Then $%
a_{n+1}\in I(w)$ because of $t_n\notin F$ and (6.3.1). Hence

\begin{center}
$\delta _{w,\eta _{p,q}(u)}(a_{n+1})\equiv (\stackunder{b\in I(w)-\{a_{n+1}\}%
}{\Box }b.true)\Box a_{n+1}.\eta _{p,q}(u)$.\qquad
\end{center}

By (6.3.1) and $t_n\stackrel{a_{n+1}}{\Rightarrow }_F\left| t_{n+1}\right. $%
, we get $t_{n+1}\stackunder{\sim }{\sqsubset }_{RS}r$ for some $r$ with

\begin{center}
$w\wedge \left( \stackunder{a\in Act}{\bigwedge }\delta _{w,\eta
_{p,q}(u)}(a)\right) \stackrel{a_{n+1}}{\Rightarrow }_F\left| r\right. $.
\end{center}

Further, due to commutative and associative laws of $\wedge $, it is not
difficult to see that $r\approx _{RS}v\wedge s$ for some $v$ and $s$ with $%
\delta _{w,\eta _{p,q}(u)}(a_{n+1})\stackrel{a_{n+1}}{\rightarrow }_F\eta
_{p,q}(u)\stackrel{\varepsilon }{\Rightarrow }_F\left| s\right. .$ Moreover,
it follows from $t_{n+1}\stackunder{\sim }{\sqsubset }_{RS}r$ and $t_k\equiv
t_{n+1}\not \sqsubseteq _{RS}q$ that

\begin{center}
$\eta _{p,q}(u)\stackrel{\tau }{\rightarrow }_Fp\wedge \left( \stackunder{%
a\in Act}{\bigwedge }\left\lceil a\right\rceil u\right) \stackrel{%
\varepsilon }{\Rightarrow }_F\left| s\right. $ .
\end{center}

Then, analogous to the induction basis, we have $s\equiv s_1\wedge s_2$ for
some $s_1$ and $s_2$ with $p\stackrel{\varepsilon }{\Rightarrow }_F\left|
s_1\right. $ and $s_2\equiv \stackunder{a\in Act}{\bigwedge }\delta
_{s_1,u}(a)$. Hence $t_{n+1}\stackunder{\sim }{\sqsubset }_{RS}r\approx
_{RS}v\wedge s\stackunder{\sim }{\sqsubset }_{RS}s\equiv s_1\wedge s_2$, as
desired. \TeXButton{End Proof}{\endproof}

\qquad

We are thus led to the following strengthening of Lemma 6.2.

\qquad

{\bf Lemma 6.4 }For any process{\bf \ }$p$ and $q$, $p\varpi q$ is the
greatest (w.r.t $\sqsubseteq _{RS}$) post-fixed point of $\eta _{p,q}$.

\TeXButton{Proof}{\proof}By Lemma 6.2, we are left with the task of
determining that $p\varpi q$ is greatest among post-fixed points of $\eta
_{p,q}$. Let $t\sqsubseteq _{RS}\eta _{p,q}(t).$ We intend to prove that $%
t\sqsubseteq _{RS}p\varpi q$. Assume that $t\stackrel{\varepsilon }{%
\Rightarrow }_F\left| t_0\right. \stackrel{a_1}{\Rightarrow }_F\left|
t_1\right. ...\stackrel{a_k}{\Rightarrow }_F\left| t_k\right. $ with $k\geq
0 $ and $\neg \exists i\leq k(t_i\sqsubseteq _{RS}q)$. By Theorem 5.1, it
suffices to show that $t_k\sqsubseteq _{RS}p$. Since $t\sqsubseteq _{RS}\eta
_{p,q}(t)$ and $\eta _{p,q}$ is monotonic w.r.t $\sqsubseteq _{RS}$, we get $%
t\sqsubseteq _{RS}\eta _{p,q}^{k+1}(t)$. Then $t_k$ $\sqsubseteq _{RS}p$
immediately follows from Lemma 6.3. \TeXButton{End Proof}{\endproof}

\qquad

The next theorem constitutes one of the two main theorems of this section.

\quad

{\bf Theorem 6.2 (}Fixed-point characterization of $\varpi ${\bf ) }For any
process{\bf \ }$p$ and $q$, $p\varpi q$ is the greatest (w.r.t $\sqsubseteq
_{RS}$) fixed point of $\eta _{p,q}$.

\TeXButton{Proof}{\proof}By Lemma 6.4 and 6.2, we only need to show that $%
\eta _{p,q}\left( p\varpi q\right) \sqsubseteq _{RS}p\varpi q$. It follows
from $p\varpi q\sqsubseteq _{RS}\eta _{p,q}\left( p\varpi q\right) $ that $%
\eta _{p,q}\left( p\varpi q\right) \sqsubseteq _{RS}\eta _{p,q}\left( \eta
_{p,q}\left( p\varpi q\right) \right) $. Then, by Lemma 6.4, we have $\eta
_{p,q}\left( p\varpi q\right) \sqsubseteq _{RS}p\varpi q$, as desired.
\TeXButton{End Proof}{\endproof}

\qquad

It is well known that, for any continuous function $\Phi $ over a complete
lattice with the top element $\top $, its greatest fixed-point is exactly
the largest lower bound of the decreasing sequence $\left\{ \Phi ^i(\top
)\right\} _{i\in \omega }$ (i.e., $\nu Z.\Phi =\stackunder{i\in \omega }{%
\sqcap }\Phi ^i(\top )$) (see for instance [21]). The next theorem gives an
analogous result for $p\varpi q$.

\qquad

{\bf Theorem 6.3 (}Approximation of $\varpi ${\bf ) }For any process $p$ and
$q$, $p\varpi q$ is the greatest lower bound of the decreasing (w.r.t $%
\sqsubseteq _{RS}$) sequence $\left\{ \eta _{p,q}^i(true)\right\} _{i\in
\omega }$.

\TeXButton{Proof}{\proof} Since $p\varpi q\sqsubseteq _{RS}true$, by Lemma
6.2, it is obvious that $p\varpi q$ is a lower bound of $\left\{ \eta
_{p,q}^i(true)\right\} _{i\in \omega }$. Let $t$ be any lower bound of $%
\left\{ \eta _{p,q}^i(true)\right\} _{i\in \omega }$. We intend to show $%
t\sqsubseteq _{RS}p\varpi q$. Assume that $t\stackrel{\varepsilon }{%
\Rightarrow }_F\left| t_0\right. \stackrel{a_1}{\Rightarrow }_F\left|
t_1\right. ...\stackrel{a_k}{\Rightarrow }_F\left| t_k\right. $ with $k\geq
0 $ and $\neg \exists i\leq k(t_i\sqsubseteq _{RS}q)$. Clearly, we have $%
t\sqsubseteq _{RS}\eta _{p,q}^{k+1}(true)$. Then $t_k\sqsubseteq _{RS}p$ due
to Lemma 6.3. Consequently, $t\sqsubseteq _{RS}p\varpi q$ follows from
Theorem 5.1. \TeXButton{End Proof}{\endproof}

\quad

{\bf Remark 6.1} It has been established in [43, 60] (see also Lemma 3.11 in
this paper) that, for any process $q,$ $p_1$ and $p_2$, ($i$) $p_1\wedge
p_2\sqsubseteq _{RS}p_i$ ($i=1,2$) and ($ii$) if $q\sqsubseteq _{RS}p_1$ and
$q\sqsubseteq _{RS}p_2$ then $q\sqsubseteq _{RS}p_1\wedge p_2$. That is, $%
p_1\wedge p_2$ is the largest lower bound of $\{p_1,p_2\}$ w.r.t $%
\sqsubseteq _{RS}$. Inspired by this, someone may try to introduce the
notion of generalized conjunction in a natural way to express the largest
lower bound of $\left\{ \eta _{p,q}^i(true)\right\} _{i\in \omega }$ by the
term $\stackunder{i\in \omega }{\bigwedge }\eta _{p,q}^i(true)$. The rule
below is one of potential candidate rules that generalize the rules (Ra-7)
and (Ra-8) to the generalized conjunction.

\begin{center}
$\dfrac{p_k\stackrel{\tau }{\rightarrow }t\text{ }}{\stackunder{i\in I}{%
\bigwedge }p_i\stackrel{\tau }{\rightarrow }\stackunder{i\in I}{\bigwedge }%
t_i}$ with $k\in I$. \qquad ($GC$)
\end{center}

Here $I$ is an arbitrary indexed set, and for $i\in I,$ if $i\neq k$ then $%
t_i\equiv p_i$ else $t_i\equiv t$. Unfortunately, it would be an
unsuccessful attempt if the rule ($GC$) is adopted as the only rule
concerning $\tau $-transition for such generalized conjunction. By this
rule, $\stackunder{i\in \omega }{\bigwedge }\eta _{p,q}^i(true)$ can not
arrive at any stable state within finitely many $\tau $-transitions. Thus $%
\stackunder{i\in \omega }{\bigwedge }\eta _{p,q}^i(true)$ is inconsistent
and $\stackunder{i\in \omega }{\bigwedge }\eta _{p,q}^i(true)=_{RS}\perp $.
In fact, the conjunction $\wedge $ in the framework of LLTS can not be
generalized in the above manner to capture the generalized conjunction in
usual logics. For instance, by ($GC$), it is easy to see that the
(generalized) idempotent law $\stackunder{i\in I}{\bigwedge }p=_{RS}p$ does
not always hold, e.g., consider $p_i\equiv a.0\vee b.0$ with $i\in \omega $,
then we have $\stackunder{i\in \omega }{\bigwedge }p_i\in F$ but $a.0\vee
b.0\notin F$, and hence $\stackunder{i\in \omega }{\bigwedge }p_i\not
=_{RS}a.0\vee b.0$. By the way, since the definition of the function $\eta
_{p,q}$ refers to the term $\stackunder{a\in Act}{\bigwedge }\left\lceil
a\right\rceil X$, we assume that $Act$ is finite in this and the next two
sections.

\qquad

Analogous to [44], some basic laws concerning $\sharp $ and $\left\lceil
a\right\rceil $ are listed below, which reveals that a few of standard
temporal laws hold in CLLT.

\quad

{\bf Corollary 6.2 }For any process $p$ and $q$, we have

(1) $\left\lceil a\right\rceil true=_{RS}true=_{RS}\stackunder{A\subseteq Act%
}{\bigvee }\left( \stackunder{a\in A}{\Box }a.true\right) $

(2) $\left\lceil a\right\rceil (p\wedge q)=_{RS}\left\lceil a\right\rceil
p\wedge \left\lceil a\right\rceil q$

(3) $\sharp (p\wedge q)=_{RS}\sharp p\wedge \sharp q$

(4) $\sharp p=_{RS}p\varpi \perp $

(5) $\sharp p=_{RS}p\wedge \left( \stackunder{a\in Act}{\bigwedge }%
\left\lceil a\right\rceil \sharp p\right) $

(6) $true\wedge p=_{RS}$ $p\wedge true=_{RS}p$

\TeXButton{Proof}{\proof}(1) Obvious. (2) is implied by Lemma 3.11(3) and
(4), Theorem 6.1 and Corollary 6.1. (3) follows from Lemma 3.11(3) and (4),
Theorem 4.1 and Corollary 4.1. (4) It is enough to show that $t\sqsubseteq
_{RS}\sharp p$ iff $t\sqsubseteq _{RS}p\varpi \perp $ for any $t,$ which is
implied by Theorem 4.1 and 5.1. (5) follows from the item (4) in this lemma
and Theorem 6.2. (6) is implied by $p\sqsubseteq _{RS}true$ and Lemma
3.11(3)(4).\TeXButton{End Proof}{\endproof}

\quad

As an easy consequence, we also obtain the following fixed point
characterization and approximation of $\sharp $.

\quad

{\bf Corollary 6.3 (}Fixed-point characterization of $\sharp ${\bf )}

(1) $\sharp p=_{RS}\eta _{p,\perp }\left( \sharp p\right) $.

(2) $\sharp p$ is the greatest (post-)fixed point of $\eta _{p,\perp }$.

(3) $\sharp p$ is the greatest lower bound of the decreasing sequence $%
\left\{ \eta _{p,\perp }^i(true)\right\} _{i\in \omega }$.

\TeXButton{Proof}{\proof}Follows from Theorem 6.2 and 6.3 and Corollary 6.2
(4). \TeXButton{End Proof}{\endproof}

\quad

We conclude this section with providing some sound inference rules
concerning $\varpi $ w.r.t $\sqsubseteq _{RS}$. As an immediate consequence
of Lemma 6.4 and Theorem 6.3, it is obvious that the rules below are sound
provided that $\leq $ is interpreted as $\sqsubseteq _{RS}$. Moreover, by
Corollary 6.3, similar rules also exist for $\sharp p$.

\begin{center}
$\dfrac{t\leq \eta _{p,q}(t)\text{ }}{t\leq p\varpi q}$ (GPF)$\qquad \dfrac{%
\forall i<\omega (t\leq \eta _{p,q}^i(true))\text{ }}{t\leq p\varpi q}$ (APP)
\end{center}

Clearly, since the premise in (APP) may be proved by induction on natural
numbers, we also have the rule below. Notice that, since it always holds
that $t\sqsubseteq _{RS}true\equiv \eta _{p,q}^0(true)$, the premise $t\leq
\eta _{p,q}^0(true)$ in (INAPP) may be omitted.

\begin{center}
\quad

$\dfrac{t\leq \eta _{p,q}^0(true),\text{ }\forall i<\omega (t\leq \eta
_{p,q}^i(true)\rightarrow t\leq \eta _{p,q}^{i+1}(true))\text{ }}{t\leq
p\varpi q}$ (INAPP)
\end{center}

\section{Graphically representing $unless$ by recursion}

In the light of the greatest fixed-point characterization of $\varpi $, this
section will consider an alternative approach to giving a graphical
representation of the temporal operator $unless$ in pure process-algebraic
style. Following Milner [45], for any process $p$ and $q$, we introduce the
constant $\overline{p\varpi q}$, which is defined by the equation below

\begin{center}
$\overline{p\varpi q}=\eta _{p,q}(\overline{p\varpi q})$.
\end{center}

Formally, two rules below are added into CLLT, which are usual rules about
recursion.

\begin{center}
(Ra$_\eta $) $\dfrac{\eta _{p,q}(\overline{p\varpi q})\stackrel{\alpha \quad
}{\rightarrow t}\quad }{\overline{p\varpi q}\stackrel{\alpha \quad }{%
\rightarrow t}}$\qquad (Rp$_\eta $)$\dfrac{\eta _{p,q}(\overline{p\varpi q}%
)F\quad }{\overline{p\varpi q}F}$
\end{center}

The resulting calculus is denoted by CLLT$_\eta $. CLLT$_\eta $ inherits the
notion of the degree of a process (see, Def. 3.3) with adding the clause $%
\left| \overline{p\varpi q}\right| =1$ for each $p$ and $q$. Then it is easy
to check that the function $S_{_\eta }$ is a stratification of CLLT$_\eta $,
where $S_{_\eta }$ is defined by

\begin{itemize}
\item  $S_{_\eta }(t\stackrel{\alpha }{\rightarrow }r)=G(t)\times \omega
+\left| t\right| $ for any literal $t\stackrel{\alpha }{\rightarrow }r$, and$%
\;$

\item  $S_{_\eta }(tF)=\omega \times 2$ for any process $t$.
\end{itemize}

Here $G(t)$ is the number of {\em unguarded} occurrences of constants with
the format $\overline{r\varpi s}$ in $t$. For instance, $G(\overline{p\varpi
q}\Box \overline{r\varpi t})=2$ and $G(\overline{p\varpi q}\vee \overline{%
r\varpi t})=0$ \footnote{%
Notice that, since the `first move' of $r\vee s$ is independent of $r$ and $%
s $, the occurrence of $r$ and $s$ are (weakly) guarded in $r\vee s$.}.
Obviously, the function $G$ can be defined inductively, and we leave it to
the reader.

Therefore CLLT$_\eta $ has a unique stable transition model, and the LTS
associated with CLLT$_\eta $, denoted by $LTS(CLLT_\eta )$, may be defined
as usual. Moreover, all results obtained in Subsection 3.3 still hold for $%
LTS(CLLT_\eta )$ and will be used in the remainder of this section. Here we
do not verify them in full detail and only illustrate that $LTS(CLLT_\eta )$
is a LLTS. To this end, the notion of $\tau -$degree (see, Def. 3.6) is
enriched by adding the clause below, for any process $p$ and $q$,

\begin{center}
$d(\overline{p\varpi q})=\max \{d(q),$ $d(p\wedge \stackunder{a\in Act}{%
\bigwedge }\left\lceil a\right\rceil \overline{p\varpi q})\}+1.$
\end{center}

Clearly, $\tau $ is the only action enabled from $\overline{p\varpi q}$ and
the target state of such $\tau $-transition is either $q$ or $p\wedge
\stackunder{a\in Act}{\bigwedge }\left\lceil a\right\rceil \overline{p\varpi
q}$. Thus the above clause also appropriately measures $\overline{p\varpi q}$
's capability of executing successive $\tau $ actions. Moreover, since $d(a.%
\overline{p\varpi q})=0$ for each $a\in Act$ (see also, Def. 3.6), the
definition above is well defined.

By (Ra$_\eta $) and Def. 6.4, it is obvious that Lemma 3.7 still holds for $%
\overline{p\varpi q}$. Then, analogous to Lemma 3.8, we can prove that the
condition (LTS2) holds for $LTS(CLLT_\eta )$. Moreover, by (Ra$_\eta $) and
(Rp$_\eta $), it can be showed without any difficulty that $LTS(CLLT_\eta )$
is $\tau -$pure and satisfies (LTS1). Summarizing, $LTS(CLLT_\eta )$ is a $%
\tau -$pure LLTS.

As mentioned above, this section aims to capture the temporal operator $%
unless$ in the recursive manner. Thus we need to show an analogue of Theorem
5.1 for any constant $\overline{p\varpi q}$. We do not intend to prove such
result from scratch. The remaining work will attend to proving that $%
\overline{p\varpi q}$ is equivalent to $p\varpi q$ modulo $=_{RS}$, which
implies one that we desire.

Although the equivalence between $\overline{p\varpi q}$ and $p\varpi q$
seems straightforward, its proof is far from trivial and requires a solid
effort. In fact, if we neglect the requirement on the consistency in the
notion of ready simulation (see, Def. 2.2), it is trivial to show that $%
\overline{p\varpi q}$ and $p\varpi q$ are matching on actions. However,
everything becomes quite troublesome when the predicate $F$ is involved. The
main difficulty in carrying out such proof is that we need to prove that $%
\overline{p\varpi q}\in F$ implies $p\varpi q\in F$. This requires a
sequence of auxiliary propositions about proof trees. Before giving these
propositions, we introduce the notion below.

\qquad

{\bf Definition 7.1} Given processes $p_i$ with $i\leq n$, a process $u$ is
said to be a conjunction of these $p_i$ if each $p_i$ occurs in $u$ and $u$
is obtained from these $p_i$ by using only the operator $\wedge $ in
arbitrary order and grouping. Similarly, we can define the analogous notion
for disjunction.

\qquad \qquad

{\bf Lemma 7.1 }Given processes $p_i$ and $p_i^{*}$ such that $p_i\stackrel{%
\varepsilon }{\Rightarrow }_Fp_i^{*}$ with $i\leq n$, and let $p$ be a
conjunction of these $p_i$. If $\Im $ is a proof tree of $pF$ then there
exists a nonempty set $K\subseteq \{0,1,2\cdots n\}$ such that $\Im $
contains a subtree with the root labelled with $wF$, in particular, such
subtree is proper provided that $p_{i_0}\stackrel{\tau }{\rightarrow }%
\stackrel{\varepsilon }{\Rightarrow }p_{i_0}^{*}$ for some $i_0\leq n$,
where $w$ is a conjunction of $p_i^{*}$ with $i\in K$.

\TeXButton{Proof}{\proof}The proof will be done by induction on the depth of
the inference by which $pF$ is inferred. We denote $p[p_1^{*}\left/
p_1\right. \cdots $ $p_n^{*}\left/ p_n\right. ]$ briefly by $p^{*}$. Then $p$
$(\stackrel{\tau }{\rightarrow })^m$ $p^{*}$ for some $m$. If $m=0$ then the
conclusion holds trivially due to $p^{*}\equiv p$. Next we consider the case
$m>0$. Then $p\stackrel{\tau }{\rightarrow }s$ $(\stackrel{\tau }{%
\rightarrow })^{m-1}$ $p^{*}$ for some $s$. Moreover, since $p_i\notin F$
for each $i$ and $p\in F$, we get $n>0$. Hence $p$ $\equiv w_1\wedge w_2$
for some $w_1$ and $w_2$. Thus the last rule applied in $\Im $ is

\begin{center}
either $\dfrac{w_iF}{pF}$ with $i\in \{1,2\}$ or $\dfrac{p\stackrel{\tau }{%
\rightarrow }u,\text{ }\left\{ tF:p\stackrel{\tau }{\rightarrow }t\right\} }{%
pF}.$
\end{center}

For the first alternative, w.l.o.g, we assume $i=1$. Then $\Im $ contains a
proper subtree $\Im _1$ with the root labelled with $w_1F$. Clearly, there
exists a nonempty set $N\subseteq \{0,1,2\cdots n\}$ such that $w_1$ is a
conjunction of $p_i$ with $i\in N$. Thus, by IH, it follows that there
exists a nonempty set $K\subseteq N\subseteq \{0,1,2\cdots n\}$ such that $%
\Im _1$ contains a node labelled with $wF$, where $w$ is a conjunction of
all $p_i^{*}$ with $i\in K$.

For the second alternative, since $p\stackrel{\tau }{\rightarrow }$, there
exists $k\leq n$ and $p_k^{^{\prime }}$ such that $p_k\stackrel{\tau }{%
\rightarrow }_Fp_k^{^{\prime }}$ $\stackrel{\varepsilon }{\Rightarrow }%
_Fp_k^{*}$. Then $\Im $ contains a proper subtree $\Im _1$ with the root
labelled with $sF$, where $s$ is a conjunction of $p_k^{^{\prime }}$ and $%
p_i $ with $k\neq i\leq n$. Further, by IH, it follows that there exists a
nonempty set $K\subseteq \{0,1,2\cdots n\}$ such that $\Im _1$ contains a
node labelled with $wF$ for some conjunction $w$ of all $p_i^{*}$ with $i\in
K$. \TeXButton{End Proof}{\endproof}

\qquad

{\bf Lemma 7.2} For any nonempty $A\subseteq Act$ and processes $r$ and $t$,
let $p$ be any conjunction of $\delta _{r,t}(a)$ with $a\in A$, then each
proof tree of $pF$ must contain a proper subtree with the root labelled with
$uF$, where $u\equiv t$ or $u$ is a conjunction of $t$ and $true$.

\TeXButton{Proof}{\proof}Prove it by induction on the depth of inference.
Let $\Im $ be any proof tree of $pF$. Since $Act$ is finite, so is $A$. If $%
\left| A\right| =1$ then $p\equiv \delta _{r,t}(a)$ for some $a$. Hence it
follows from $\delta _{r,t}(a)\in F$ that $a\in I(r)$ and $\delta
_{r,t}(a)\equiv (\stackunder{b\in I(r)-\{a\}}{\Box }b.true)\Box a.t$.
Moreover, since $\stackunder{b\in I(r)-\{a\}}{\Box }b.true\notin F$, it is
easy to see that $\Im $ contains a proper subtree with the root labelled
with $tF$. In the following, we consider the case where $\left| A\right| >1$%
. In such situation, $p\equiv p_1\wedge p_2$ for some $p_1$, $p_2$. Since $%
I(p_1)=I(p_2)=I(r)$, the last rule applied in $\Im $ is

\begin{center}
either $\dfrac{p_kF}{pF}$ with $k\in \{1,2\}$ or $\dfrac{p\stackrel{\alpha }{%
\rightarrow }s,\text{ }\left\{ wF:p\stackrel{\alpha }{\rightarrow }w\right\}
}{pF}$ for some $\alpha $.
\end{center}

The proof for the first case is immediate by applying IH. For the second
one, we must have $\alpha \neq \tau $, for otherwise it immediately follows
that $\{\tau \}=I(r)$ and $p=_{RS}true\notin F$, a contradiction. Moreover,
we also have $\alpha \in A$, for otherwise a contradiction arises as $p%
\stackrel{\alpha }{\rightarrow }p^{^{\prime }}=_{RS}true\notin F$ for some $%
p^{^{\prime }}$. Then it follows from $\delta _{r,t}(\alpha )\stackrel{%
\alpha }{\rightarrow }t$ and $\delta _{r,t}(b)\stackrel{\alpha }{\rightarrow
}true$ with $b(\neq \alpha )\in A$ that the only $\alpha -$labelled
transition from $p$ is $p\stackrel{\alpha }{\rightarrow }u\equiv p[t\left/
\delta _{r,t}(\alpha ),\right. \ true/$ $\delta _{r,t}(b_1),\cdots ,$ $%
true/\delta _{r,t}(b_n)]$ with $\{b_1,$ $\cdots ,$ $b_n\}=A-\{\alpha \}$.
Clearly, $u$ is a conjunction of $t$ and a number of $true$, and $\Im $
contains a proper subtree with the root labelled with $uF$.
\TeXButton{End Proof}{\endproof}

\qquad

By the lemma above, it is obvious that $\stackunder{a\in A}{\bigwedge }%
\delta _{r,t}(a)\in F$ implies $t\in F$. In fact, the converse also holds if
$I(r)\cap A\neq \emptyset $. The result below is analogous to the well-known
fact that the sentence $\stackunder{i\leq n}{\bigwedge }(\stackunder{j\leq
m_i}{\bigvee \beta _{ij}})$ is inconsistent in classical logics if and only
if, for any set $\{\beta _{0j_0}$, $\beta _{1j_1}\cdots \beta _{nj_n}\}$
with $j_i\leq m_i$ for each $i\leq n$, there exists a nonempty set $%
N\subseteq \{0,1,\cdots n\}$ such that $\stackunder{k\in N}{\bigwedge }\beta
_{kj_k}$ is inconsistent.

\qquad

{\bf Lemma 7.3} Assume that $p$ is a conjunction of $p_i$ with $0\leq i\leq
n $ and for any $i\leq n$, there exist $p_{ij}$ with $j\leq m_i$ such that $%
p_i $ is a disjunction of $p_{ij}$'s\footnote{%
Notice that if $m_i=0$ then $p_i\equiv p_{i0}$.}. Then, for any proof tree $%
\Im $ of $pF$ and $n+1-$tuple $\overrightarrow{p_{ik_i}}$ such that $\forall
i\leq n(k_i\leq m_i)$, there exists a nonempty set $K\subseteq \{0,1,2\cdots
n\}$ such that $\Im $ contains a subtree with the root labelled with $wF$
for some conjunction $w$ of $p_{ik_i}$ with $i\in K$, in particular, such
subtree is proper whenever $\exists i\leq n(m_i>0)$.

\TeXButton{Proof}{\proof}Proceeding by induction on the depth of $\Im $.
Suppose that $\overrightarrow{p_{ik_i}}$ is any $n+1-$tuple such that $%
\forall i\leq n(k_i\leq m_i)$. If $m_i=0$ for each $i\leq n$ then there
exists exactly one such $n+1-$tuple and $p$ is a conjunction of $%
\overrightarrow{p_{ik_i}}$. Hence the conclusion holds trivially. In the
following, we consider the case where $\exists i\leq n(m_i>0)$.

If $n=0$ then $p\equiv p_0$, and hence $p$ is a disjunction of $p_{0j}$ with
$j\leq m_0$. Moreover, due to $m_0>0$, it is obvious that $\Im $ contains a
proper subtree with the root labelled with $p_{0k_0}F$. We next consider the
case where $n>0$. In such situation, we may assume that $p$ $\equiv
w_1\wedge w_2$ for some $w_1$ and $w_2$. Moreover, it is not difficult to
see that the last rule applied in $\Im $ is

\begin{center}
either $\dfrac{w_iF}{pF}$ with $i\in \{1,2\}$ or $\dfrac{p\stackrel{\tau }{%
\rightarrow }w,\text{ }\left\{ tF:p\stackrel{\tau }{\rightarrow }t\right\} }{%
pF}.$
\end{center}

For the first alternative, w.l.o.g, we assume $i=1$. Thus $\Im $ contains a
proper subtree $\Im _1$ with the root labelled with $w_1F$. Clearly, there
exists a nonempty set $N\subset \{0,1,2\cdots n\}$ such that $w_1$ is a
conjunction of $p_i$ with $i\in N$. For $\left| N\right| -$tuple $%
\overrightarrow{p_{ik_i}}$ with $i\in N$, by IH, there exists a nonempty set
$K\subseteq N\subset \{0,1,2\cdots n\}$ such that $\Im _1$ contains a node
labelled with $wF$, where $w$ is a conjunction of $p_{ik_i}$ with $i\in K$,
as desired.

For the second alternative, it follows from $\exists i\leq n(m_i>0)$ that $p$
$\stackrel{\tau }{\rightarrow }s\stackrel{\varepsilon }{\Rightarrow }$ $p[\
\overrightarrow{p_{ik_i}}/$ $\overrightarrow{p_i}]$ for some $s$. Thus $\Im $
contains a proper subtree $\Im _1$ with the root labelled with $sF$.
Obviously, for some $j_0\leq n$ and $p_{j_0}^{^{\prime }}$ with $p_{j_0}%
\stackrel{\tau }{\rightarrow }p_{j_0}^{^{\prime }}$, $s$ is a conjunction of
$p_{j_0}^{^{\prime }}$ and $p_i$ with $j_0\neq i\leq n$. Moreover, there
exists a nonempty set $N\subset \{0,1,2\cdots m_{j_0}\}$ such that $%
p_{j_0}^{^{\prime }}$ is a disjunction of $p_{j_0i}$ with $i\in N$. In
particular, $p_{j_0k_{j_0}}\equiv p_{j_0l}$ for some $l\in N$ due to $p$ $%
\stackrel{\tau }{\rightarrow }s\stackrel{\varepsilon }{\Rightarrow }$ $p[\
\overrightarrow{p_{ik_i}}/$ $\overrightarrow{p_i}]$. Then, by IH, there
exists a nonempty set $K\subseteq \{0,1,2\cdots n\}$ such that $\Im _1$
contains a node labelled with $wF$, where $w$ is a conjunction of $p_{ik_i}$
with $i\in K$. \TeXButton{End Proof}{\endproof}

\qquad

Now we are ready to show that $\overline{p\varpi q}\in F$ implies $p\varpi
q\in F$ by induction on inference. The lemma below contains four assertions
which state the links between consistency of some processes in the
transition system generated by $\overline{p\varpi q}$ and consistency of
corresponding processes in the transition system generated by $p\varpi q$.

\quad \qquad

{\bf Lemma 7.4} Assume that $u$ $\in F$. Then

(1) If $u\equiv \overline{p\varpi q}$ or $u$ is a conjunction of $\overline{%
p\varpi q}$ and a number of $true$ then $p\varpi q\in F$.

(2) If $u$ is a conjunction of $p\wedge \stackunder{a\in Act}{\bigwedge }%
\left\lceil a\right\rceil \overline{p\varpi q}$ and $p_i$ with $i<n$ then $%
t\odot (p\varpi q)\in F$ for any conjunction $t$ of $p$ and $p_i$'s.

(3) If $u$ is a conjunction of $\delta _{p_0,\overline{p\varpi q}}(a)$ and
stable $p_i$ with $a\in A$ and $i\leq n$, then $t\odot (p\varpi q)\in F$ for
any conjunction $t$ of $p_i$'s, where $\emptyset \neq A\subseteq Act$.

(4) If $u$ is a conjunction of $p_i$ with $i\leq n$, $\overline{p\varpi q}$
and a number of $true$ then $(t\wedge p)\odot (p\varpi q)\in F$ for any
conjunction $t$ of $p_i$'s.

\TeXButton{Proof}{\proof}Let $\Im $ be any proof tree of $uF$. We will prove
item (1)-(4) simultaneously by induction on the depth of $\Im $. The
argument splits into five cases based on the format of $u$.

\begin{description}
\item  Case 1 $u\equiv \overline{p\varpi q}$.
\end{description}

It is obvious that the last two inference steps in $\Im $ are

\begin{center}
$\dfrac{\dfrac{qF\text{,\quad }(p\wedge \stackunder{a\in Act}{\bigwedge }%
\left\lceil a\right\rceil \overline{p\varpi q})F}{\eta _{p,q}(\overline{%
p\varpi q})F}}{\overline{p\varpi q}F}.$
\end{center}

Thus $q\in F$. Moreover, by IH about item (2) with $n=0$, we also get $%
p\odot (p\varpi q)\in F$. Then $p\varpi q\in F$ by Lemma 3.3 (5).

\begin{description}
\item  Case 2 $u$ is a conjunction of $\overline{p\varpi q}$ and $p_i$ ($%
\equiv $ $true$) with $i\leq n$.
\end{description}

In this situation, we may assume that $u\equiv u_1\wedge u_2$. Since $u%
\stackrel{\tau }{\rightarrow }$, the last rule applied in $\Im $ is

\begin{center}
either $\dfrac{u_iF}{uF}$ with $i\in \{1,2\}$ or $\dfrac{u\stackrel{\tau }{%
\rightarrow }w,\text{ }\left\{ rF:u\stackrel{\tau }{\rightarrow }r\right\} }{%
uF}.$
\end{center}

For the first alternative, w.l.o.g, we assume $i=1$. Since $u_1\in F$, the
process $\overline{p\varpi q}$ must occur in $u_1$. So, by IH about item
(1), we have $p\varpi q\in F$, as desired.

For the second alternative, since $\overline{p\varpi q}\stackrel{\tau }{%
\rightarrow }q$ and $\overline{p\varpi q}\stackrel{\tau }{\rightarrow }%
p\wedge \stackunder{a\in Act}{\bigwedge }\left\lceil a\right\rceil \overline{%
p\varpi q}$, there exist two proper subtrees of $\Im $ whose roots are
labelled with $v_1F$ and $v_2F$ respectively, where $v_1$ (or, $v_2$) is a
conjunction of $q$ ( respectively, $p\wedge \stackunder{a\in Act}{\bigwedge }%
\left\lceil a\right\rceil \overline{p\varpi q}$) and $p_i$'s. Then $q\in F$
by Corollary 6.2(6). Moreover, by IH about item (2), Corollary 6.2(6) and
Theorem 5.2, we also get $p\odot (p\varpi q)\in F$. Hence $p\varpi q\in F$.

\begin{description}
\item  Case 3 $u$ is a conjunction of $p\wedge \stackunder{a\in Act}{%
\bigwedge }\left\lceil a\right\rceil \overline{p\varpi q}$ and $p_i$ with $%
i<n$.
\end{description}

Let $t$ be any conjunction of $p$ and $p_i$'s. If $t\in F$ then it
immediately follows that $t\odot (p\varpi q)\in F$. In the following, we
consider the nontrivial case where $t\notin F$. If it were true that $s\odot
(p\varpi q)\in F$ for any $s$ with $t\stackrel{\varepsilon }{\Rightarrow }%
_F\left| s\right. $, we would have $t\odot (p\varpi q)\in F$ by Lemma
3.1(7), 3.3(6) and 3.8. Thus we assume that $t\stackrel{\varepsilon }{%
\Rightarrow }_F\left| t_0\right. $ and intend to prove $t_0\odot (p\varpi
q)\in F$ . Clearly, $\stackunder{a\in Act}{\bigwedge }\left\lceil
a\right\rceil \overline{p\varpi q}\notin F$\footnote{%
This follows from $\stackunder{a\in Act}{\bigwedge }\left\lceil
a\right\rceil \overline{p\varpi q}\stackrel{\varepsilon }{\Rightarrow }%
\left| \stackunder{a\in Act}{\bigwedge }0\right. \notin F$ and Lemma 3.6.}
and there exist $p^{*}$ and $w_i$ $(i<n)$ with properties below:

\begin{center}
$t_0$ is a conjunction of $p^{*}$ and $w_i$, $p\stackrel{\varepsilon }{%
\Rightarrow }_F\left| p^{*}\right. $ and $p_i\stackrel{\varepsilon }{%
\Rightarrow }_F\left| w_i\right. $ for each $i<n$.
\end{center}

Then, by Lemma 7.1, there is a nonempty set $\Gamma \subseteq \{p^{*}$, $w_i$%
, $\stackunder{a\in Act}{\bigwedge }\left\lceil a\right\rceil \overline{%
p\varpi q}:i<n\}$ such that $\Im $ contains a node labelled with $wF$, where
$w$ is a conjunction of processes within $\Gamma $. Moreover, due to $%
\stackunder{a\in Act}{\bigwedge }\left\lceil a\right\rceil \overline{p\varpi
q}\notin F$ and $t_0\notin F$, we have $\stackunder{a\in Act}{\bigwedge }%
\left\lceil a\right\rceil \overline{p\varpi q}\in \Gamma $ and $\Gamma \cap
\{p^{*}$, $w_i:i<n\}\neq \emptyset $.

On the other hand, by Def. 6.3, for each $a\in Act$, the process $%
\left\lceil a\right\rceil \overline{p\varpi q}$ is a disjunction of
processes $S_A$ with $A\subseteq Act$, where

\begin{center}
$S_A\equiv \left\{
\begin{array}{c}
\stackunder{b\in A}{\Box }b.true\qquad \qquad \text{if }a\notin A \\  \\
(\stackunder{b\in A-\{a\}}{\Box }b.true)\Box a.\overline{p\varpi q}\quad
\text{otherwise}
\end{array}
\right. .$
\end{center}

In particular, by setting $A=I(p^{*})$ for each $a\in Act$, we get a tuple $%
\overrightarrow{\delta _{p^{*},\overline{p\varpi q}}(a)}$ with $a\in Act$.
Moreover, each process in $\Gamma \cap \{p^{*}$, $w_i:i<n\}$ may be regarded
as a disjunction of itself. Thus, by Lemma 7.3, there exists a nonempty set $%
\Theta \subseteq (\Gamma \cap \{p^{*},w_i:i<n\})\cup \{\delta _{p^{*},%
\overline{p\varpi q}}(a):a\in Act\}$ such that $\Im $ contains a proper
subtree $\Im _1$ with the root labelled with $sF$ for some conjunction $s$
of all processes in $\Theta $. Due to $t_0\notin F$, $\Theta $ must contain $%
\delta _{p^{*},\overline{p\varpi q}}(a)$ for some $a\in Act$. We distinguish
two cases below.

\begin{description}
\item  Case 3.1 $\Theta \subseteq \{$ $\delta _{p^{*},\overline{p\varpi q}%
}(a):a\in Act\}$.
\end{description}

Then $s$ is a conjunction of some processes with the format $\delta _{p^{*},%
\overline{p\varpi q}}(a)$. By Lemma 7.2, $\Im _1$ contains a proper subtree
with the root labelled with $rF$, where either $r\equiv \overline{p\varpi q}$
or $r$ is a conjunction of $\overline{p\varpi q}$ and a number of $true$.
So, by IH about item (1), we have $p\varpi q\in F$. Hence $p\odot (p\varpi
q)\in F$ by Lemma 3.3(5). Moreover, by Lemma 3.6 (2), it follows from $%
p\odot (p\varpi q)\stackrel{\varepsilon }{\Rightarrow }\left| p^{*}\odot
(p\varpi q)\right. $ that $p^{*}\odot (p\varpi q)\in F$. Further, by Theorem
5.2 and $t_0$ $\sqsubseteq _{RS}$ $p^{*}$, we get $t_0\odot (p\varpi q)\in F$%
, as desired.

\begin{description}
\item  Case 3.2 $\Theta \not \subseteq \{$ $\delta _{p^{*},\overline{p\varpi
q}}(a):a\in Act\}$.
\end{description}

In such situation, $\Theta \cap \{p^{*},w_i:i<n\}\neq \emptyset $. Let $t_1$
be any conjunction of processes within $\Theta \cap \{p^{*},w_i:i<n\}$. Then
$t_1\odot (p\varpi q)\in F$ due to IH about item (3)\footnote{%
Notice that, due to $t_0$ $\notin F$, $I(p^{*})=I(w_i)$ for each $i<n$.
Hence, for any $i<n$ and $a\in Act$, $\delta _{p^{*},\overline{p\varpi q}%
}(a) $ is identical with $\delta _{w_i,\overline{p\varpi q}}(a)$.}. Further,
by Theorem 5.2 and $t_0$ $\sqsubseteq _{RS}$ $t_1$, we get $t_0\odot
(p\varpi q)\in F$.

\begin{description}
\item  Case 4 $u$ is a conjunction of stable processes $p_i$ and $\delta
_{p_0,\overline{p\varpi q}}(a)$ with $i\leq n$ and $a\in A\neq \emptyset $.
\end{description}

Let $t$ be any conjunction of $p_i$ with $i\leq n$. Clearly, $u\equiv
u_1\wedge u_2$ for some $u_1$ and $u_2$. In the following, we consider only
the nontrivial case $t\notin F$. In such situation, it is obvious that all $%
p_i$ have the same ready set. Since $u$ is stable and $I(u_1)=I(u_2)=I(p_i)$
for each $i\leq n$, we may distinguish two cases based on the last rule
applied in $\Im $.

\begin{description}
\item  Case 4.1$\dfrac{u_iF}{uF}$ with $i\in \{1,2\}$.
\end{description}

W.l.o.g, we assume $i=1$. Since $t\notin F$, $\delta _{p_0,\overline{p\varpi
q}}(a)$ occurs in $u_1$ for some $a\in A$. If $u_1$ also contains $p_i$ for
some $i\leq n$ then, by IH about item (3), we have $t_1\odot (p\varpi q)\in
F $, where $t_1$ is any conjunction of all $p_i$ occurring in $u_1$. By $%
t\sqsubseteq _{RS}t_1$ and Theorem 5.2, we have $t\odot (p\varpi
q)\sqsubseteq _{RS}t_1\odot (p\varpi q)$. Then it follows from $t_1\odot
(p\varpi q)\in F$ that $t\odot (p\varpi q)\in F$, as desired.

Next we consider another case where none of $p_i$ ($i\leq n$) occurs in $u_1$%
. Then there exists a nonempty set $B\subseteq A$ such that $u_1$ is a
conjunction of $\delta _{p_0,\overline{p\varpi q}}(a)$ with $a\in B$. Thus,
by Lemma 7.2, $\Im $ contains a proper subtree with the root labelled with $%
wF$, where $w$ is either $\overline{p\varpi q}$ or a conjunction of $%
\overline{p\varpi q}$ and a number of $true$. Hence $p\varpi q\in F$ due to
IH about item (1). Further, by Lemma 3.3(5), we obtain

\begin{center}
$q\in F$ and $p\odot (p\varpi q)\in F$.
\end{center}

On the other hand, since $u_1$ is a conjunction of $\delta _{p_0,\overline{%
p\varpi q}}(a)$ with $a\in B$, we must have $I(p_0)\neq \emptyset $, for
otherwise a contradiction arises due to $u_1=_{RS}\stackunder{a\in B}{%
\bigwedge }\delta _{p_0,\overline{p\varpi q}}(a)\equiv \stackunder{a\in B}{%
\bigwedge }0\notin F$. Let $b$ be any action in $I(p_0)$. Since all $p_i$
have the same ready set and $t$ is a conjunction of $p_i$'s, we have $b\in
I(t)=I(t\odot (p\varpi q))$. In order to prove that $t\odot (p\varpi q)\in F$%
, by Lemma 3.5, it is enough to show that $v\in F$ for each $v$ with $t\odot
(p\varpi q)\stackrel{b}{\rightarrow }v$. Let $r$ be any target state of $b-$%
labelled transitions from $t\odot (p\varpi q)$. Then $r\equiv t_1\wedge q$
or $r\equiv (t_1\wedge p)\odot (p\varpi q)$ for some $t_1$ with $t\stackrel{b%
}{\rightarrow }t_1$. For the former, it follows from $q\in F$ that $r\in F$.
For the latter, by Lemma 3.11(3) and Theorem 5.2, we have $r\equiv
(t_1\wedge p)\odot (p\varpi q)\sqsubseteq _{RS}p\odot (p\varpi q)$, and
hence $r\in F$ because of $p\odot (p\varpi q)\in F$.

\begin{description}
\item  Case 4.2 $\dfrac{u\stackrel{b}{\rightarrow }s,\text{ }\left\{ rF:u%
\stackrel{b}{\rightarrow }r\right\} }{uF}$ for some $b\in Act$.
\end{description}

Since $u$ is a conjunction of $p_i$ and $\delta _{p_0,\overline{p\varpi q}%
}(a)$ with $i\leq n$ and $a\in A$, we get $b\in I(p_i)$ for each $i\leq n$.
Thus $b\in I(t)=I(t\odot (p\varpi q))$. Analogous to Case 4.1, in order to
prove that $t\odot (p\varpi q)\in F$, it is enough to show that each $b-$%
derivative of $t\odot (p\varpi q)$ is inconsistent. Let $r$ be any process
such that $t\odot (p\varpi q)\stackrel{b}{\rightarrow }r$. Then $r\equiv
t_1\wedge q$ or $r\equiv (t_1\wedge p)\odot (p\varpi q)$ for some $t_1$ with
$t\stackrel{b}{\rightarrow }t_1$. In the following, we intend to prove that
both $t_1\wedge q$ and $(t_1\wedge p)\odot (p\varpi q)$ are inconsistent.

We first prove that $b\in A$. On the contrary, suppose that $b\notin A$.
Hence $\delta _{p_0,\overline{p\varpi q}}(a)\stackrel{b}{\rightarrow }true$
for each $a\in A$. Moreover, since $\stackunder{i\leq n}{\bigwedge }%
p_i=_{RS}t\notin F$, by Lemma 3.5, $\stackunder{i\leq n}{\bigwedge }p_i%
\stackrel{b}{\rightarrow }s$ for some $s\notin F$. Then a contradiction
arises as $s=_{RS}r$ for some $r$ with $u\stackrel{b}{\rightarrow }r$.

Since $t\stackrel{b}{\rightarrow }t_1$, there exist $w_i$ with $p_i\stackrel{%
b}{\rightarrow }w_i$ for each $i\leq n$ and $t_1$ is a conjunction of these $%
w_i$. Moreover, since $b\in A$ and $u$ is a conjunction of $p_i$ and $\delta
_{p_0,\overline{p\varpi q}}(a)$ with $i\leq n$ and $a\in A$, there is a
process $w$ such that $u\stackrel{b}{\rightarrow }w$ and $w$ is a
conjunction of $w_i$ with $i\leq n$, $\overline{p\varpi q}$ and a number of $%
true$. Thus $\Im $ contains a proper subtree with the root labelled with $wF
$. Hence $(t_1\wedge p)\odot (p\varpi q)\in F$ due to IH about item (4).

On the other hand, since $\overline{p\varpi q}\stackrel{\tau }{\rightarrow }%
q $, there exists $v$ such that $w\stackrel{\tau }{\rightarrow }v$ and $v$
is a conjunction of $w_i$ with $i\leq n$, $q$ and a number of $true$. So, by
Lemma 3.6 (1), it follows from $w\in F$ that $v\in F$. Moreover, by
Corollary 6.2 (6) and the idempotent, commutative and associative laws of $%
\wedge $, it is easy to see that $v=_{RS}t_1\wedge q$. Hence $t_1\wedge q\in
F$.

\begin{description}
\item  Case 5 $u$ is a conjunction of processes $p_i$, $\overline{p\varpi q}$
and $t_j$ ($\equiv $ $true$) with $i\leq n$ and $j<m$.
\end{description}

Let $t$ be any conjunction of $p_i$ with $i\leq n$. Similarly, we consider
only the nontrivial case $t\wedge p\notin F$, and assume that $u\equiv
u_1\wedge u_2$. Since $u\stackrel{\tau }{\rightarrow }$, we may distinguish
two cases based on the last rule applied in $\Im $.

\begin{description}
\item  Case 5.1 $\dfrac{u\stackrel{\tau }{\rightarrow }s,\text{ }\left\{ rF:u%
\stackrel{\tau }{\rightarrow }r\right\} }{uF}$
\end{description}

Since $\overline{p\varpi q}\stackrel{\tau }{\rightarrow }p\wedge \stackunder{%
a\in Act}{\bigwedge }\left\lceil a\right\rceil \overline{p\varpi q}$, there
exists $w$ such that $u\stackrel{\tau }{\rightarrow }w$ and $w$ is a
conjunction of $p_i$, $p\wedge \stackunder{a\in Act}{\bigwedge }\left\lceil
a\right\rceil \overline{p\varpi q}$ and $t_j$ with $i\leq n$ and $j<m$. So, $%
\Im $ contains a proper subtree with the root labelled with $wF$. Let $v$ be
any conjunction of $p_i$, $p$ and $t_j$ with $i\leq n$ and $j<m$. Then $%
v\odot (p\varpi q)\in F$ by IH about item (2). On the other hand, by
Corollary 6.2 (6) and the idempotent, commutative and associative laws of $%
\wedge $, we have $t\wedge p=_{RS}v$. Further, by Theorem 5.2, it follows
from $v\odot (p\varpi q)\in F$ that $(t\wedge p)\odot (p\varpi q)\in F$.

\begin{description}
\item  Case 5.2 $\dfrac{u_iF}{uF}$ with $i\in \{1,2\}$.
\end{description}

W.l.o.g, we assume $i=1$. Since $t\wedge p\notin F$, $\overline{p\varpi q}$
must occur in $u_1$. We consider two cases below.

If $p_i$ does not occur in $u_1$ for each $i\leq n$, then $u_1$ is a
conjunction of $\overline{p\varpi q}$ and a number of $true$. Thus $p\varpi
q\in F$ by applying IH about item (1). So, by Lemma 3.3(5), we have $p\odot
(p\varpi q)\in F$. On the other hand, by Theorem 5.2(2), it follows from $%
t\wedge p\sqsubseteq _{RS}p$ that $(t\wedge p)\odot (p\varpi q)\sqsubseteq
_{RS}p\odot (p\varpi q)$. Hence $(t\wedge p)\odot (p\varpi q)\in F$.

If there exist some $p_i$ occurring in $u_1$, then $(t_1\wedge p)\odot
(p\varpi q)\in F$ by IH about item (4), where $t_1$ is any conjunction of
all $p_i$ occurring in $u_1$. Similarly, by Theorem 5.2(2) and $t\wedge
p\sqsubseteq _{RS}t_1\wedge p$, it follows that $(t\wedge p)\odot (p\varpi
q)\in F$. \TeXButton{End Proof}{\endproof}

\quad

The preceding result guarantees that a series of processes are consistent
under certain circumstance. We will encounter such processes and
circumstance in the next lemma, which will be used in demonstrating the main
result of this section.

\qquad \qquad \qquad

{\bf Lemma 7.5} Suppose that $v\sqsubseteq _{RS}p\varpi q$ and the relation $%
R$ exactly consists of all pairs $<t,w\wedge (\stackunder{a\in Act}{%
\bigwedge }\delta _{w,\overline{p\varpi q}}(a))>$ such that there exist $%
n<\omega $, $p_i,v_i$, $a_j$ and $u_j$ with $i\leq n$ and $j\leq n-1$
satisfying the conditions below

($a$) $p\stackrel{\varepsilon }{\Rightarrow }_F\left| p_0\right. $ and $v%
\stackrel{\varepsilon }{\Rightarrow }_F\left| v_0\right. $,

($b$) for each $i$ with $0\leq i\leq n-1$, $v_i\stackrel{a_i}{\Rightarrow }%
_F\left| v_{i+1}\right. $, $p_i\stackrel{a_i}{\rightarrow }_Fu_i$ and $%
u_i\wedge p\stackrel{\varepsilon }{\Rightarrow }_F\left| p_{i+1}\right. $,

($c$) for each $i$ with $0\leq i\leq n$, $v_i\stackunder{\sim }{\sqsubset }%
_{RS}p_i\odot (p\varpi q)$ and $v_i\not \sqsubseteq _{RS}q$, and

($d$) $t\equiv v_n$ and $w\equiv p_n$.

Then $R\bigcup \stackunder{\sim }{\sqsubset }_{RS}$ is a stable ready
simulation relation up to $\stackunder{\sim }{\sqsubset }_{RS}$.

\TeXButton{Proof}{\proof}Let $<r,s\wedge (\stackunder{a\in Act}{\bigwedge }%
\delta _{s,\overline{p\varpi q}}(a))>$ be any pair in $R$. Thus there exist $%
p_i,v_i$, $a_j$ and $u_j$ with $i\leq n$ and $j\leq n-1$ satisfying the
conditions ($a$)- ($d$). In particular, $r\equiv v_n$ and $s\equiv p_n$. We
intend to check that this pair satisfies four conditions in Def. 4.1.
Amongst, it is straightforward for (RS1) and (RS4). Moreover, due to $v_n%
\stackunder{\sim }{\sqsubset }_{RS}p_n\odot (p\varpi q)$ and $v_n\notin F$,
we have $p_n\odot (p\varpi q)\notin F$. Then, by Lemma 7.4 (3), it follows
that

\begin{center}
$p_n\wedge (\stackunder{a\in Act}{\bigwedge }\delta _{p_n,\overline{p\varpi q%
}}(a))\notin F$.\qquad (7.5.1)
\end{center}

Hence (RS2) holds. Next we verify (RS3-upto). Let $r(\equiv v_n)\stackrel{b}{%
\Rightarrow }_F\left| v_{n+1}\right. $. Since $v\sqsubseteq _{RS}p\varpi q$
and $v\notin F$, by Lemma 7.4(1), it follows that $\overline{p\varpi q}%
\notin F$. Then, due to $b\in I(v_n)=I(p_n\odot (p\varpi q))=I(p_n)$, we
have, for any $a\in Act$,

\begin{center}
$\delta _{p_n,\overline{p\varpi q}}(a)\stackrel{b}{\rightarrow }_F\left\{
\begin{array}{c}
true\qquad
\text{if }a\neq b \\  \\
\overline{p\varpi q}\qquad \text{if }a=b
\end{array}
\right. .$\quad (7.5.2)
\end{center}

To complete the proof, we want to find $t$ such that $p_n\wedge (\stackunder{%
a\in Act}{\bigwedge }\delta _{p_n,\overline{p\varpi q}}(a))\stackrel{b}{%
\Rightarrow }_F\left| t\right. $ and $\left\langle v_{n+1},t\right\rangle
\in (R\bigcup \stackunder{\sim }{\sqsubset }_{RS})\circ \stackunder{\sim }{%
\sqsubset }_{RS}$. We distinguish two cases below.

\begin{description}
\item  Case 1 $v_{n+1}\sqsubseteq _{RS}q$.
\end{description}

Due to $v_n\stackunder{\sim }{\sqsubset }_{RS}p_n\odot (p\varpi q)$ and
Lemma 5.1 (1), we have $v_n\stackunder{\sim }{\sqsubset }_{RS}p_n$. Further,
we get $v_{n+1}\stackunder{\sim }{\sqsubset }_{RS}p_n^{^{\prime }}$ for some
$p_n^{^{\prime }}$, $u_n$ with $p_n\stackrel{b}{\rightarrow }_Fu_n\stackrel{%
\varepsilon }{\Rightarrow }_F\left| p_n^{^{\prime }}\right. $. On the other
hand, it follows from $v_{n+1}\sqsubseteq _{RS}q$ that $v_{n+1}\stackunder{%
\sim }{\sqsubset }_{RS}q_1$ for some $q_1$ with $q\stackrel{\varepsilon }{%
\Rightarrow }_F\left| q_1\right. $. By Lemma 3.11(2), since $v_{n+1}%
\stackunder{\sim }{\sqsubset }_{RS}p_n^{^{\prime }}$ and $v_{n+1}\stackunder{%
\sim }{\sqsubset }_{RS}q_1$, we obtain

\begin{center}
$v_{n+1}\stackunder{\sim }{\sqsubset }_{RS}p_n^{^{\prime }}\wedge q_1$.
\quad (7.5.3)
\end{center}

Hence $p_n^{^{\prime }}\wedge q_1\notin F$ because of $v_{n+1}\notin F$.
Further, by Lemma 3.6 (2) and 3.1(8), it follows that

\begin{center}
$u_n\wedge \overline{p\varpi q}\stackrel{\tau }{\rightarrow }_Fu_n\wedge q%
\stackrel{\varepsilon }{\Rightarrow }_F\left| p_n^{^{\prime }}\wedge
q_1\right. .$
\end{center}

By (7.5.2), Lemma 3.2(2), Corollary 6.2(6) and the idempotent, commutative
and associative laws of $\wedge $, we obtain

\begin{center}
$p_n\wedge (\stackunder{a\in Act}{\bigwedge }\delta _{p_n,\overline{p\varpi q%
}}(a))\stackrel{b}{\rightarrow }_Fu=_{RS}u_n\wedge \overline{p\varpi q}$ for
some $u.$
\end{center}

Hence there exists $t$ such that

\begin{center}
$p_n\wedge (\stackunder{a\in Act}{\bigwedge }\delta _{p_n,\overline{p\varpi q%
}}(a))\stackrel{b}{\rightarrow }_Fu\stackrel{\varepsilon }{\Rightarrow }%
_F\left| t\right. $ and $p_n^{^{\prime }}\wedge q_1\stackunder{\sim }{%
\sqsubset }_{RS}t$.
\end{center}

Further, by (7.5.3), we have $v_{n+1}\stackunder{\sim }{\sqsubset }%
_{RS}p_n^{^{\prime }}\wedge q_1\stackunder{\sim }{\sqsubset }_{RS}t$. Thus
the process $t$ is exactly the one that we seek.

\quad

\begin{description}
\item  Case 2 $v_{n+1}\not \sqsubseteq _{RS}q$.
\end{description}

Since $v_n\stackunder{\sim }{\sqsubset }_{RS}p_n\odot (p\varpi q)$ and $v_n%
\stackrel{b}{\Rightarrow }_F\left| v_{n+1}\right. $, there exists $u$ such
that $v_{n+1}\stackunder{\sim }{\sqsubset }_{RS}u$ and $p_n\odot (p\varpi q)$
$\stackrel{b}{\Rightarrow }_F\left| u\right. $. Further, due to $v_{n+1}\not
\sqsubseteq _{RS}q$, it is not difficult to see that there exist $p_{n+1}$, $%
u_n$, $p^{^{\prime }}$ and $p_n^{^{\prime }}$ such that

\begin{center}
$p_n\stackrel{b}{\rightarrow }_Fu_n\stackrel{\varepsilon }{\Rightarrow }%
_F\left| p_n^{^{\prime }}\right. $, $p$ $\stackrel{\varepsilon }{\Rightarrow
}_F\left| p^{^{\prime }}\right. $, $p_{n+1}\equiv p_n^{^{\prime }}\wedge
p^{^{\prime }}$, and

$p_n\odot (p\varpi q)\stackrel{b}{\rightarrow }_F(u_n\wedge p)\odot (p\varpi
q)$ $\stackrel{\varepsilon }{\Rightarrow }_F\left| p_{n+1}\odot (p\varpi
q)\right. \equiv u$.
\end{center}

Then it follows that

\begin{center}
$<v_{n+1},p_{n+1}\wedge (\stackunder{a\in Act}{\bigwedge }\delta
_{p_{_{n+1}},\overline{p\varpi q}}(a))>\in R$. \qquad (7.5.4)
\end{center}

On the other hand, by Lemma 3.2 (2), Corollary 6.2(6) and the idempotent,
commutative and associative laws of $\wedge $, it follows from (7.5.2) that
there exists $t$ such that

\begin{center}
$p_n\wedge (\stackunder{a\in Act}{\bigwedge }\delta _{p_n,\overline{p\varpi q%
}}(a))\stackrel{b}{\rightarrow }t=_{RS}u_n\wedge \overline{p\varpi q}.\quad $%
(7.5.5)
\end{center}

Moreover, it is obvious that

\begin{center}
$u_n\wedge \overline{p\varpi q}\stackrel{\varepsilon }{\Rightarrow }\left|
p_n^{^{\prime }}\wedge (p^{^{\prime }}\wedge \stackunder{a\in Act}{\bigwedge
}\delta _{p_{_{n+1}},\overline{p\varpi q}}(a))\right. \approx
_{RS}p_{n+1}\wedge (\stackunder{a\in Act}{\bigwedge }\delta _{p_{_{n+1}},%
\overline{p\varpi q}}(a)).$
\end{center}

By Lemma 7.4 (3), it follows from $p_{n+1}\odot (p\varpi q)\notin F$ that

\begin{center}
$p_{n+1}\wedge (\stackunder{a\in Act}{\bigwedge }\delta _{p_{_{n+1}},%
\overline{p\varpi q}}(a))\notin F$.
\end{center}

Hence $p_n^{^{\prime }}\wedge (p^{^{\prime }}\wedge \stackunder{a\in Act}{%
\bigwedge }\delta _{p_{_{n+1}},\overline{p\varpi q}}(a))\notin F$. Thus, by
Lemma 3.6 (2), we have

\begin{center}
$u_n\wedge \overline{p\varpi q}\stackrel{\varepsilon }{\Rightarrow }_F\left|
p_n^{^{\prime }}\wedge (p^{^{\prime }}\wedge \stackunder{a\in Act}{\bigwedge
}\delta _{p_{_{n+1}},\overline{p\varpi q}}(a))\right. .$
\end{center}

So, it follows from (7.5.1) and (7.5.5) that there exists $w$ such that

\begin{center}
$p_n\wedge (\stackunder{a\in Act}{\bigwedge }\delta _{p_n,\overline{p\varpi q%
}}(a))\stackrel{b}{\rightarrow }_Ft\stackrel{\varepsilon }{\Rightarrow }%
_F\left| w\right. $ and $p_{n+1}\wedge (\stackunder{a\in Act}{\bigwedge }%
\delta _{p_{_{n+1}},\overline{p\varpi q}}(a))\stackunder{\sim }{\sqsubset }%
_{RS}w.$
\end{center}

Moreover, due to (7.5.4), we get $\left\langle v_{n+1},w\right\rangle \in
(R\bigcup \stackunder{\sim }{\sqsubset }_{RS})\circ \stackunder{\sim }{%
\sqsubset }_{RS}$, as desired. \TeXButton{End Proof}{\endproof}

\quad

We now have the below assertion of the equivalence of $\overline{p\varpi q}$
and $p\varpi q$.

\qquad

{\bf Theorem 7.1} $\overline{p\varpi q}=_{RS}p\varpi q$ for any process $p$
and $q$.

\TeXButton{Proof}{\proof}Since $\overline{p\varpi q}=_{RS}\eta _{p,q}(%
\overline{p\varpi q})$, by Theorem 6.2, it is enough to prove that $p\varpi
q\sqsubseteq _{RS}\overline{p\varpi q}$. To this end, we intend to show that
$v\sqsubseteq _{RS}\overline{p\varpi q}$ for any $v$ such that $v\sqsubseteq
_{RS}p\varpi q$. Assume that $v\sqsubseteq _{RS}p\varpi q$ and $v\stackrel{%
\varepsilon }{\Rightarrow }_F\left| v_0\right. $. Then $p\varpi q\notin F$.
By Lemma 7.4(1), we have $\overline{p\varpi q}\notin F$. In the following,
we want to find $s$ such that $\overline{p\varpi q}\stackrel{\varepsilon }{%
\Rightarrow }_F\left| s\right. $ and $v_0\stackunder{\sim }{\sqsubset }%
_{RS}s $. In the situation that $v_0\sqsubseteq _{RS}q$, this is
straightforward. We next consider the case where $v_0\not \sqsubseteq _{RS}q$%
. In such case, it follows from $v\sqsubseteq _{RS}p\varpi q$ and $v%
\stackrel{\varepsilon }{\Rightarrow }_F\left| v_0\right. $ that $v_0%
\stackunder{\sim }{\sqsubset }_{RS}p_0\odot (p\varpi q)$ for some $p_0$ such
that $p\stackrel{\varepsilon }{\Rightarrow }_F\left| p_0\right. $. Hence, by
Lemma 7.4 (3), we have $p_0\wedge (\stackunder{a\in Act}{\bigwedge }\delta
_{p_0,\overline{p\varpi q}}(a))\notin F$. Then, by the rule (Ra$_\eta $) and
Lemma 3.6(2), it follows that $\overline{p\varpi q}\stackrel{\varepsilon }{%
\Rightarrow }_F\left| p_0\wedge (\stackunder{a\in Act}{\bigwedge }\delta
_{p_0,\overline{p\varpi q}}(a))\right. $. Moreover, by Lemma 7.5, we also
have $v_0\stackunder{\sim }{\sqsubset }_{RS}p_0\wedge (\stackunder{a\in Act}{%
\bigwedge }\delta _{p_0,\overline{p\varpi q}}(a))$. Thus the process $%
p_0\wedge (\stackunder{a\in Act}{\bigwedge }\delta _{p_0,\overline{p\varpi q}%
}(a))$ is indeed the one that we need. \TeXButton{End Proof}{\endproof}

\qquad

It is obvious that the temporal operator $always$ can also be handled in the
recursive manner. Formally, we have the result below.

\quad

{\bf Corollary 7.1 }$\sharp p=_{RS}\overline{p\varpi \bot }$ for any process
$p$.

\TeXButton{Proof}{\proof}Immediately follows from Corollary 6.2(4) and
Theorem 7.1. \TeXButton{End Proof}{\endproof}

\qquad

Hitherto this paper has provided two approaches to dealing with the temporal
modal operator $unless$ in pure process algebraic style. One approach is to
introduce the operators $\varpi $ and $\odot $, and provide SOS rules to
describe their behavior. The other is to define constants $\overline{p\varpi
q}$ in terms of $\eta _{p,q}$. The latter resorts to only usual rules about
recursion, but depends on the finiteness of $Act$ as the definition of $\eta
_{p,q}$ refers to the process having the format $\stackunder{a\in Act}{%
\bigwedge }\left\lceil a\right\rceil p$, which can not be generalized
smoothly to the situation involving infinitely many actions (see, Remark
6.1).

\section{Connections between CLLT and ACTL}

As mentioned in Section 1, the links between process algebras and (modal)
logics have been of concern in the literature. Amongst, Pnueli points out
that [54], given a logic language and a process algebra, interesting
connections between them at least include (see, Section 1):

\begin{itemize}
\item  Hennessy-Milner-style characterization

\item  expressivity of the logic language w.r.t the process algebra

\item  expressivity of the process algebra w.r.t the logic language
\end{itemize}

This section will study the links between two specification formalisms,
namely CLLT and a fragment of ACTL[49], from these three angles. Following
[44], the fragment of ACTL considered in this section, denoted by $\ell $,
consists of all formulas generated by BNF below

\begin{center}
$\phi ::=tt\left| ff\right| en(a)\left| dis(a)\right| \phi \vee \phi \left|
\phi \wedge \phi \right| [a]\phi \left| \Box \phi \right| \phi W\phi $,
where $a\in Act$.
\end{center}

As noticed by L\"uttgen and Vogler, $\ell $ contains essentially the safety
properties of the universal fragment of ACTL [44]. The satisfaction relation
$p\models $ $\phi $, to be read as ``the process $p$ satisfies the formula $%
\phi $'', is given as follows.

\qquad

{\bf Definition 8.1}([44]) The satisfaction relation $\models $ $\subseteq $
$T(\Sigma _{CLLT})\times \ell $ is defined inductively by:

$p\models tt$

\qquad

$p\models $ $ff$ \quad \qquad iff\quad $p\in F$.

\qquad

$p\models $ $en(a)\qquad $iff $\quad \forall p_0(p\stackrel{\varepsilon }{%
\Rightarrow }_F\left| p_0\right. $ $\Rightarrow $ $a\in I(p_0))$.

\qquad

$p\models $ $dis(a)\qquad $iff $\quad \forall p_0(p\stackrel{\varepsilon }{%
\Rightarrow }_F\left| p_0\right. $ $\Rightarrow $ $a\notin I(p_0))$.

\qquad

$p\models $ $\phi \vee \varphi \qquad $iff $\quad \forall p_0(p\stackrel{%
\varepsilon }{\Rightarrow }_F\left| p_0\right. $ $\Rightarrow $ $p_0\models $
$\phi $ or $p_0\models \varphi )$.

\qquad

$p\models $ $\phi \wedge \varphi \qquad $iff $\quad \forall p_0(p\stackrel{%
\varepsilon }{\Rightarrow }_F\left| p_0\right. $ $\Rightarrow $ $p_0\models $
$\phi $ and $p_0\models \varphi )$.

\qquad

$p\models $ $[a]\phi \qquad \quad $iff $\quad \forall p_0$, $p_1(p\stackrel{%
\varepsilon }{\Rightarrow }_F\left| p_0\right. $ $\stackrel{a}{\Rightarrow }%
_F\left| p_1\right. \Rightarrow $ $p_1\models $ $\phi $ $)$.

\qquad

$p\models \Box \phi \qquad \quad $iff $\quad \forall p_0,p_1,...p_k(p%
\stackrel{\varepsilon }{\Rightarrow }_F\left| p_0\right. \stackrel{Act}{%
\Rightarrow }_F\left| p_1\right. ...\stackrel{Act}{\Rightarrow }_F\left|
p_k\right. $ $\Rightarrow $ $p_k\models \phi )$.

\quad

$p\models \phi W\varphi \qquad $iff $\quad \forall p_0,p_1,...p_k\left(
\begin{array}{c}
p
\stackrel{\varepsilon }{\Rightarrow }_F\left| p_0\right. \stackrel{Act}{%
\Rightarrow }_F\left| p_1\right. ...\stackrel{Act}{\Rightarrow }_F\left|
p_k\right. \text{ }\Rightarrow \\ \text{ }p_k\models \phi \text{ or }\exists
i\leq k(p_i\models \varphi )
\end{array}
\right) $.

\qquad

Two simple results immediately follows from the above definition:

\qquad

{\bf Lemma 8.1} For any $p\in T(\Sigma _{CLLT})$ and $\phi \in \ell $, $%
p\models \phi $ if and only if $\forall p_0(p\stackrel{\varepsilon }{%
\Rightarrow }_F\left| p_0\right. $ $\Rightarrow $ $p_0\models \phi )$. In
particular, $p\models \phi $ whenever $p\in F$.

\TeXButton{Proof}{\proof}Easily by induction on $\phi $.
\TeXButton{End Proof}{\endproof}

\quad \qquad

{\bf Lemma 8.2} If $p\sqsubseteq _{RS}q$ then $q\models \phi $ implies $%
p\models \phi $ for each $\phi \in \ell $.

\TeXButton{Proof}{\proof}Straightforward by induction on $\phi $.
\TeXButton{End Proof}{\endproof}

\qquad

The converse of Lemma 8.2 can be proved in the standard manner. Hence we can
get a Hennessy-Milner-style characterization of $\sqsubseteq _{RS}$. In
fact, to obtain such characterization, a fragment of $\ell $ is enough [44].

As argued by Pnueli, Hennessy-Milner-style characterization presents only
the weakest requirement of compatibility between a process calculus and a
logic [54]. The remainder of this section will devote itself to explore
stronger associations between ($T(\Sigma _{CLLT})$, $\sqsubseteq _{RS}$) and
($\ell $, $\models $). Firstly, we consider the expressivity of ($T(\Sigma
_{CLLT})$, $\sqsubseteq _{RS}$) w.r.t $(\ell ,\models )$. The starting point
of our discussion is the notion of a characteristic process.

\qquad

{\bf Definition 8.2} Given a formula $\phi \in \ell ,$ a process $t_\phi \in
T(\Sigma _{CLLT})$ is said to be a characteristic process for $\phi $ if $%
\forall p\in T(\Sigma _{CLLT})(p\models \phi $ $\Leftrightarrow $ $%
p\sqsubseteq _{RS}t_\phi )$. Moreover, ($T(\Sigma _{CLLT})$, $\sqsubseteq
_{RS}$) is said to be expressive w.r.t $(\ell ,\models )$ if there exists a
translation function from $\ell $ to $T(\Sigma _{CLLT})$ which associates
each formula $\phi \in \ell $ with a characteristic process $t_\phi $ in
syntactic manner.

\qquad

Intuitively, the characteristic process $t_\phi $ represents the most loose
process that realizes $\phi $. If such $t_\phi $ exists, verifying the
validity of an assertion $p\models \phi $ may be reduced to the
implementation verification of $p\sqsubseteq _{RS}t_\phi $. It can be showed
without any difficulty that, for any $\phi $, it has at most one
characteristic process modulo $=_{RS}$. In the following, a function $\left[
\cdot \right] :\ell \rightarrow T(\Sigma _{CLLT})$ is provided, which
associates each formula $\phi \in \ell $ with a characteristic process $%
\left[ \phi \right] $.

\quad

{\bf Definition 8.3} The translation function $\left[ \cdot \right] :\ell
\rightarrow T(\Sigma _{CLLT})$ is defined by

\begin{description}
\item  $\left[ ff\right] =$ $\perp \qquad \left[ tt\right] =$ $true\quad
\left[ \phi \wedge \varphi \right] =\left[ \phi \right] \wedge \left[
\varphi \right] \qquad \left[ \phi \vee \varphi \right] =\left[ \phi \right]
\vee \left[ \varphi \right] $

\item  $\left[ en(a)\right] =\stackunder{a\in A\subseteq Act}{\bigvee }(%
\stackunder{b\in A}{\Box }b.true)\qquad \left[ dis(a)\right] =\stackunder{%
a\notin A\subseteq Act}{\bigvee }(\stackunder{b\in A}{\Box }b.true)\qquad $

\item  $\left[ [a]\phi \right] =\left\lceil a\right\rceil \left[ \phi
\right] \qquad \left[ {\bf \Box }\phi \right] =\sharp \left[ \phi \right]
\qquad \left[ \phi W\varphi \right] =\left[ \phi \right] \varpi \left[
\varphi \right] $
\end{description}

\quad

The above definition is motivated by L\"uttgen and Vogler's construction. In
the framework of LLTS, they have given the method of embedding of formulas
(in $\ell $) into LLTS [44].

\qquad

{\bf Lemma 8.3} If $p\sqsubseteq _{RS}p_i$ for some $i\in \{1,2\}$ then $%
p\sqsubseteq _{RS}p_1\vee p_2$. Moreover, the converse also holds whenever $%
p $ is stable.

\TeXButton{Proof}{\proof}Straightforward. \TeXButton{End Proof}{\endproof}

\quad

Notice that the assumption that $p$ is stable is necessary for the converse
implication in the above. For instance, $a.0\vee b.0\sqsubseteq _{RS}a.0\vee
b.0$ but neither $a.0\vee b.0\sqsubseteq _{RS}a.0\ $nor $a.0\vee
b.0\sqsubseteq _{RS}b.0$. Next we intend to show that, given a $\varphi \in
\ell $, $\left[ \varphi \right] $ indeed is the characteristic process of $%
\varphi $, which, as the most important result in [44], have been obtained
by L\"uttgen and Vogler in the framework of LLTS.

\qquad

{\bf Lemma 8.4} For any $\varphi \in \ell $, $\left[ \varphi \right] $ is
the characteristic process of $\varphi $.

\TeXButton{Proof}{\proof}It is enough to prove that, $p\models \varphi $ if
and only if $p\sqsubseteq _{RS}\left[ \varphi \right] $ for any $p\in
T(\Sigma _{CLLT})$ and $\varphi \in \ell $. This can be proved by induction
on $\varphi $. Here we do not present them in full detail but handle three
cases as samples. In particular, for the case where $\varphi $ has one of
formats $[a]\phi $, $\Box \phi $ and $\phi _1W\phi _2$, the proof is
straightforward by applying Theorem 6.1, 4.1 and 5.1 respectively.

\begin{itemize}
\item  $\varphi \equiv tt$
\end{itemize}

The implication from right to left follows trivially from Definition 8.1.
For the converse implication, it suffices to prove $p\sqsubseteq _{RS}true$.
Let $p\stackrel{\varepsilon }{\Rightarrow }_F\left| p_0\right. $. Clearly, $%
true\stackrel{\tau }{\rightarrow }_F|\stackunder{a\in I(p_0)}{\Box }a.true$,
moreover, we also have $p_0\stackunder{\sim }{\sqsubset }_{RS}\stackunder{%
a\in I(p_0)}{\Box }a.true$ by Lemma 6.1.

\begin{itemize}
\item  $\varphi \equiv en(a)$
\end{itemize}

(Left implies Right) Let $p\stackrel{\varepsilon }{\Rightarrow }_F\left|
p_0\right. $. Then it follows from $p\models en(a)$ that $a\in I(p_0)$. Thus
$\left[ en(a)\right] \equiv \stackunder{a\in A\subseteq Act}{\bigvee }(%
\stackunder{b\in A}{\Box }b.true)\stackrel{\varepsilon }{\Rightarrow }_F|%
\stackunder{b\in I(p_0)}{\Box }b.true$. Moreover, by Lemma 6.1, $p_{_0}$ $%
\stackunder{\sim }{\sqsubset }_{RS}\stackunder{b\in I(p_0)}{\Box }b.true$.

(Right implies Left) Let $p\stackrel{\varepsilon }{\Rightarrow }_F\left|
p_0\right. $. It suffices to show that $a\in I(p_0)$. Since $p\sqsubseteq
_{RS}\left[ en(a)\right] \equiv \stackunder{a\in A\subseteq Act}{\bigvee }(%
\stackunder{b\in A}{\Box }b.true)$, we get $p_0\stackunder{\sim }{\sqsubset }%
_{RS}\stackunder{b\in A_0}{\Box }b.true$ for some $A_0$ with $a\in A_0$.
Then, due to $p_0\notin F$, we have $I(p_0)=I(\stackunder{b\in A_0}{\Box }%
b.true)=A_0$. Hence $a\in I(p_0)$.

\begin{itemize}
\item  $\varphi \equiv $ $\phi _1\vee \phi _2$
\end{itemize}

$p\models $ $\phi _1\vee \phi _2$

$\Leftrightarrow \forall p_0(p\stackrel{\varepsilon }{\Rightarrow }_F\left|
p_0\right. $ $\Rightarrow $ $p_0\models $ $\phi _1$ or $p_0\models \phi _2)$

$\Leftrightarrow \forall p_0(p\stackrel{\varepsilon }{\Rightarrow }_F\left|
p_0\right. $ $\Rightarrow $ $p_0\sqsubseteq _{RS}\left[ \phi _1\right] $ or $%
p_0\sqsubseteq _{RS}\left[ \phi _2\right] )$ \quad $\qquad \quad $(by IH)

$\Leftrightarrow \forall p_0(p\stackrel{\varepsilon }{\Rightarrow }_F\left|
p_0\right. $ $\Rightarrow $ $p_0\sqsubseteq _{RS}\left[ \phi _1\right] $ $%
\vee \left[ \phi _2\right] )\qquad \qquad \qquad \qquad $(by Lemma 8.3)

$\Leftrightarrow \forall p_0(p\stackrel{\varepsilon }{\Rightarrow }_F\left|
p_0\right. $ $\Rightarrow $ $p_0\sqsubseteq _{RS}\left[ \phi _1\vee \phi
_2\right] $ $)$

$\Leftrightarrow p\sqsubseteq _{RS}\left[ \phi _1\vee \phi _2\right] $%
.\qquad \qquad \qquad \qquad \qquad \qquad \TeXButton{End Proof}{\endproof}

\qquad

As usual, for any formula $\phi $ and $\varphi $, $\varphi $ is said to be a
logic consequence of $\phi $, in symbols $\phi \models $ $\varphi $, if for
any process $p$, $p\models $ $\phi $ implies $p\models $ $\varphi $.
Moreover, $\phi $ and $\varphi $ are said to be logic equivalent if $\phi
\models $ $\varphi $ and $\varphi \models $ $\phi $. As an immediate
consequence of the above theorem, we have the result below.

\qquad

{\bf Corollary 8.1} For any formula $\phi $ and $\varphi $ in $\ell $,

(1) $\left[ \phi \right] \models $ $\phi .$

(2) $\phi \models $ $\varphi $ if and only if $\left[ \phi \right] $ $%
\sqsubseteq _{RS}\left[ \varphi \right] $.

\TeXButton{Proof}{\proof}(1) immediately follows from $\left[ \phi \right]
\sqsubseteq _{RS}\left[ \phi \right] $ and Lemma 8.4. (2) follows from (1),
the transitivity of $\sqsubseteq _{RS}$ and Lemma 8.4. \TeXButton{End Proof}
{\endproof}

\qquad

Moreover, since the function $\left[ \cdot \right] :\ell \rightarrow
T(\Sigma _{CLLT})$ is given in syntactic manner, we have the result below.

\quad

{\bf Theorem 8.1 }($T(\Sigma _{CLLT})$, $\sqsubseteq _{RS}$) is expressive
w.r.t $(\ell ,\models )$.

\qquad

We next deal with another stronger connection between CLLT and $(\ell
,\models )$, which involves the fragment $T(\Sigma _{CLLT})^{-}$ of $%
T(\Sigma _{CLLT})$ defined below.

\qquad

{\bf Definition 8.4} $T(\Sigma _{CLLT})^{-}$ consists of processes generated
by BNF below, where $A\subseteq Act$ and $a\in Act$.

\begin{description}
\item  \begin{center}
$p::=0\mid \bot \mid true\mid a.p\mid p\vee p\mid p\wedge p\mid \stackunder{%
b\in A}{\Box }b.true\mid \sharp p\mid p\varpi p\mid (\stackunder{a\neq b\in A%
}{\Box }b.true)\Box a.p$
\end{center}
\end{description}

\qquad

In the following, we intend to prove that $(\ell ,\models )$ is expressive
w.r.t ($T(\Sigma _{CLLT})^{-}$, $\sqsubseteq _{RS}$). Analogous to [54],
such notion is defined formally as follows.

\qquad

{\bf Definition 8.5} $(\ell ,\models )$ is said to be expressive w.r.t ($%
T(\Sigma _{CLLT})^{-}$, $\sqsubseteq _{RS}$) if for any process $p$ in $%
T(\Sigma _{CLLT})^{-}$, there exists a formula $\phi _p$ in $\ell $ such that

\begin{description}
\item  {\bf (E1)} $\forall q\in T(\Sigma _{CLLT})($ $q\sqsubseteq
_{RS}p\Leftrightarrow q\models \phi _p)$, and

\item  {\bf (E2)} $\forall \varphi \in \ell (p\models \varphi $ $%
\Leftrightarrow $ $\phi _p\models \varphi )$.
\end{description}

\quad

Obviously, given a process $p$, $\phi _p$ (if it exists) is a characteristic
formula for $p$ due to (E1), moreover, it is the strongest logic formula $%
\phi $ in $\ell $ such that $p\models $ $\phi $ due to (E2). In order to
prove that $(\ell ,\models )$ is expressive w.r.t ($T(\Sigma _{CLLT})^{-}$, $%
\sqsubseteq _{RS}$), we will introduce the function $*$ below, and show that
it is exactly the lower adjoint of the function $\left[ \cdot \right] $ and
associates each process $p\in $ $T(\Sigma _{CLLT})^{-}$ with a
characteristic formula $p^{*}$.

\qquad

{\bf Definition 8.6} The translation function $*:T(\Sigma
_{CLLT})^{-}\rightarrow \ell $ is defined inductively by

\begin{description}
\item  $\perp ^{*}=ff\quad \quad true^{*}=tt\quad \quad
(a.p)^{*}=en(a)\wedge [a]p^{*}\wedge \stackunder{a\neq b\in Act}{\bigwedge }%
dis(b)$\qquad

\item  $0^{*}=\stackunder{a\in Act}{\bigwedge }dis(a)\quad \quad (%
\stackunder{b\in A}{\Box }b.true)^{*}=(\stackunder{b\in A}{\bigwedge }%
en(b))\wedge (\stackunder{a\in Act-A}{\bigwedge }dis(a))$\quad

\item  $(p\wedge q)^{*}=p^{*}\wedge q^{*}\quad \quad (p\vee q)^{*}=p^{*}\vee
q^{*}\quad \quad (\sharp p)^{*}=\Box p^{*}\quad \quad (p\varpi
q)^{*}=p^{*}Wq^{*}$

\item  $(\stackunder{a\neq b\in A}{\Box }b.true\Box a.p)^{*}=(\stackunder{%
b\in A\cup \{a\}}{\Box }b.true)^{*}\wedge [a]p^{*}$
\end{description}

\qquad

{\bf Lemma 8.5} $p\sqsubseteq _{RS}q$ if and only if $p\models q^{*}$ for
any $p\in T(\Sigma _{CLLT})$ and $q\in T(\Sigma _{CLLT})^{-}$.

\TeXButton{Proof}{\proof}Clearly, it holds trivially whenever $p\in F$. In
the following, we consider the nontrivial case $p\notin F$, and proceed by
induction on $q$.

\begin{itemize}
\item  $q\equiv \bot $
\end{itemize}

It follows from $p\notin F$ that $p\not \sqsubseteq _{RS}\bot $ and $p\not
\models ff$ . Hence $p\sqsubseteq _{RS}\bot $ $\Leftrightarrow $ $p\models
ff $.

\begin{itemize}
\item  $q\equiv true$
\end{itemize}

Immediately follows from $p\sqsubseteq _{RS}true$ and $p\models tt$ for each
$p$ .

\begin{itemize}
\item  $q\equiv 0$
\end{itemize}

$p\sqsubseteq _{RS}0$

$\Leftrightarrow \forall p_0(p\stackrel{\varepsilon }{\Rightarrow }_F\left|
p_0\right. $ $\Rightarrow $ $I(p_0)=\emptyset )$.

$\Leftrightarrow $ $\forall p_0(p\stackrel{\varepsilon }{\Rightarrow }%
_F\left| p_0\right. $ $\Rightarrow $ $\forall a\in Act(a\notin I(p_0))$ $%
)\qquad \qquad \qquad \qquad \quad $(due to $p_0\stackrel{\tau }{\not
\rightarrow }$ )

$\Leftrightarrow $ $\forall p_0(p\stackrel{\varepsilon }{\Rightarrow }%
_F\left| p_0\right. $ $\Rightarrow $ $\forall a\in Act(p_0\models dis(a)$ $%
)) $

$\Leftrightarrow $ $\forall p_0(p\stackrel{\varepsilon }{\Rightarrow }%
_F\left| p_0\right. $ $\Rightarrow $ $p_0\models \stackunder{a\in Act}{%
\bigwedge }dis(a)$ $)$

$\Leftrightarrow $ $p\models \stackunder{a\in Act}{\bigwedge }dis(a)$\qquad $%
\ \qquad \qquad \ \qquad \qquad \ \qquad \ \qquad \qquad \qquad $(by Lemma
8.1)

\begin{itemize}
\item  $q\equiv \stackunder{b\in A}{\Box }b.true$
\end{itemize}

$p\sqsubseteq _{RS}\stackunder{b\in A}{\Box }b.true$

$\Leftrightarrow \forall p_0(p\stackrel{\varepsilon }{\Rightarrow }_F\left|
p_0\right. $ $\Rightarrow $ $I(p_0)=A)$

$\Leftrightarrow \forall p_0(p\stackrel{\varepsilon }{\Rightarrow }_F\left|
p_0\right. $ $\Rightarrow $ $A\subseteq I(p_0)$ and $(Act-A)\cap
I(p_0)=\emptyset )$

$\Leftrightarrow \forall p_0(p\stackrel{\varepsilon }{\Rightarrow }_F\left|
p_0\right. $ $\Rightarrow $ $p_0\models \stackunder{a\in A}{\bigwedge }en(a)$
and $p_0\models \stackunder{b\in Act-A}{\bigwedge }dis(b))$

$\Leftrightarrow \forall p_0(p\stackrel{\varepsilon }{\Rightarrow }_F\left|
p_0\right. $ $\Rightarrow $ $p_0\models \stackunder{a\in A}{\bigwedge }en(a)$
$\wedge \stackunder{b\in Act-A}{\bigwedge }dis(b))$

$\Leftrightarrow $ $p\models \stackunder{a\in A}{\bigwedge }en(a)\wedge
\stackunder{b\in Act-A}{\bigwedge }dis(b)$\qquad $\ \qquad \qquad \ \qquad \
\qquad \qquad $(by Lemma 8.1)

\begin{itemize}
\item  $q\equiv a.q_1$
\end{itemize}

$p\sqsubseteq _{RS}a.q_1$

$\Leftrightarrow $ $\forall p_0(p\stackrel{\varepsilon }{\Rightarrow }%
_F\left| p_0\right. $ $\Rightarrow $ $p_0\stackunder{\sim }{\sqsubset }%
_{RS}a.q_1)$

\quad

$\stackrel{(\clubsuit )}{\Leftrightarrow }$ $\forall p_0\left( p\stackrel{%
\varepsilon }{\Rightarrow }_F\left| p_0\right. \Rightarrow \left(
\begin{array}{c}
a\in I(p_0)
\text{ and }\forall b\in Act(a\neq b\Rightarrow b\notin I(p_0)) \\ \text{and
}\forall p_1(p_0\stackrel{a}{\Rightarrow }_F\left| p_1\right. \Rightarrow
p_1\sqsubseteq _{RS}q_1)
\end{array}
\right) \right) $

\quad

$\Leftrightarrow $ $\forall p_0\left( p\stackrel{\varepsilon }{\Rightarrow }%
_F\left| p_0\right. \Rightarrow \left(
\begin{array}{c}
p_0\models en(a)
\text{ and }p_0\models \stackunder{a\neq b\in Act}{\bigwedge }dis(b)) \\
\text{and }\forall p_1(p_0\stackrel{a}{\Rightarrow }_F\left| p_1\right.
\Rightarrow p_1\models q_1^{*})
\end{array}
\right) \right) $

\quad

$\Leftrightarrow $ $\forall p_0\left( p\stackrel{\varepsilon }{\Rightarrow }%
_F\left| p_0\right. \Rightarrow \left(
\begin{array}{c}
p_0\models en(a)
\text{ and }p_0\models \stackunder{a\neq b\in Act}{\bigwedge }dis(b)) \\
\text{and }p_0\models [a]q_1^{*})
\end{array}
\right) \right) $

$\Leftrightarrow $ $\forall p_0\left( p\stackrel{\varepsilon }{\Rightarrow }%
_F\left| p_0\right. \Rightarrow \left( p_0\models en(a)\wedge
[a]q_1^{*}\wedge \stackunder{a\neq b\in Act}{\bigwedge }dis(b)\right)
\right) $

$\Leftrightarrow $ $p\models en(a)\wedge [a]q_1^{*}\wedge \stackunder{a\neq
b\in Act}{\bigwedge }dis(b)$\qquad $\ \qquad \qquad \ \qquad \qquad \ $(by
Lemma 8.1)

($\clubsuit $) For the implication from right to left, we need to show that $%
a.q_1\notin F$ under the assumption $p\stackrel{\varepsilon }{\Rightarrow }%
_F\left| p_0\right. $. By Lemma 3.8 and 3.5, it follows from $p_0\notin F$
and $a\in I(p_0)$ that $p_0\stackrel{a}{\Rightarrow }_F\left| p_1\right. $
for some $p_1$. Hence $p_1\sqsubseteq _{RS}q_1$. Then $q_1\notin F$ because
of $p_1\notin F$. Thus $a.q_1\notin F$ by Lemma 3.3(2).

\begin{itemize}
\item  $q\equiv \sharp q_1$ or $q_1\varpi q_2$
\end{itemize}

Immediately follows from Theorem 4.1, 5.1 and IH.

\begin{itemize}
\item  $q\equiv q_1\wedge q_2$
\end{itemize}

$p\sqsubseteq _{RS}$ $q_1\wedge q_2$

$\Leftrightarrow \forall p_0(p\stackrel{\varepsilon }{\Rightarrow }_F\left|
p_0\right. $ $\Rightarrow $ $p_0\sqsubseteq _{RS}$ $q_1\wedge q_2)$

$\Leftrightarrow \forall p_0(p\stackrel{\varepsilon }{\Rightarrow }_F\left|
p_0\right. $ $\Rightarrow $ $p_0\sqsubseteq _{RS}$ $q_1$ and $p_0\sqsubseteq
_{RS}q_2)$ $\ $ $\qquad $ $\qquad \ \quad $(by Lemma 3.11)

$\Leftrightarrow \forall p_0(p\stackrel{\varepsilon }{\Rightarrow }_F\left|
p_0\right. $ $\Rightarrow $ $p_0\models q_1^{*}$ and $p_0\models q_2^{*})$
\quad $\ \qquad \qquad \ \qquad \quad $(by IH)

$\Leftrightarrow \forall p_0(p\stackrel{\varepsilon }{\Rightarrow }_F\left|
p_0\right. $ $\Rightarrow $ $p_0\models q_1^{*}\wedge q_2^{*})\qquad $

$\Leftrightarrow \forall p_0(p\stackrel{\varepsilon }{\Rightarrow }_F\left|
p_0\right. $ $\Rightarrow $ $p_0\models (q_1\wedge q_2)^{*}$ $)$

$\Leftrightarrow p\models (q_1\wedge q_2)^{*}$.\qquad \qquad \qquad \qquad
\qquad $\ \qquad \qquad \ \qquad \ \qquad $(by Lemma 8.1)\qquad

\begin{itemize}
\item  $q\equiv $ $q_1\vee q_2$
\end{itemize}

$p\sqsubseteq _{RS}$ $q_1\vee q_2$

$\Leftrightarrow \forall p_0(p\stackrel{\varepsilon }{\Rightarrow }_F\left|
p_0\right. $ $\Rightarrow $ $p_0\sqsubseteq _{RS}$ $q_1$ $\vee q_2)$

$\Leftrightarrow \forall p_0(p\stackrel{\varepsilon }{\Rightarrow }_F\left|
p_0\right. $ $\Rightarrow $ $p_0\sqsubseteq _{RS}$ $q_1$ or $p_0\sqsubseteq
_{RS}q_2)\ \qquad \ \qquad \qquad $(by Lemma 8.3)

$\Leftrightarrow \forall p_0(p\stackrel{\varepsilon }{\Rightarrow }_F\left|
p_0\right. $ $\Rightarrow $ $p_0\models q_1^{*}$ or $p_0\models q_2^{*})$
\quad $\ \qquad \qquad \ \qquad \quad $(by IH)

$\Leftrightarrow \forall p_0(p\stackrel{\varepsilon }{\Rightarrow }_F\left|
p_0\right. $ $\Rightarrow $ $p_0\models q_1^{*}\vee q_2^{*})\qquad $

$\Leftrightarrow \forall p_0(p\stackrel{\varepsilon }{\Rightarrow }_F\left|
p_0\right. $ $\Rightarrow $ $p_0\models (q_1\vee q_2)^{*}$ $)$

$\Leftrightarrow p\models (q_1\vee q_2)^{*}$.\qquad \qquad \qquad \qquad $\
\qquad \qquad \ \qquad \qquad \ \ \qquad $(by Lemma 8.1)

\begin{itemize}
\item  $q\equiv \stackunder{a\neq b\in A}{\Box }b.true\Box a.q_1$
\end{itemize}

$p\sqsubseteq _{RS}\stackunder{a\neq b\in A}{\Box }b.true\Box a.q_1$

$\Leftrightarrow \forall p_0(p\stackrel{\varepsilon }{\Rightarrow }_F\left|
p_0\right. $ $\Rightarrow $ $p_0\stackunder{\sim }{\sqsubset }_{RS}%
\stackunder{a\neq b\in A}{\Box }b.true\Box a.q_1$ $)$

$\stackrel{(\spadesuit )}{\Leftrightarrow }\forall p_0(p\stackrel{%
\varepsilon }{\Rightarrow }_F\left| p_0\right. $ $\Rightarrow $ $%
I(p_0)=A\cup \{a\}$ and $\forall $ $p_1(p_0\stackrel{a}{\Rightarrow }%
_F\left| p_1\right. $ $\Rightarrow $ $p_1\sqsubseteq _{RS}q_1))$

$\Leftrightarrow \forall p_0(p\stackrel{\varepsilon }{\Rightarrow }_F\left|
p_0\right. $ $\Rightarrow $ $p_0\stackunder{\sim }{\sqsubset }_{RS}%
\stackunder{b\in A\cup \{a\}}{\Box }b.true$ and $\forall p_1(p_0\stackrel{a}{%
\Rightarrow }_F\left| p_1\right. \Rightarrow p_1\models q_1^{*}))$

$\Leftrightarrow \forall p_0(p\stackrel{\varepsilon }{\Rightarrow }_F\left|
p_0\right. $ $\Rightarrow $ $p_0\models (\stackunder{b\in A\cup \{a\}}{\Box }%
b.true)^{*}$ and $p_0\models [a]q_1^{*})$

$\Leftrightarrow $ $p\models (\stackunder{b\in A\cup \{a\}}{\Box }%
b.true)^{*} $ $\wedge [a]q_1^{*}$\qquad

($\spadesuit $) For the implication from right to left, it is required to
verify $\stackunder{a\neq b\in A}{\Box }b.true\Box a.q_1$ $\notin F$ under
the assumption $p\stackrel{\varepsilon }{\Rightarrow }_F\left| p_0\right. $.
Clearly, it suffices to prove that $q_1\notin F$, which can be proved
analogously to ($\clubsuit $). \TeXButton{End Proof}{\endproof}

\qquad

As an immediate consequence of the above result, we have

\qquad

{\bf Corollary 8.2} For any process $p$ and $q$ in $T(\Sigma _{CLLT})^{-}$,

(1) $p\models $ $p^{*}$

(2) $p\sqsubseteq _{RS}q$ if and only if $p^{*}\models $ $q^{*}$ .

\TeXButton{Proof}{\proof}(1) immediately follows from $p\sqsubseteq _{RS}p$
and Lemma 8.5. (2) follows from (1), the transitivity of $\sqsubseteq _{RS}$
and Lemma 8.5. \TeXButton{End Proof}{\endproof}

\qquad

In order to prove that $(\ell ,\models )$ is expressive w.r.t ($T(\Sigma
_{CLLT})^{-}$, $\sqsubseteq _{RS}$), the only point remaining concerns (E2),
that is, $p\models \varphi $ iff $p^{*}\models \varphi $ for any $p\in
T(\Sigma _{CLLT})^{-}$ and $\varphi \in \ell $. Before proving it, let we
recall the well-known notion of a Galois connection between two preordered
sets.

\qquad

{\bf Definition 8.7} A Galois connection between two preordered sets $(A$, $%
\preceq _A)$ and $(B$, $\preceq _B)$ is a pair of function $F:B\rightarrow A$
and $G:A\rightarrow B$ satisfying that, for any $x\in B$ and $y\in A$, $%
F(x)\preceq _Ay$ if and only if $x\preceq _BG(y)$.

\qquad

It is well known that $(F,G)$ is a Galois connection if and only if $F$ and $%
G$ are monotonic and satisfy the {\em cancellation} laws below (see for
instance [6])

\begin{description}
\item  (C1) $x\preceq _BG(F(x))$ for all $x\in B$, and

\item  (C2) $F(G(y))\preceq _Ay$ for all $y\in A$.
\end{description}

Using Lemma 8.4, it is easy to see that (E2) holds if and only if the pair $%
(*,$ $\left[ \cdot \right] )$ is a Galois connection between preordered sets
$\left\langle \ell ,\text{ }\models \right\rangle $ and $\left\langle
T(\Sigma _{CLLT})^{-},\text{ }\sqsubseteq _{RS}\right\rangle $. Next we
shall prove the latter.

\qquad \quad

{\bf Theorem 8.2 (Galois connection)} The pair of functions $*:T(\Sigma
_{CLLT})^{-}\rightarrow \ell $ and $\left[ \cdot \right] :\ell \rightarrow
T(\Sigma _{CLLT})^{-}$ is a Galois connection between preordered sets $%
\left\langle \ell ,\text{ }\models \right\rangle $ and $\left\langle
T(\Sigma _{CLLT})^{-},\text{ }\sqsubseteq _{RS}\right\rangle $. That is, $%
p^{*}\models \phi $ if and only if $p\sqsubseteq _{RS}\left[ \phi \right] $
for any $p\in T(\Sigma _{CLLT})^{-}$ and $\phi \in \ell $.

\TeXButton{Proof}{\proof}By Definition 8.3, 8.4 and 6.3, it is easy to check
that $\left[ \phi \right] \in T(\Sigma _{CLLT})^{-}$ for any $\phi \in \ell $%
. Thus the function $\left[ \cdot \right] $ may be regarded as a function
from $\ell $ to $T(\Sigma _{CLLT})^{-}$. On the other hand, by Corollary 8.1
and 8.2, both the function $*$ and $\left[ \cdot \right] $ are monotonic.
Thus it suffices to prove that cancellation laws (C1) and (C2) hold.

For (C1), suppose $p\in T(\Sigma _{CLLT})^{-}$. By Corollary 8.2, we get $%
p\models $ $p^{*}$. Then $p\sqsubseteq _{RS}\left[ p^{*}\right] $ by Lemma
8.4. Hence (C1) holds.

For (C2), let $\phi \in \ell $. We intend to prove that $\left[ \phi \right]
^{*}\models \phi $. Let $q$ be any process such that $q\models \left[ \phi
\right] ^{*}$. To complete the proof, it is enough to verify that $q\models
\phi $. By Corollary 8.1, we obtain $\left[ \phi \right] \models $ $\phi $.
Moreover, by Lemma 8.5, it follows from $q\models \left[ \phi \right] ^{*}$
that $q\sqsubseteq _{RS}\left[ \phi \right] $. Hence $q\models \phi $ by
Lemma 8.2, as desired. \TeXButton{End Proof}{\endproof}

\qquad

Roughly speaking, the above theorem says that the function $*$ is exactly
the lower adjoint of the function $\left[ \cdot \right] $. That is, for each
process $p\in T(\Sigma _{CLLT})^{-}$, $p^{*}$ is the strongest logic formula
$\phi $ in $\ell $ such that $p\sqsubseteq _{RS}\left[ \phi \right] $,
dually, the function $\left[ \cdot \right] $ associates with each formula $%
\phi $ in $\ell $ the most loose process $p\in T(\Sigma _{CLLT})^{-}$ such
that $p^{*}\models \phi $. As an immediate consequence of Theorem 8.2, we
obtain the assertion below.

\qquad

{\bf Theorem 8.3 }$(\ell ,\models )$ is expressive w.r.t ($T(\Sigma
_{CLLT})^{-}$, $\sqsubseteq _{RS}$).

\TeXButton{Proof}{\proof}Let $p\in T(\Sigma _{CLLT})^{-}$. It suffices to
illustrate that $p^{*}$ satisfies (E1) and (E2) in Definition 8.5. Clearly,
(E1) holds due to Lemma 8.5, and (E2) comes from Theorem 8.2 and Lemma 8.4.
\TeXButton{End Proof}{\endproof}

\qquad

By the way, it is obvious that, for CLLT$_\eta $, all results obtained in
this section also hold by making a few slight modifications.

\section{Conclusions and future work}

This paper gives two distinct methods of representing the loosest (modulo $%
\sqsubseteq _{RS}$) implementations that realize logic specifications ``$%
always$ $p$'' or ``$p$ $unless$ $q$'' in terms of algebraic expressions. One
method is to introduce nonstandard process-algebraic operators $\sharp $, $%
\varpi $, $\bigtriangleup $ and $\odot $ to capture L\"uttgen and Vogler's
constructions in [44] directly. The other is to apply the greatest
fixed-point characterization of $\varpi $ and $\sharp $ obtained in this
paper (see, Theorem 6.2 and Corollary 6.3) and provide graphical
representing of temporal operators $always$ and $unless$ in a recursive
manner. The latter is independent of L\"uttgen and Vogler's constructions,
and its advantage lies in the fact that it makes no appeal to any
nonstandard operational operators, but it depends on the mild assumption
that $Act$ is finite. In a word, this paper not only lifts L\"uttgen and
Vogler's work in [44] to a pure process algebraic setting but also provides
another more succinct method to realize their intention.

This work brings the process calculuses CLLT in which usual operational
operators ($prefix$, $external$ $choice$ and $parallel$ $operator$), logic
connectives ($conjunction$ and $disjunction$) and standard temporal
operators ($always$ and $unless$) may be freely mixed without any
restriction, and compositional reasoning is admitted. Such calculus allows
one to capture desired operational behavior and describe intended safety
properties in the same framework. Moreover, the links between CLLT and the
fragment $\ell $ of ACTL are explored from angles suggested by Pnueli in
[54]. These links reveal that there exist intimate relationships among
distinct verification activities including model checking, implementation
verification and validity problem within $\ell $. We summarize the
reductions among these verification activities in Fig.1, where dashed lines
are used to indicate that the process term involved in the corresponding
reduction is required to be in $T(\Sigma _{CLLT})^{-}$.

\begin{figure}
\begin{center}
\setlength{\unitlength}{1cm}
\begin{picture}(15,4)
\put(0,4){\QTR{bf}{Implementation Verification  }}
\put(9,4){\QTR{bf}{Model Checking}}
\put(3.5,0){\QTR{bf}{Validity Problem within $\ell $}}
\put(6.5,4.4){Lemma 8.4}
\put(6.5,3.5){Lemma 8.5}

\put(8.8,4.2){\vector(-1,0){3.5}}
\multiput(5.3,4)(0.6,0){5}{\line(1,0){0.5}}
\put(8.3,4){\vector(1,0){0.5}}

\put(5,0.5){\vector(-1,1){3.1}}
\put(7,0.5){\vector(1,1){3.1}}

\multiput(9.6,3.5)(-0.5,-0.5){5}{\line(-1,-1){0.4}}
\put(7.15,1.05){\vector(-1,-1){0.5}}

\put(5.25,1.7){\makebox{$\begin{array}{c}\text{Corollary 8.2(2)} , \\ \text{Lemma 8.4}\end{array}$}}

\put(8.5,1.7){\makebox{$\begin{array}{c}\text{Corollary 8.1(2)} , \\ \text{Lemma 8.5}\end{array}$}}
\put(1,1.7){\makebox{Corollary 8.1(2)}}
\end{picture}
\end{center}
\caption{\QTR{bf}{connections between distinct verification activities}}
 \end{figure}
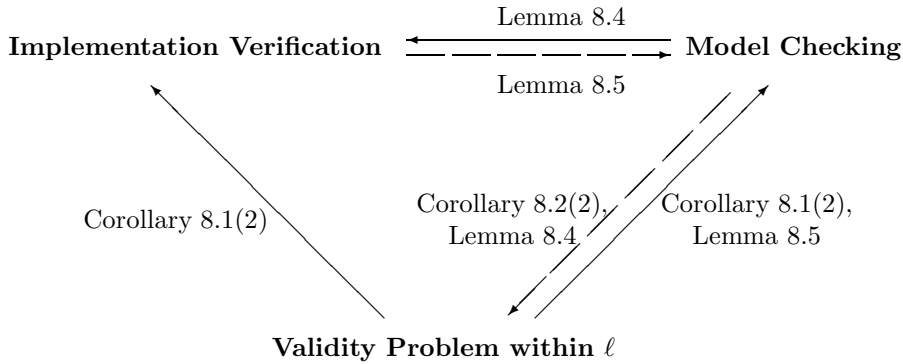

In the literature, various work on combining operational operators with
logic operators have been reported [33, 41, 50]. Olderog provides a
framework in which operational operators may be combined with trace formula
[50]. But such framework does not allow one to freely mix operational and
logic specifications. Guerra and Costa enrich a simple process algebra with
a modal operator which can express some liveness property [33]. However, due
to adopting trace semantics, this system is not deadlock-sensitive, and
hence it is inadequate in the situation where concurrency is involved. In
[41], based on the notion of modal LTS, Larsen et al. consider the operator
conjunction over independent processes and obtain the result analogous to
Lemma 3.11. Moreover, in such framework, it is shown that conjunction may
distribute over parallel composition. However, an algebraic theory of mixing
operational and logic operators is not considered in [41]. There also exist
investigations of operational behavior involving logic ingredient but
without admitting the free mixing of operational and logic operators, see,
e.g., [7, 22].

We conclude this paper with giving several possible avenues for further
work. Firstly, finding a complete proof system for CLLT would be the next
task. Secondly, although this paper provides recursive constants to
represent the ``loosest'' implementations realizing logic specifications ``$%
always$ $p$'' or ``$p$ $unless$ $q$'', no attempt has made here to develop
general theory concerning recursion for LLTS and a few fundamental problems
are still open. For instance, whether $\sqsubseteq _{RS}$ is precongruent in
the presence of (nested) recursive operator? Under usual conditions (see,
e.g., [45]), whether equations containing (nested) recursive operator still
have a unique solution? Notice that, since LLTS involve consideration of
inconsistencies, the answers for these questions can not be trivially
inferred from existent results in the literature. Thirdly, it would also be
interesting to develop a general view of the connections between process
algebras and modal logics. We leave these further developments for further
work.

\end{document}